\documentclass[preprint2]{aastex62}
\graphicspath{ {./} }
\usepackage{xspace, url, amsmath, multirow}

\newcommand{\FSPS}{{\sc FSPS}\xspace}
\newcommand{\pFSPS}{{\tt \textbf{python-fsps}}\xspace}
\newcommand{\CloudyFSPS}{{\tt \textbf{CloudyFSPS}}\xspace}

\newcommand{\Cloudy}{\textsc{Cloudy}\xspace}

\newcommand{\hii}{\ion{H}{2}}
\newcommand{\nii}{[\ion{N}{2}]}

\newcommand{\sii}{[\ion{S}{2}]}

\newcommand{\oiii}{[\ion{O}{3}]}
\newcommand{\oii}{[\ion{O}{2}]}

\newcommand{\heii}{\ion{He}{2}}

\newcommand{\civ}{\ion{C}{4}}

\newcommand{\SiuIII}{[\ion{Si}{3}]}

\newcommand{\ciii}{\ion{C}{3}]}

\newcommand\vs{\ensuremath{\mathrm{vs.}}\xspace}

\newcommand\Msun{\ensuremath{\mathrm{M_{\sun}}}\xspace}
\newcommand{\ha}{\ensuremath{\mathrm{H\alpha}}\xspace}
\newcommand{\hb}{\ensuremath{\mathrm{H\beta}}\xspace}
\newcommand{\Myr}{$\,$Myr\xspace}
\newcommand{\logten}{\ensuremath{\log_{10}}}
\newcommand{\logZ}{\ensuremath{\logten \mathrm{Z}/\mathrm{Z}_{\sun}}\xspace}
\newcommand{\logOH}{\ensuremath{\logten (\mathrm{O}/\mathrm{H})}\xspace}
\newcommand{\logZeq}[1]{\ensuremath{\logten \mathrm{Z}/\mathrm{Z}_{\sun} = #1}}
\newcommand{\ang}{\ensuremath{\mbox{\AA}}\xspace}

\newcommand{\U}{\ensuremath{\mathcal{U}_{0}}\xspace}
\newcommand{\logU}{\ensuremath{\logten \mathcal{U}_0}}
\newcommand{\logUeq}[1]{\ensuremath{\logten \mathcal{U}_0 = #1}}

\newcommand{\mage}{{\sc Meg}a{\sc S}a{\sc ura}\xspace}
\newcommand{\Te}{\ensuremath{T_{\mathrm{e}}}\xspace}

\newcommand{\SN}{\ensuremath{\mathrm{S}/\mathrm{N}}\xspace}
\defcitealias{Byler+2018}{B18}
\accepted{March 9, 2020}
\shorttitle{}
\shortauthors{Byler et al.}
\begin{document}
\title{A comparison of UV and optical metallicities in star-forming galaxies}
\correspondingauthor{Nell Byler}
\email{nell.byler@anu.edu.au}
\author[0000-0002-7392-3637]{Nell Byler}
\affil{Research School of Astronomy and Astrophysics, The Australian National University, ACT, Australia}
\affil{ARC Centre of Excellence for All Sky Astrophysics in 3 Dimensions (ASTRO 3D), Australia}
\author[0000-0001-8152-3943]{Lisa J. Kewley}
\affil{Research School of Astronomy and Astrophysics, The Australian National University, ACT, Australia}
\affil{ARC Centre of Excellence for All Sky Astrophysics in 3 Dimensions (ASTRO 3D), Australia}
\author[0000-0002-7627-6551]{Jane R. Rigby}
\affil{Observational Cosmology Lab, NASA's Goddard Space Flight Center, 8800 Greenbelt Road, Greenbelt, MD 20771, USA}
\author[0000-0003-4804-7142]{Ayan Acharyya}
\affil{Research School of Astronomy and Astrophysics, The Australian National University, ACT, Australia}
\affil{ARC Centre of Excellence for All Sky Astrophysics in 3 Dimensions (ASTRO 3D), Australia}

\author[0000-0002-4153-053X]{Danielle A. Berg}
\affil{Center for Gravitation, Cosmology and Astrophysics, Department of Physics, University of Wisconsin Milwaukee, 1900 East Kenwood Boulevard, Milwaukee, WI 53211, USA}
\affil{Department of Astronomy, The Ohio State University, 140 W. 18th Avenue, Columbus, OH 43202, USA}

\author[0000-0003-1074-4807]{Matthew Bayliss}
\affil{MIT Kavli Institute for Astrophysics and Space Research, 77 Massachusetts Ave., Cambridge, MA 02139, USA}

\author[0000-0002-7559-0864]{Keren Sharon}
\affiliation{Department of Astronomy, University of Michigan, 1085 South University Ave, Ann Arbor, MI 48109, USA}

\begin{abstract}

Our ability to study the properties of the interstellar medium (ISM) in the earliest galaxies will rely on emission line diagnostics at rest-frame ultraviolet (UV) wavelengths. In this work, we identify metallicity-sensitive diagnostics using UV emission lines. We compare UV-derived metallicities with standard, well-established optical metallicities using a sample of galaxies with rest-frame UV and optical spectroscopy. We find that the He2-O3C3 diagnostic (\heii$\,\lambda$1640\ang{} / \ciii$\,\lambda$1906,1909\ang{} \vs \oiii$\,\lambda$1666\ang{} / \ciii$\,\lambda$1906,9\ang{}) is a reliable metallicity tracer, particularly at low metallicity ($12+\logOH \leq 8$), where stellar contributions are minimal. We find that the Si3-O3C3 diagnostic (\SiuIII$\,\lambda$1883\ang{} / \ciii$\,\lambda$1906\ang{}  \vs \oiii$\,\lambda$1666\ang{} / \ciii$\,\lambda$1906,9\ang{}) is a reliable metallicity tracer, though with large scatter (0.2-0.3 dex), which we suggest is driven by variations in gas-phase abundances. We find that the C4-O3C3 diagnostic (\civ$\,\lambda$ 1548,50\ang{} / \oiii $\,\lambda$\,1666\ang{} \vs \oiii $\,\lambda$\,1666\ang{} / \ciii $\,\lambda$ 1906,9\ang{}) correlates poorly with optically-derived metallicities. We discuss possible explanations for these discrepant metallicity determinations, including the hardness of the ionizing spectrum, contribution from stellar wind emission, and non-solar-scaled gas-phase abundances. Finally, we provide two new UV oxygen abundance diagnostics, calculated from polynomial fits to the model grid surface in the He2-O3C3 and Si3-O3C3 diagrams.
\end{abstract}


\section{Introduction} \label{sec:intro}

Quantifying the build-up of stellar mass and metals in galaxies over cosmic time is critical in understanding how galaxies form and evolve. Changes in mass and metallicity are intertwined through the star formation process, as gas is converted into stars and subsequently enriched within stars through nucleosynthetic processes. Some of this enriched gas is eventually expelled via stellar winds or supernovae (SNe), and thereby returned to the interstellar medium (ISM). As star formation continues and this enrichment process repeats, the metal content of the galaxy increases \citep[e.g.,][]{Nomoto+2013}.

The specific pattern of elemental abundances in a given galaxy depends on the history of star formation \citep{Pagel+1995}. In turn, the duration and efficiency of star formation depends on both the availability of gas and its ability to collapse and form stars. For a given gas cloud, the balance between gravity and thermal pressure is mediated by the temperature, density, and chemical composition of the gas itself \citep[][]{Evans+1999}. Thus, understanding the mechanisms that drive and sustain star formation relies on our ability to measure the physical properties of the ISM (see recent reviews by \citealt{Peimbert+2017}; \citealt{Maiolino+2018}; \citealt{Kewley+2019ARAA}; and references therein).

Fortunately, emission from ionized gas (nebular emission) in galaxies is easily observable via bright emission lines such as the hydrogen recombination transitions (e.g., \ha) and the radiative de-excitation of collisionally excited metal ions (e.g., \oiii\,$\lambda$\,5007). These emission lines encode information about dust attenuation and the local gas conditions, including the temperature, chemical composition, density, and ionization state of the gas. We can extract the properties of the ISM by comparing the relative strengths of various emission lines with predictions from atomic physics \citep[e.g.,][]{Sargent+1970, Searle+1972, Pagel+1979, McKee+1977, McGaugh+1991, Zaritsky+1994, Kewley+2002, Pilyugin+2005, Berg+2015}.

Ratios of emission lines that have proven to be particularly sensitive to properties like metallicity or density are called ``emission line diagnostics''. The best-known emission line diagnostics use transitions at optical wavelengths. There is a long history of using these emission line diagnostics to constrain the physical properties of the ISM in galaxies with redshifts between $0<z<4$ \citep[e.g.,][and many others]{Tremonti+2004, Erb+2006, Brinchmann+2008, Shapley+2011, Steidel+2014, Zahid+2014, Bian+2017}. 

However, for observations of distant galaxies ($z{\gtrsim}4$), these well-established optical emission line diagnostics become inaccessible to ground-based optical and infrared facilities, as the expansion of the universe redshifts these lines out of observational windows. To probe the physical conditions of the gas in the earliest galaxies, we must instead rely on emission line diagnostics that originate in the rest-frame ultraviolet (UV) part of the spectrum. Many studies have identified promising emission lines in the UV that can be used to probe ISM conditions in the most distant galaxies \citep[e.g.,][]{Kinney+1993, Garnett+1995, Heckman+1998, Shapley+2003, Leitherer+2011, James+2014, Bayliss+2014, Stark+2014, Zetterlund+2015, Steidel+2016, Feltre+2016, Du+2017, Byler+2018}.

Unfortunately, the properties of both nebular and stellar emission in the UV are still poorly understood, driven by our comparative lack of UV spectroscopy. UV photons are more difficult to detect than in the optical and generally require space-based observatories. Moreover, observations of star-forming galaxies in the rest-frame UV have revealed complex and overlapping features, including broad emission features (e.g., \heii$\,\lambda$1640\ang; \citealt{Leitherer+2018}), emission associated with stellar winds (\civ$\,\lambda$1550, \ion{Si}{4}$\,\lambda$1400\ang; \citealt{Pettini+2000}, also \citealt{Chisholm+2019}), and emission from resonant transitions or transitions with a non-stellar ionization source (e.g., continuum upscattering in \ion{Mg}{2}$\,\lambda$2796\ang; \citealt{Rigby+2014}). In many cases, these non-nebular features overlap with nebular emission lines. Line profiles can be further complicated by interstellar absorption features, making the interpretation of UV emission lines challenging \citep{Vidal-Garcia+2017}.

Overcoming these challenges is crucial for studying the most distant galaxies. Future surveys using the multi-object Near Infrared Spectrograph (NIRSpec) on the {\it James Webb Space Telescope} (JWST) will provide rest-UV spectra for thousands of galaxies at redshifts above $z{\sim}5$. The Mid-Infrared Instrument (MIRI) on JWST extends to longer wavelengths, 5-30$\mu$m, compared to the 0.6-5$\mu$m covered by NIRSpec. However, MIRI is not multiplexed, and it will be impossible to obtain rest-optical spectroscopy for \emph{most} of the galaxies observed with NIRSpec. Similar challenges exist for ground-based spectra: 30m-class telescopes will observe the rest-UV for distant galaxies in the infrared, but Earth's atmosphere makes it impossible to obtain the rest-optical at far-infrared wavelengths. Thus, our most cost-effective means of understanding the gas in these objects will come from their rest-UV emission. 

Before UV emission line diagnostics can be applied to large samples of high redshift galaxies, we must test whether or not the diagnostics can accurately recover key astrophysical parameters like the gas-phase metallicity. Moreover, to robustly compare the properties of high redshift galaxies with the results from lower redshift studies, we must first establish that predictions from UV emission line diagnostics are consistent with predictions from optical emission line diagnostics. However, calibrating UV and optical emission line diagnostics is a non-trivial task for two reasons. First, the task requires a sample of star-forming galaxies with rest-UV and rest-optical emission line spectroscopy. Second, to properly study the evolution of ISM properties over cosmic time, the sample should span a range of redshifts, so that metallicity calibrations are not biased to local ISM conditions.

In recent years, significant effort has gone into compiling samples of galaxies with rest-UV and rest-optical spectroscopy. \citet{Berg+2016, Berg+2019} and \citet{Senchyna+2017, Senchyna+2019} used the \emph{Hubble Space Telescope} (HST) Cosmic Origins Spectrograph (COS) to obtain UV spectroscopy for nearby galaxies that already have optical spectroscopy from SDSS. Samples are still small (26 and 16 objects, respectively), because it is difficult to predict the \SN of UV emission lines without preliminary UV spectra.

Above $z\gtrsim 1$, the rest-frame UV is redshifted into the observed optical and infrared, wavelength ranges that are accessible from ground-based telescopes. However, the improved observational access comes at the expense of detectability, since distant galaxies are, in general, fainter. Thus, rest-frame UV spectroscopic surveys of high-redshift objects have generally taken three approaches: (1) probe extreme objects with emission line fluxes high enough for direct detection \citep[e.g., ][]{Erb+2010, Stark+2014}; (2) target galaxies that have been magnified via gravitational lensing \citep[e.g.,][]{Bayliss+2014, Rigby+2018a}; or (3) create composite spectra from stacks of individual galaxy spectra \citep[e.g.,][]{Shapley+2003, Steidel+2016}. The calibration of UV metallicity diagnostics should be based on spectra from individual galaxies because stacking can be influenced by outliers, and we thus focus on objects from the first two approaches. However, both approaches rely on relatively rare objects and samples are small (of order 10 galaxies).

In this work, we test the UV emission line diagnostics presented in \citet{Byler+2018} using a sample of local galaxies ($z < 0.1$) and moderate-redshift galaxies ($z=2-3$) with rest-frame UV and optical spectra. We first calculate metallicities using UV emission lines. We then compare the metallicities calculated from UV emission lines with metallicities calculated from optical emission lines to identify which UV diagnostics are most consistent with optical diagnostics. For future studies where only rest-frame UV spectroscopy is available, this comparison provides a crucial link between abundances derived using UV emission lines and optical emission lines.

The structure of the paper is as follows. We describe the stellar and nebular model in \S\ref{sec:model:stars} \& \S\ref{sec:model:neb}, respectively. We introduce the sample in \S\ref{sec:data}, including local galaxies (\S\ref{sec:data:BCDs}) and moderate-redshift galaxies (\S\ref{sec:data:mage}) with rest-UV and rest-optical spectra. We discuss abundance determinations in \S\ref{sec:Z}. We determine theoretical abundance calibrations in \S\ref{sec:Z:corr} and calculate UV-based metallicities for the comparison samples in \S\ref{sec:Z:UV}. We compare UV and optical abundance metallicities in \S\ref{sec:ZZ}. In \S\ref{sec:discussion} we discuss problematic UV metallicity diagnostics and sources of significant uncertainty, including the contribution from stellar wind emission, and the use of rotating and binary star models. Finally, we summarize our conclusions in \S\ref{sec:conclusions}.

\section{Description of Model} \label{sec:model}

The stellar and nebular models are described at length in \citet{Byler+2018} (hereafter B18); we briefly summarize the most relevant information here.

\subsection{Stellar Model} \label{sec:model:stars}

For stellar population synthesis, we use the Flexible Stellar Population Synthesis package \citep[\FSPS;][]{Conroy+2009, Conroy+2010} via the Python interface, \pFSPS \citep{pythonFSPSdfm}\footnote{GitHub commit hash \texttt{d1bb5d5}}.

We use the MESA Isochrones \& Stellar Tracks \citep[MIST;][]{Dotter+2016, Choi+2016}, single-star stellar evolutionary models which include the effect of stellar rotation. The evolutionary tracks are computed using the publicly available stellar evolution package Modules for Experiments in Stellar Astrophysics \citep[MESA v7503;][]{Paxton+2011,Paxton+2013, Paxton+2015}. The MIST models cover ages from $10^5$ to $10^{10.3}$ years, initial masses from $0.1$ to $300\,$\Msun, and metallicities between $-2.0 \leq$ $[\mathrm{Z}/\mathrm{H}]$ $\leq 0.5$ in steps of 0.25\,dex. MIST adopts the protosolar abundances recommended by \citet{Asplund+2009} as the reference scale for all metallicities, such that [Z/H] is computed with respect to $Z=Z_{\odot,\mathrm{protosolar}}=0.0142$ rather than $Z=Z_{\odot,\mathrm{photosphere}}=0.0134$, the present-day photospheric abundances.

We combine the MIST tracks with a high resolution theoretical spectral library (C3K; Conroy, Kurucz, Cargile, Castelli, \emph{in prep.}) based on Kurucz stellar atmosphere and spectral synthesis routines \citep[ATLAS12 and SYNTHE,][]{Kurucz+2005}. The C3K library is supplemented with alternative spectral libraries for very hot stars and stars in rapidly evolving evolutionary phases. For main sequence stars with temperatures above 25,000$\,$K (O- and B-type stars), we use WM-Basic \citep{Pauldrach+2001} spectra, as described in \citet{Eldridge+2017}. For Wolf-Rayet (W-R) stars, we use the spectral library from \citet{Smith+2002}, computed using CMFGEN \citep{Hillier+2001}.

We note that even though the stellar masses in the MIST models extend to $300\,$\Msun, we adopt a Kroupa initial mass function \citep[IMF;][]{Kroupa+2001} with an upper and lower mass limit of 120\Msun and 0.08\Msun, respectively.

In \S\ref{sec:discussion} we consider the effect of binary stars. FSPS includes pre-computed simple stellar populations (SSPs) from the Binary Population and Spectral Synthesis code \citep[BPASS, v2.2;][]{Eldridge+2017}. All population synthesis parameters are summarized in Table~\ref{tab:sps}.

\begin{deluxetable*}{llll}
\tabletypesize{\footnotesize}
\tablecolumns{4}
\tablecaption{Population Synthesis model parameters.}\label{tab:sps}
\tablehead{
\colhead{Stellar Model} &
\colhead{Hot Star Spectral Libraries} &
\colhead{IMF} &
\colhead{SFH}
}
\startdata
MIST    & WM-Basic (O- and B-type stars; \citealt{Eldridge+2017}) & \citealt{Kroupa+2001};   & Constant SFR \\
\; & CMFGEN (W-R stars; \citealt{Hillier+2001}) & $M_{\mathrm{lower}} = 0.08$\Msun, & \; \\
\; & \; & $M_{\mathrm{upper}} = 120$\Msun. & \; \\
BPASS   & WM-Basic (O- and B-type stars; \citealt{Eldridge+2017}) & \citealt{Kroupa+2001};   & Constant SFR \\
\; & PoWR (W-R stars; \citealt{Hamann+2003}) & $M_{\mathrm{lower}} = 0.08$\Msun, & \; \\
\; & \; & $M_{\mathrm{upper}} = 120$\Msun. & \; 
\enddata
\end{deluxetable*}

\subsection{Nebular Model} \label{sec:model:neb}

We use the nebular model implemented within \FSPS, \CloudyFSPS \citep{cloudyFSPSv1}, to generate spectra that include nebular line and nebular continuum emission. Calculations were performed with the photoionization code \Cloudy \citep[v13.03; ][]{Ferland+2013}. 

The nebular model is a grid in (1) SSP age, (2) SSP and gas-phase metallicity, and (3) ionization parameter, \U, a dimensionless quantity that gives the ratio of ionizing photons to the total hydrogen density. We use the \Cloudy definition of \U, which is computed at the illuminated inner-face of the gas cloud.

The model uses \FSPS to generate single-age, single-metallicity stellar populations. Using the photoionization code \Cloudy, the SSP is used as the ionization source for the gas cloud and the gas-phase metallicity is scaled to the metallicity of the SSP. For each SSP of age $t$ and metallicity $Z$, photoionization models are run at different ionization parameters, \U, from \logUeq{-4} to \logUeq{-1} in steps of 0.5 dex. \U is a free parameter, but the nebular line fluxes are scaled to the ionizing photon flux of the input spectrum. For a detailed discussion of this intensity scaling, see \S 2.1.4 of \citep{Byler+2017}.

Star clusters do not form instantaneously, and may be better modeled by a population with a range of ages spanning a few million years. To account for more extended, complex star formation histories (SFHs), we also generate stellar populations assuming a continuous star formation rate (CSFR). For continuous star formation models, the rate of stars forming and the rate of stars evolving off the main sequence eventually reaches an equilibrium. As noted in \citetalias{Byler+2018}, the MIST models with continuous star formation reach a ``steady state'' between 7 and 10\Myr, after which the ionizing properties change very little.

A full comparison of the instantaneous burst and CSFR models can be found in \citetalias{Byler+2018}. In what follows, we assume stellar populations with continuous star formation over 10\Myr (1\Msun{}/year). We note that the use of CSFR models may not be appropriate for massive \hii{} regions and galaxies with extremely bursty SFHs. 

Reported emission line strengths always reflect the pure nebular emission line intensities. However, in \S\ref{sec:discussion:broad} - \S\ref{sec:discussion:CIV}, we discuss possible contamination from stellar emission.

\subsubsection{Gas Phase Abundances}\label{sec:model:neb:z}

The abundances used in this work follow those used in \citetalias{Byler+2018}. We assume that the gas phase metallicity scales with the metallicity of the stellar population (i.e., $Z_{\mathrm{gas}} \approx Z_{\mathrm{stars}}$), given that the metallicity of the most massive stars should be identical to the metallicity of the gas cloud from which the stars formed. Both the gas phase and stellar abundances are solar-scaled, such that that individual elemental abundances are monolithically scaled up or down with metallicity, [Z/H]. In practice, [Z/H] scales with [Fe/H] for the stellar models, and with [O/H] for the gas phase abundances. 

In this work, we use models with stellar metallicities between $-2.0 \leq [\mathrm{Fe}/\mathrm{H}] \leq 0.25$ in steps of 0.25 dex, which corresponds to gas-phase oxygen abundances between $6.69 \leq 12+\logOH \leq 8.94$ in steps of 0.25 dex. In what follows, we often use the terms gas phase metallicity and oxygen abundance interchangeably, since these quantities scale equivalently in our model. 

For most elements we use the solar abundances from \citet{Grevesse+2010}, based on the results from \citet{Asplund+2009}, and adopt the dust depletion factors specified by \citet{Dopita+2013}. The abundance for each element and dust depletion factors at solar metallicity are given in Table~\ref{tab:solarAbunds}. Notably, there are a few elements (C, N) with gas phase abundances that deviate from perfect solar-scaling, due to additional production mechanisms that operate at high metallicity (secondary or pseudo-secondary nucleosynthetic production; for details, see \citealt{Berg+2016}). We describe the scaling for these elements below.

To set the relationship between N/H and O/H, we use the following equation from \citetalias{Byler+2018}:
\begin{equation}\label{eq:nitrogen}
\begin{aligned}
    \log_{10}&(\mathrm{N}/\mathrm{O}) = \\
    & -1.5 + \log\left( 1 + e^{\frac{12 + \log_{10}(\mathrm{O}/\mathrm{H})-8.3}{0.1}}\right),
\end{aligned}
\end{equation}

and for C/H and O/H:
\begin{equation}\label{eq:carbon}
\begin{aligned}
    \log_{10}&(\mathrm{C}/\mathrm{O}) = \\
    & -0.8 + 0.14\cdot\left(12 + \log_{10}(\mathrm{O}/\mathrm{H})-8.0\right) \\
    & + \log\left( 1 + e^{\frac{12 + \log_{10}(\mathrm{O}/\mathrm{H})-8.0}{0.2}}\right).
\end{aligned}
\end{equation}

\noindent The relationships from \citetalias{Byler+2018} for N/H and C/H with O/H were modified from the empirically calculated \citet{Dopita+2013} relationships to better match observations below $12+\logOH = 8$, which did not exist when the original relationships were published. For $12+\log_{10} (\mathrm{O}/\mathrm{H}) = 8.69$ (solar metallicity), this corresponds to $\log_{10} (\mathrm{N}/\mathrm{O}) = -1.09$ and $\log_{10} (\mathrm{C}/\mathrm{O}) = -0.26$, including the effects of dust depletion, typical of star-forming galaxies with $12 + \log \mathrm{O}/\mathrm{H} = 8.7$ \citep[e.g.,][]{Belfiore+2017b}. For a more complete discussion of N/O and C/O ratios used in photoionization models, we refer the reader to Appendix B of \citetalias{Byler+2018}. We note that empirically-derived relationships are always limited by the calibration sample, and detailed gas-phase abundance studies are only feasible in the local universe. As such, these locally-derived relations may not be appropriate for high-redshift systems.

\begin{deluxetable}{lcc}[b!]
\tabletypesize{\footnotesize}
\tablecolumns{3}
\tablecaption{Elemental abundances and adopted depletion factors $D$ for each element in the nebular model at solar metallicity, which has $Z=0.0142$ ($\log_{10}$(O/H) = $-3.31$ or $12+\log_{10}$(O/H) = $8.69$).}
\tablehead{
\colhead{Element} &
\colhead{$\log_{10}(\mathrm{E}/\mathrm{H})$} &
\colhead{$\log_{10}(D)$}
}
\startdata
H   & 0	& 0 \\
He  & -1.01 & 0 \\
C   & -3.57 & -0.30 \\
N   & -4.60 & -0.05 \\
O   & -3.31 & -0.07 \\
Ne  & -4.07 & 0 \\
Na  & -5.75 & -1.00 \\
Mg  & -4.40 & -1.08 \\
Al  & -5.55 & -1.39 \\
Si  & -4.49 & -0.81 \\
S   & -4.86 & 0 \\
Cl  & -6.63 & -1.00 \\
Ar  & -5.60 & 0 \\
Ca  & -5.66 & -2.52 \\
Fe  & -4.50 & -1.31 \\
Ni  & -5.78 & -2.00 \\
\enddata
\tablecomments{Solar abundances are from \citet{Grevesse+2010} and depletion factors are from \citet{Dopita+2013}.}
\label{tab:solarAbunds}
\end{deluxetable}

\paragraph{C/O variations} The \ciii{}$\,\lambda$1906,1909 emission lines are the brightest UV emission lines after Ly$-\alpha$. As such, the \ciii{} lines are optimal candidates for emission line diagnostics. However, it has not yet been established whether or not these lines provide a robust tracer for the gas phase oxygen abundance. Observed C/O ratios vary by more than $0.6$ dex between $7\lesssim 12+\logOH \lesssim 8$ \citep{Berg+2019}. Using detailed chemical evolution models, \citet{Berg+2019} found that the C/O ratio is sensitive to both the detailed SFH and supernova feedback. This implies that the UV \ciii{} and oxygen emission lines \emph{alone} may not provide a reliable indicator of the gas phase oxygen abundance. Robust metallicity diagnostics may require additional spectral features.

\paragraph{Decoupled stellar and gas-phase abundances} Recent work has suggested decoupling the stellar metallicity from the gas phase metallicity, to approximate scenarios in which high star formation rates rapidly enrich the gas in $\alpha$-elements \citep[e.g.,][]{Steidel+2016}. In practice, this involves pairing gas of a given oxygen abundance with a slightly more metal-poor (i.e., lower iron abundance) stellar ionizing spectrum. This ultimately increases the excitation of the nebula, since ionizing spectra are harder with decreasing metallicity. Observationally, the prevalence and scale of this $\alpha$-enrichment has yet to be determined. Recent work by \citet{Senchyna+2019} compared stellar iron abundances and gas-phase oxygen abundances in nearby galaxies based on UV spectra, and did not find significantly $\alpha-$enhanced gas.

\section{Data} \label{sec:data}

We compare our models to two observational samples: (1) nearby star-forming galaxies ($z<0.04$) with UV and optical spectroscopy, and (2) moderate-redshift star-forming galaxies ($z \gtrsim 1.5$) with optical and near-infrared (NIR) spectroscopy that probes rest-frame UV and optical wavelengths. References for all galaxies used in the sample can be found in Table~\ref{tab:logOH}. We briefly describe the two samples below.

\begin{figure*}
  \begin{center}
    \includegraphics[width=\linewidth]{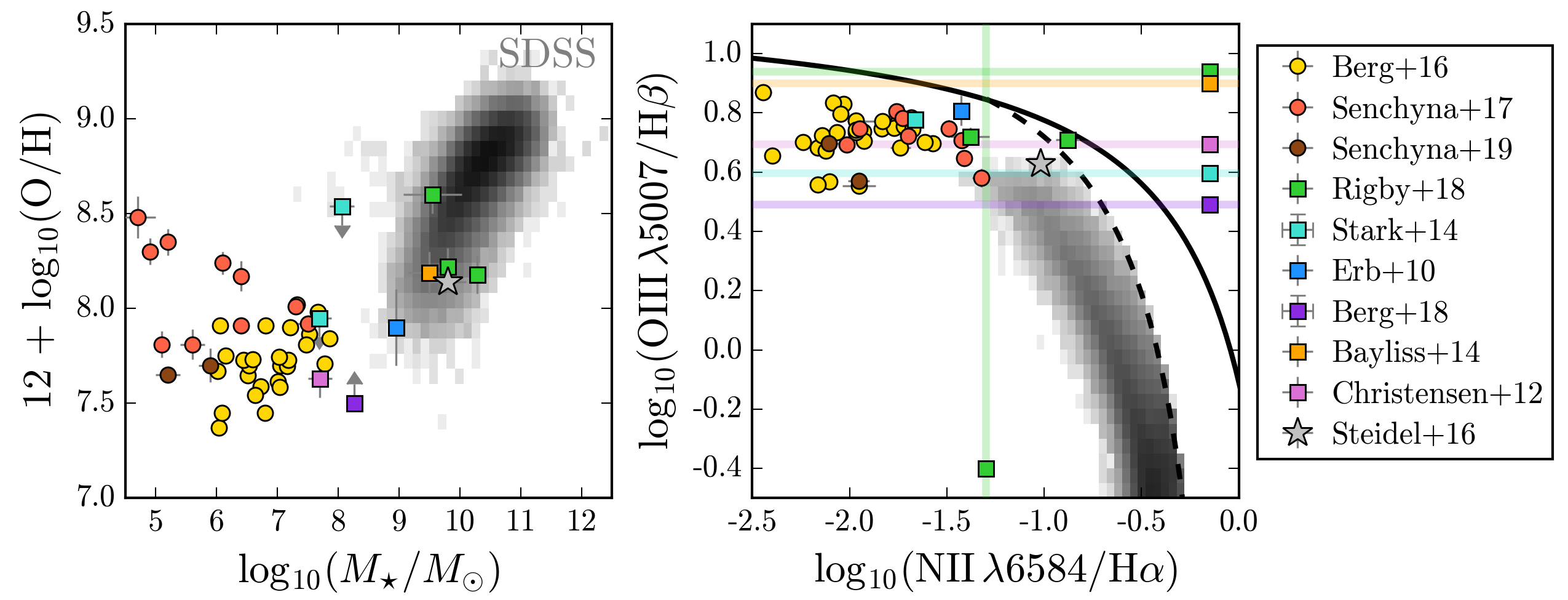}
    \caption{\emph{Left:} Stellar mass ($x$-axis) and optical gas-phase metallicity ($y$-axis) for the galaxies used in this work. The grey 2D histogram shows local star-forming galaxies from SDSS. For the galaxies considered in this work, the color of each marker indicates the source of the observation, compiled from the literature. Circular markers show local galaxies ($z<0.1$) with rest-UV observations from HST/COS, while square markers show moderate-redshift galaxies with rest-UV spectra observed in the optical. For clarity, we only show those galaxies with robust UV and optical metallicities. The galaxies considered in this work have lower masses than typical star-forming galaxies in the local universe, but span a wide range in stellar mass and gas-phase metallicity. Overall, the low- and moderate-redshift galaxies occupy different regions of parameter space, with local galaxies having lower masses and metallicities. \emph{Right:} The standard BPT diagram. The dashed line shows the \citet{Kauffmann+2003b} separation between star formation and composite regions, and the solid black line shows the \citet{Kewley+2001} separation between AGN and star formation. Horizontal lines are shown for objects without \nii{}/\ha{} measurements, and vertical lines are shown for objects without \oiii{}/\hb{} measurements. In general, the galaxies considered in this work have highly excited gas and lower metallicities than the typical star-forming galaxy in the local universe.}
    \label{fig:sampleMZ}
  \end{center}
\end{figure*}

\subsection{Local blue compact dwarf galaxies}\label{sec:data:BCDs}

\citet{Berg+2016} presented UV and optical spectra for a sample of 7 nearby, low-metallicity, high-ionization blue compact dwarf galaxies (BCDs). We include 19 additional galaxies from \citet{Berg+2019}; the combined sample of 26 galaxies is hereafter referred to as the \emph{Berg sample}. The galaxies are nearby ($0.003 < z < 0.040$), UV-bright ($m_{\mathrm{FUV}} \leq 19.5$ AB), compact (D $< 5$''), low-metallicity ($7.2 \leq 12 + \logOH \leq 8.0$), and low extinction ($0.05 < E(B-V) < 0.2$. These galaxies have relatively low masses (${\sim}10^7$\Msun) and high specific star formation rates (sSFRs; ${\sim}10^{-8}$ yr$^{-1}$). 

We use the dereddened UV emission line fluxes and optical oxygen abundances published in \citet{Berg+2016, Berg+2019}. All of the galaxies in the sample have auroral line detections for direct-method calculations of the nebular temperature, density, and metallicity. The UV spectra were obtained with HST COS using the G140L grating and cover roughly $1300-2000$\ang. This wavelength coverage includes a number of emission lines, including \civ$\,\lambda$\,1548,1551, \heii$\,\lambda$\,1640, \oiii$\,\lambda$1661,1666, \SiuIII$\,\lambda$1883,1892, and \ciii $\,\lambda$1906,1909\footnote{For convenience, in this work \ciii $\,\lambda$1906,1909 represents the combination of the forbidden [\ion{C}{3}]$\,\lambda$1906 line and the semi-forbidden \ion{C}{3}]$\,\lambda$1909 line.}.

\paragraph{Other local BCD observations} Where possible, we also compare our models to the sample of local BCDs from \citet{Senchyna+2017, Senchyna+2019}, which have similar properties to the \citet{Berg+2016} galaxies and were also observed with HST COS. However, the \citet{Senchyna+2017} observations used the G185M and G160M gratings, which provide increased spectral resolution at the expense of wavelength coverage. As a result, the \citet{Senchyna+2017} BCD sample has fewer emission lines observed than the \citet{Berg+2016} sample, but the higher spectral resolution (typical FWHM of 0.6\ang, compared to 3\ang in \citealt{Berg+2016}) allows the authors to simultaneously fit for broad and narrow emission line components, when present. The published emission line fluxes are not corrected for galactic or intrinsic extinction, but the authors provide their derived $E(B-V)$ for each object. We correct line fluxes for galactic and intrinsic extinction using  \citet{Fitzpatrick+1999} and \citet{Cardelli+1989} reddening laws respectively, with $R_{\mathrm{v}} = 3.1$. The optical emission line ratios for the \citet{Senchyna+2019} objects (used in Fig.~\ref{fig:sampleMZ}) were obtained via private communication.

\subsection{Moderate-redshift star-forming galaxies}\label{sec:data:mage}

We use strongly lensed galaxies from Project \mage: the Magellan Evolution of Galaxies Spectroscopic and Ultraviolet Reference Atlas \citep{Rigby+2018a, Rigby+2018b}, spanning the redshift range $1.68 < z < 3.6$.

The \mage galaxies have rest-UV spectroscopy taken with the MagE instrument on the Magellan telescopes. The spectra cover the wavelength range $3200 < \lambda < 8280$\ang in the observed frame (approximately $1000 \lesssim \lambda \lesssim 3000$\ang in the rest frame), with average spectral resolving power of $\mathrm{R}\sim3300$ and $\SN=21$ per resolution element in the median spectrum.

We include 4 of the 19 \mage galaxies. Galaxies were excluded from the sample based on the following criteria: 
\begin{itemize}
    \item Galaxies suspected to harbor low luminosity active galactic nuclei (AGN; $N=2$).
    \item Galaxies without rest-frame optical spectra ($N=10$). 
    \item We require at least 3 UV emission lines to calculate the UV metallicity, and remove galaxies with fewer than three UV emission lines ($N=3$).
\end{itemize}

The remaining four galaxies have rest-frame optical spectra from Keck NIRSPEC \citep{Rigby+2011, Wuyts+2012a}, Keck OSIRIS \citep{Wuyts+2014}, HST/WFC3 \citep{Whitaker+2014}, LBT/LUCIFER \citep{Bian+2010}, and Magellan FIRE spectrograph \citep{Rivera+2017}. 

UV emission line fluxes for the \mage galaxies are measured following \citet{Acharyya+2019} and will be published in a future \mage paper (Rigby et al., \emph{in prep}). The \mage spectra have been corrected for foreground extinction using the galactic extinction from the \citet{Schlafly+2011} recalibration of the \citet{Schlegel+1998} infrared-based dust map, assuming a \citet{Fitzpatrick+1999} reddening law with $R_{\mathrm{v}} = 3.1$. Emission line fluxes are dereddened using a \citet{Cardelli+1989} dust curve using values of $E(B-V)$ derived from SED fitting (Rigby et al., \emph{in prep}).

\paragraph{Other observations of moderate-redshift galaxies} When possible, we also compare our models to the published emission line fluxes for the four lensed galaxies from \citet{Stark+2014}, and the single lensed galaxies from \citet{Erb+2010}, \citet{Christensen+2012}, \citet{Bayliss+2014}, and \citet{Berg+2018}. We also include the stacked spectrum of lensed galaxies from \citet{Steidel+2016}, however, we note that it is much more difficult to interpret the metallicity derived from a stacked spectrum. Galaxies included in the sample are given in Table~\ref{tab:logOH}. For all objects, we use dereddened emission line fluxes. In cases where dereddened line fluxes were not available in a published table, we dereddened emission line fluxes following the original source's description.

Our requirement of 3 distinct UV emission line detections limits the total number of objects in our sample, and we note that some of these references include additional objects with rest-UV and rest-optical spectra \citep[e.g.,][]{Christensen+2012} that we do not include in this work. A more complete list of objects with rest-UV and rest-optical spectra can be found in \citet{Patricio+2019} and \citet{Plat+2019}. 

\subsection{Sample global properties and caveats}\label{sec:data:caveats}

In the left panel of Fig.~\ref{fig:sampleMZ}, we show the stellar mass and gas phase metallicity for all of the galaxies included in our sample (see Table~\ref{tab:logOH}). The grey 2D-histogram shows star-forming galaxies from SDSS \citep[DR7;][]{Abazajian+2009}. The color of each marker indicates the literature source for the observation and the shape of the marker separates nearby galaxies (circles) from moderate-redshift galaxies (squares). The UV-Optical sample is comprised of galaxies with stellar masses between $10^{4.7} - 10^{10.3}$\,\Msun and gas phase metallicities between $7.3 < 12+\logOH < 8.6$.

The right panel of Fig.~\ref{fig:sampleMZ} shows the standard Baldwin, Phillips, \& Terlevich \citep[BPT;][]{BPT} diagram, which uses the \nii{}/\ha{} and \oiii{}/\hb{} emission line ratios. We include empirically derived relationships used to separate objects with different ionizing sources; the dashed line shows the \citet{Kauffmann+2003b} separation between star formation and composite regions, and the solid black line shows the \citet{Kewley+2001} separation between AGN and star formation. Some of the moderate-redshift galaxies are sufficiently distant such that the \nii{}$\,\lambda$6584 and \ha{}$\,\lambda$6563 emission lines have redshifted out of optical wavelengths, and only have \oiii{}$\,\lambda$5007 and \hb{}$\,\lambda$4861 measurements. For these objects, we include a horizontal line at the measured \oiii{}/\hb{} ratio on the BPT diagram. For those objects without observations of \oiii{}$\,\lambda$5007 and \hb{}$\,\lambda$4861, we include a vertical line at the measured \nii{}/\ha{} ratio.

In general, the UV-Optical sample has higher \oiii{}/\hb{} ratios and lower \nii{}/\ha{} ratios than the sample of local star-forming galaxies from SDSS. We note that both the \citet{Berg+2016, Berg+2019} and \citet{Senchyna+2017, Senchyna+2019} samples were designed to target high ionization, high excitation dwarf galaxies, to maximize the detection of the UV C and O emission lines. Thus, their gas conditions are not representative of the local galaxy population as a whole. Optical emission line diagnostic diagrams demonstrate that these objects have the expected properties of typical, metal-poor photoionized galaxies. We refer the reader to each of these publications for a comprehensive comparison of optical emission properties.

We utilize these samples for comparisons among UV and optical diagnostics only, and our conclusions cannot be inferred to samples outside the parameter ranges of these original samples. We also emphasize that the comparison of local and moderate-redshift samples in this work cannot be used to infer any meaningful cosmic evolution in ISM properties. There are a number of key differences between the local and moderate-redshift samples that may complicate our conclusions, which we state here.

First, the local BCDs have lower metallicities than the moderate-redshift galaxies. The local sample has oxygen abundances between $7.3 < 12+\logOH < 8.2$ with a median of $12+\logOH = 7.7$ and dispersion of 0.2\,dex. The moderate-redshift sample has oxygen abundances between $7.6 < 12+\logOH < 8.6$ with a median of $12+\logOH = 8.2$ and a dispersion of 0.3\,dex.

The second major difference between the low- and moderate-redshift galaxies is the typical stellar mass. The low-redshift galaxies are all of very low mass, $M \lesssim 10^{7.5}$\,\Msun\footnote{We note that the mass estimates from SDSS for the lowest-mass systems ($M \lesssim 10^{7}$\,\Msun) are likely underestimated, due to the reduced accuracy in redshift as a distance indicator (e.g., \citealt{Mamon+2019})}. In contrast, the moderate-redshift galaxies have higher stellar masses, $M \gtrsim 10^{7.5}$\,\Msun. Thus, we are not comparing the same types of galaxies in this analysis, and any perceived correlations with redshift will not actually reveal any information about the evolution of the ISM. In general, however, the galaxies considered in this work have lower masses than typical star-forming galaxies from the SDSS survey. Both the low- and moderate-redshift samples have masses more typical of local dwarf galaxies \citep[e.g.,][]{Lee+2006, Berg+2012}.

We note that there are a few objects with $M \lesssim 10^{5.5}$\,\Msun from \citet{Senchyna+2017} (e.g., SB 179, 191, 198). These objects are giant \hii{} regions embedded within larger disk systems, with physical scales of order 100\,pc.

\section{Metallicity determinations}\label{sec:Z}

\subsection{Theoretical metallicity calibrations}\label{sec:Z:corr}

There is a known offset among theoretical abundances (i.e., the true specified gas-phase abundances in photoionization models) and the abundances calculated from strong line and direct-temperature methods (i.e., the gas phase abundances one would compute from the models based on emission line strengths), as discussed in \citealt{Stasinska+2005, Kewley+2008} (and references therein). It has been suggested that this offset is the result of temperature gradients within the nebulae that bias the characteristic temperature of a given line transition away from the mean ionic temperature \citep[e.g.,][]{Stasinska+2005, Bresolin+2007, Kewley+2008}; though it has also been attributed to an unknown issue with photoionization models \citep[e.g.,][]{Kennicutt+2003}.

In practice, we put our model metallicities onto the same scale as the observed metallicities by applying the same analysis techniques used on observations, which we describe below. In general, the correction to model metallicities is small, $\lesssim 0.1$\,dex for \logOH$ \lesssim 8$. The correction is larger at metallicities \logOH$\gtrsim 8.7$, between 0.2-1.0\,dex, depending on the model ionization parameter, where models with higher ionization parameters require larger corrections. We describe the correction calculation for direct-\Te and strong-line abundances below.

\subsubsection{Direct-\Te theoretical calibration}\label{sec:Z:corr:Te}

At each age, metallicity, and ionization parameter point in the \CloudyFSPS model grid, we calculate an optical direct-method abundance to determine the offset between the true oxygen abundance in the \Cloudy model and the measured oxygen abundance. Using a least-squares minimization ({\tt numpy.polyfit}), we fit a third-order polynomial to the direct temperature oxygen abundance as a function of the true \Cloudy oxygen abundance at each model age and ionization parameter. We provide the polynomial fits in Appendix~\ref{sec:appdx:poly}.

For the closest comparison between the models and the observational samples, we calculate direct-\Te abundances for our models following the method used in \citet{Garnett+1992}, as modified by \citet{Berg+2015} and applied to the \citet{Berg+2016} observations. We briefly describe the process below.

We calculate gas phase oxygen abundances using a two temperature zone approximation with {\tt PyNeb} \citep{PyNeb} and collision strengths from \citet{Aggarwal+1999}. We use the \citet{Aggarwal+1999} collision strengths rather than the newer \citet{Storey+2014} collision strengths, since the \citet{Aggarwal+1999} collision strengths are calculated for a six-level oxygen atom, which is needed for the UV \oiii$\,\lambda$1661,1666 lines (${}^5\mathrm{S}_2 \rightarrow {}^3\mathrm{P}_2$ and ${}^5\mathrm{S}_2 \rightarrow {}^3\mathrm{P}_1$ transitions, respectively).

Following \citet{Garnett+1992} and \citet{Berg+2015}, we approximate the \hii{} region with a high and low temperature zone for the $\mathrm{O}^{++}$ and $\mathrm{O}^{+}$ regions, respectively.

For the O$^{+}$ zone, we calculate the density using the \sii$\,\lambda 6731$ / \sii$\,\lambda 6716$ ratio. For the O$^{+}$ zone temperature, we use the (\nii$\,\lambda$6548 $+$ \nii$\,\lambda$ 6584 ) / \nii$\,\lambda 5755$ following the recommendation of \citet{Berg+2015}. The O$^{+}$ ionic abundance is then calculated with {\tt PyNeb}, using the line intensities:
\begin{equation}
    \left[ \frac{\mathrm{O}^{+}}{\mathrm{H}^{+}} \right] = \frac{\mathrm{I}_{\lambda\lambda\,3727,\,3729} + \mathrm{I}_{\lambda\lambda \,7319,\,7332}}{\mathrm{I}_{\mathrm{H}\beta}} \cdot \frac{j_{\mathrm{H}\beta}}{j_{\lambda}}
\end{equation}

For the O$^{++}$ zone, we calculate the density using the \sii$\,\lambda 6731$ / \sii$\,\lambda 6716$ ratio. For the O$^{++}$ zone temperature, we use the \oiii$\lambda$4363/(\oiii$\lambda$4959 + \oiii$\lambda$5007) ratio. The O$^{++}$ ionic abundance is then calculated with {\tt PyNeb}, using the line intensities:
\begin{equation}
    \left[ \frac{\mathrm{O}^{++}}{\mathrm{H}^{+}} \right] = \frac{\mathrm{I}_{\lambda\,4363} + \mathrm{I}_{\lambda\,4959} + \mathrm{I}_{\lambda\,5007}}{\mathrm{I}_{\mathrm{H}\beta}} \cdot \frac{j_{\mathrm{H}\beta}}{j_{\lambda}}.
\end{equation}

The total oxygen abundance is then calculated as the sum of the ionic abundances:
\begin{equation}
    \log_{10}(\mathrm{O}/\mathrm{H}) = \left[ \frac{\mathrm{O}^{++}}{\mathrm{H}^{+}} \right] + \left[ \frac{\mathrm{O}^{+}}{\mathrm{H}^{+}} \right],
\end{equation}
assuming a negligible contribution from O$^{0}$, O$^{+++}$ ions, appropriate for most \hii{} regions \citep{Berg+2016}. We note that the model calibrations do not change if we include the contribution from O$^{+++}$ ions. In the models used here, O$^{+++}$/H$^{+}$ is of order $10^{-8}$ and is never larger than $10^{-7}$.

In Appendix~\ref{appdx:UVdirectTe} we provide a comparison of theoretical direct-\Te abundances using UV and optical emission lines.

\subsubsection{Strong line theoretical calibration}\label{sec:Z:corr:strong}

Some of the galaxies in the moderate-redshift sample have metallicities calculated from optical strong line methods. To avoid introducing additional uncertainties into the comparison from using abundances that have been empirically-corrected to a ``standard'' abundance scale, we re-compute the optical abundances using the published optical emission line strengths and the \citet{Pettini+2004} (hereafter PP04) N2 abundance scale.

We have chosen the PP04-N2 abundance scale because it maximizes the number of objects in the moderate-redshift sample that have optical metallicities that can be computed using the same method. As noted in \S\ref{sec:data:caveats}, several galaxies in the moderate redshift sample do not have \nii{}/\ha{} measurements. However, even fewer galaxies have observations of the \oii{}$\lambda$3726,3729 doublet required for $R_{23}$ metallicities. 

It is not clear which, if any, of the metallicity scales is correct. As such, only relative metallicities (i.e., metallicities calculated using the same method) provide a reliable comparison. We note that the PP04-N2 abundance scale does not account for ionization parameter changes, which is particularly important for low-metallicity galaxies and galaxies at high redshift where the ionization parameter is typically high \citep[e.g.,][]{Kewley+2013a, Kewley+2013b, Masters+2014, Sanders+2016, Bian+2017, Strom+2017}, and likely larger than those in the \hii{} regions used by PP04 to make their N2 calibration.

At each metallicity and ionization parameter point in the \CloudyFSPS model grid, we use the model emission line fluxes to calculate a PP04-N2 abundance using the equation from PP04:
\begin{equation}\label{eq:O3N2}
\begin{aligned}
12 +& \logOH = \\
& 0.57 \cdot \log_{10}([\mathrm{NII}]\lambda6584/\mathrm{H}\alpha)\\
& + 8.90.
\end{aligned}
\end{equation}
To determine the offset between the PP04-N2 oxygen abundance and the true \Cloudy oxygen abundance, we fit a linear function to the PP04-N2 abundance as a function of the \Cloudy oxygen abundance for each of the ionization parameters in the model. The best-fit parameters for the linear function are determined using least-squares minimization with ({\tt numpy.polyfit}).

We show the resultant PP04-N2 calibration for the 10\Myr constant SFR models used in this work in Fig.~\ref{fig:offset}, following Fig.~1 of \citet{Stasinska+2005}. The $x$-axis shows the ``true'' oxygen abundance, as input to \Cloudy, and the $y$-axis shows the abundance calculated from the direct-\Te method (left) and the PP04-N2 method (right). In both panels, the blue lines show the fit used for the theoretical correction, color-coded by ionization parameter. The fitted line is a third-order polynomial for direct-\Te abundances (left) and a linear function for the strong line abundances (right). These fits are provided in Appendix~\ref{sec:appdx:poly}, Tables~\ref{tab:polyA}~\&~\ref{tab:polyB}.

The Cosmic Eye galaxy, SGASJ105039, and A1689\_860\_359 are the only objects with $3+$ UV emission lines where we could not calculate a PP04-N2 abundance, because the necessary [\ion{N}{2}] and \ha lines have been redshifted out of the optical observing window\footnote{This is also true for the \mage galaxy S1226$+$2152; however this object does not have enough UV emission line detections to derive a UV metallicity and could not be included in the UV-optical comparison.}. Instead, we repeat the above process using the \citet{Kobulnicky+2004} R23 abundance scale (hereafter KK04-R23), assuming the upper branch for Cosmic Eye and SGASJ105039, and the lower branch for A1689\_860\_359. The fits for the KK04-R23 are provided in Appendix~\ref{sec:appdx:poly}, Table~\ref{tab:polyC}. For each object in the sample, Table~\ref{tab:logOH} includes the method used for the optical abundance determination.

\begin{figure*}
  \begin{center}
    \includegraphics[width=\linewidth]{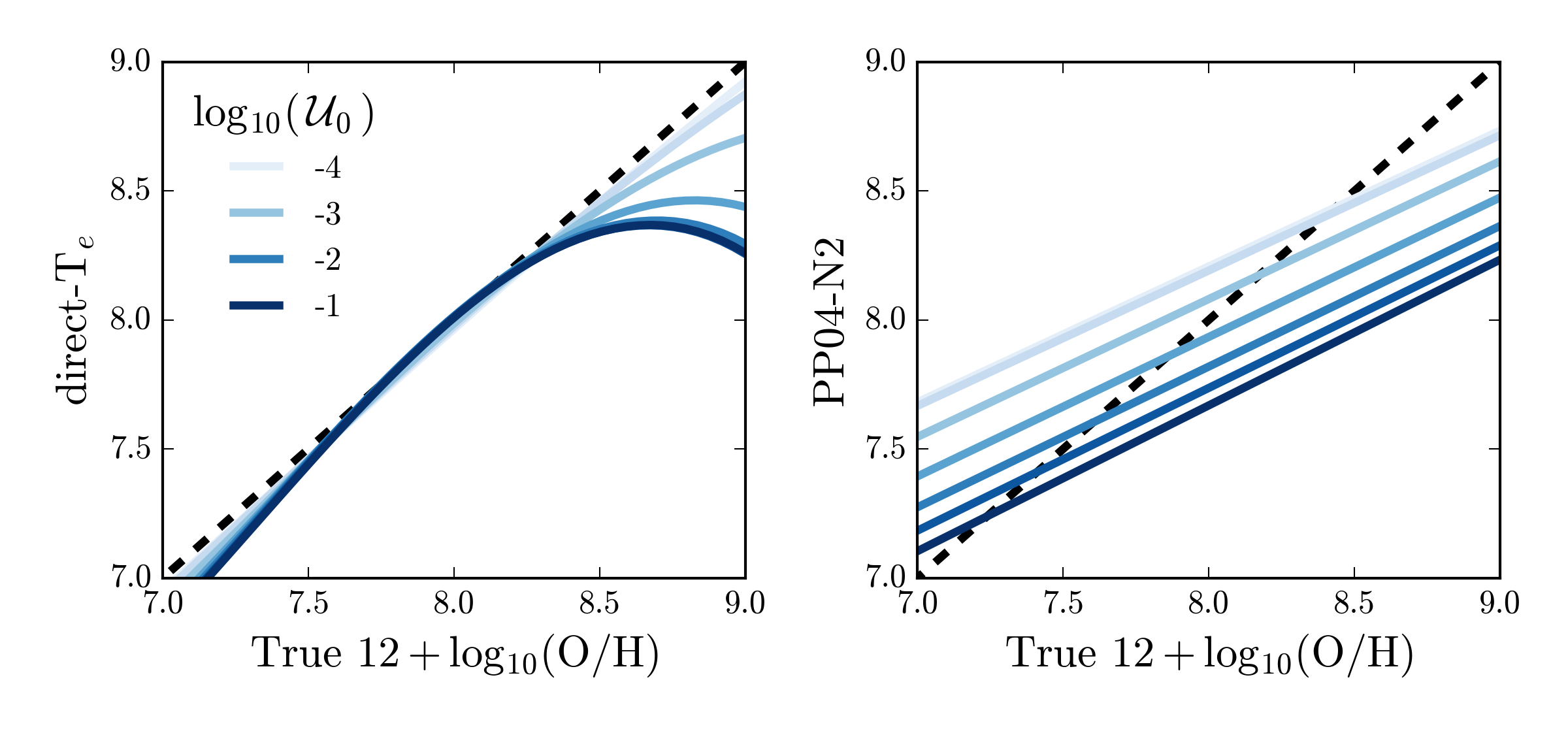}
    \caption{Metallicity offsets for the direct-\Te method (left) and the PP04-N2 method (right). The $x$-axis shows the true gas-phase oxygen abundance from \Cloudy. The $y$-axis shows the metallicity derived from emission line strengths using the respective metallicity calculations. Each line represents the polynomial fit to the models at different ionization parameters, for a model with constant SFR over 10 Myr. The lines are color-coded by ionization parameter, from \logUeq{-4} in light blue to \logUeq{-1} in dark blue. The black dashed line shows a one-to-one relationship.}
    \label{fig:offset}
  \end{center}
\end{figure*}

\subsection{Deriving abundances from UV diagnostic diagrams}\label{sec:Z:UV}

In this section, we derive gas-phase abundances for the galaxies in the sample using rest-frame UV emission lines. We use different combinations of predicted emission line ratios to construct ``diagnostic diagrams''. At a given model age, variations in ionization parameter and metallicity change the predicted emission line ratios, producing a grid or surface in the diagnostic diagram. Then, for each galaxy, we compare the observed emission line ratios to the model emission line ratios to calculate a gas-phase abundance. Specifically, the emission line ratios ($x$, $y$) for a given galaxy are matched to the surface of the model grid by interpolating between points in ionization parameter and metallicity using the {\tt scipy.interpolate.griddata} cubic spline interpolation.

This approach can be sensitive to small changes in the observed emission line ratios, so we use a Monte-Carlo method to estimate errors on the derived metallicity. For each object, we draw $N=1000$ samples from a Gaussian distribution centered at the observed line ratio ($x$,$y$), and width equal to the reported emission line ratio errors. We then recalculate the metallicity at each of these 1000 samples, and use the spread in the resultant metallicity distribution to estimate errors, where the 16$^{th}$ and 84$^{th}$ percentiles of the metallicity distribution provide the upper and lower error limits, respectively.

We then rescale the metallicity using the theoretical abundance calibrations derived in \S\ref{sec:Z:corr}, so that we can compare the UV-derived abundance to the abundances derived using optical emission lines. For the nearby galaxies with optical direct-\Te metallicities, we apply the direct-\Te correction described in \S\ref{sec:Z:corr:Te}. For the moderate-redshift galaxies with optical strong line metallicities, we apply the relevant strong line correction described in \S\ref{sec:Z:corr:strong}. In general, the metallicity correction changes the UV-derived abundance by $<0.1$\,dex at $12 + \logOH < 8$, and by ${\sim}0.1$\,dex at $8.0 < 12 + \logOH < 8.5$.

The calculated gas-phase oxygen abundances for the observational comparison sample are given in Table~\ref{tab:logOH}.

\startlongtable
\begin{deluxetable*}{ccccccc}
\tabletypesize{\footnotesize}
\tablecaption{Literature (optical) and derived (UV) gas-phase oxygen abundances.\label{tab:logOH}}
\tablehead{\colhead{\small{Ref.}} & \colhead{\small{Object}} & \multicolumn{2}{c}{\small{Optical Abundance}} & \multicolumn{3}{c}{\small{UV $12+\log_{10}(\mathrm{O}/\mathrm{H})$ Abundance}}\\ \colhead{ } & \colhead{ } & \colhead{$12+\log_{10}(\mathrm{O}/\mathrm{H})$} & \colhead{Method} & \colhead{Si3-O3C3} & \colhead{He2-O3C3} & \colhead{C4-O3C3}}
\startdata
1 & J223831 & $7.59\pm0.02$ & direct & ${8.05}^{+0.31}_{-0.15}$ & $\cdots$ & ${8.93}^{+0.26}_{-0.50}$ \\
1 & J141851 & $7.54\pm0.02$ & direct & ${8.00}^{+0.05}_{-0.14}$ & ${7.59}^{+0.07}_{-0.08}$ & ${7.99}^{+0.09}_{-0.08}$ \\
1 & J120202 & $7.63\pm0.02$ & direct & ${8.05}^{+0.15}_{-0.17}$ & ${7.97}^{+0.11}_{-0.24}$ & $\cdots$ \\
1 & J121402 & $7.67\pm0.02$ & direct & ${8.21}^{+0.18}_{-0.13}$ & $\cdots$ & ${8.49}^{+0.09}_{-0.39}$ \\
1 & J084236 & $7.61\pm0.02$ & direct & $\cdots$ & $\cdots$ & ${7.94}^{+0.50}_{-0.66}$ \\
1 & J171236 & $7.70\pm0.02$ & direct & ${7.55}^{+0.72}_{-0.38}$ & ${7.87}^{+0.16}_{-0.09}$ & ${8.72}^{+0.23}_{-0.46}$ \\
1 & J113116 & $7.65\pm0.02$ & direct & $\cdots$ & ${7.76}^{+0.19}_{-0.19}$ & ${8.19}^{+0.25}_{-0.55}$ \\
1 & J133126 & $7.69\pm0.02$ & direct & ${8.07}^{+0.17}_{-0.11}$ & ${7.91}^{+0.09}_{-0.06}$ & $\cdots$ \\
1 & J132853 & $7.73\pm0.02$ & direct & $\cdots$ & $\cdots$ & $\cdots$ \\
1 & J095430 & $7.70\pm0.02$ & direct & ${8.08}^{+0.26}_{-0.13}$ & ${7.95}^{+0.06}_{-0.09}$ & ${8.38}^{+0.15}_{-0.38}$ \\
1 & J132347 & $7.58\pm0.02$ & direct & ${7.63}^{+0.59}_{-0.20}$ & ${7.45}^{+0.09}_{-0.10}$ & ${8.19}^{+0.36}_{-0.36}$ \\
1 & J094718 & $7.73\pm0.02$ & direct & ${8.23}^{+0.16}_{-0.10}$ & $\cdots$ & ${8.92}^{+0.21}_{-0.19}$ \\
1 & J150934 & $7.71\pm0.02$ & direct & ${8.04}^{+0.37}_{-0.20}$ & ${7.86}^{+0.15}_{-0.11}$ & ${8.11}^{+0.13}_{-0.18}$ \\
1 & J100348 & $7.74\pm0.02$ & direct & ${8.27}^{+0.82}_{-0.25}$ & $\cdots$ & ${8.36}^{+0.17}_{-0.18}$ \\
1 & J025346 & $7.91\pm0.02$ & direct & ${7.86}^{+0.42}_{-0.27}$ & $\cdots$ & ${8.42}^{+0.11}_{-0.19}$ \\
1 & J015809 & $7.75\pm0.02$ & direct & ${8.14}^{+0.65}_{-0.21}$ & $\cdots$ & ${8.44}^{+0.19}_{-0.18}$ \\
1 & J104654 & $7.91\pm0.02$ & direct & ${8.16}^{+0.22}_{-0.05}$ & $\cdots$ & $\cdots$ \\
1 & J093006 & $8.02\pm0.02$ & direct & $\cdots$ & $\cdots$ & $\cdots$ \\
1 & J092055 & $7.87\pm0.02$ & direct & ${7.77}^{+0.36}_{-0.24}$ & ${7.95}^{+0.07}_{-0.08}$ & ${8.29}^{+0.10}_{-0.09}$ \\
1 & J082555 & $7.37\pm0.01$ & direct & ${8.16}^{+0.03}_{-0.02}$ & ${8.09}^{+0.02}_{-0.02}$ & ${8.27}^{+0.02}_{-0.02}$ \\
1 & J104457 & $7.45\pm0.02$ & direct & ${8.02}^{+0.03}_{-0.03}$ & ${7.74}^{+0.03}_{-0.03}$ & ${7.94}^{+0.00}_{-0.00}$ \\
1 & J120122 & $7.45\pm0.03$ & direct & $\cdots$ & $\cdots$ & $\cdots$ \\
1 & J124159 & $7.73\pm0.04$ & direct & ${7.97}^{+0.11}_{-0.09}$ & $\cdots$ & $\cdots$ \\
1 & J122622 & $7.90\pm0.01$ & direct & $\cdots$ & $\cdots$ & $\cdots$ \\
1 & J122436 & $7.84\pm0.02$ & direct & $\cdots$ & $\cdots$ & $\cdots$ \\
1 & J124827 & $7.81\pm0.03$ & direct & $\cdots$ & ${7.72}^{+0.09}_{-0.10}$ & ${8.67}^{+0.20}_{-0.46}$ \\
2 & rcs0327-B & $8.11\pm0.10$ & PP04-N2 & ${8.17}^{+0.20}_{-0.08}$ & ${7.57}^{+0.07}_{-0.07}$ & $\cdots$ \\
2 & rcs0327-E & $8.40\pm0.10$ & PP04-N2 & ${8.30}^{+0.01}_{-0.11}$ & ${7.86}^{+0.07}_{-0.06}$ & $\cdots$ \\
2 & rcs0327-G & $\cdots$ & $\cdots$ & ${8.17}^{+0.12}_{-0.07}$ & ${7.91}^{+0.09}_{-0.04}$ & $\cdots$ \\
2 & rcs0327-U & $8.16\pm0.10$ & PP04-N2 & $\cdots$ & $\cdots$ & $\cdots$ \\
2 & S0004-0103 & $\leq 8.10$ & PP04-N2 & ${7.99}^{+0.51}_{-0.21}$ & ${7.58}^{+0.13}_{-0.15}$ & $\cdots$ \\
2 & S0108+0624 & $\cdots$ & $\cdots$ & $\cdots$ & $\cdots$ & $\cdots$ \\
2 & S0900+2234 & $8.12\pm0.10$ & PP04-N2 & $\cdots$ & $\cdots$ & $\cdots$ \\
2 & S0957+0509 & $\cdots$ & $\cdots$ & ${8.03}^{+0.08}_{-0.08}$ & ${7.82}^{+0.09}_{-0.08}$ & $\cdots$ \\
2 & Cosmic Horseshoe & $8.45\pm0.10$ & PP04-N2 & $\cdots$ & $\cdots$ & $\cdots$ \\
2 & S1226+2152 & $7.89\pm0.30$ & KK04-R23-l & $\cdots$ & $\cdots$ & $\cdots$ \\
2 & S1429+1202 & $\cdots$ & $\cdots$ & $\cdots$ & ${6.92}^{+0.10}_{-0.11}$ & $\cdots$ \\
2 & S1458-0023 & $\cdots$ & $\cdots$ & $\cdots$ & $\cdots$ & $\cdots$ \\
2 & S1527+0652 & $8.53\pm0.10$ & PP04-N2 & $\cdots$ & $\cdots$ & $\cdots$ \\
2 & S2111-0114 & $\cdots$ & $\cdots$ & $\cdots$ & $\cdots$ & $\cdots$ \\
2 & Cosmic Eye & $8.02\pm0.30$ & KK04-R23-u & $\cdots$ & $\cdots$ & $\cdots$ \\
2 & Planck Arc & $8.16\pm0.09$ & PP04-N2 & $\cdots$ & ${7.52}^{+0.04}_{-0.03}$ & $\cdots$ \\
2 & PSZ0441 & $\cdots$ & $\cdots$ & $\cdots$ & $\cdots$ & $\cdots$ \\
2 & SPT0310 & $\cdots$ & $\cdots$ & $\cdots$ & $\cdots$ & $\cdots$ \\
2 & SPT2325 & $\cdots$ & $\cdots$ & $\cdots$ & $\cdots$ & $\cdots$ \\
3 & A1689\_876\_330 & $\cdots$ & $\cdots$ & $\cdots$ & $\cdots$ & $\cdots$ \\
3 & A1689\_863\_348 & $\cdots$ & $\cdots$ & ${7.98}^{+0.62}_{-0.09}$ & $\cdots$ & ${8.58}^{+0.26}_{-0.58}$ \\
3 & A1689\_860\_359 & $\leq 8.54$ & KK04-R23-l & ${8.19}^{+0.75}_{-0.25}$ & $\cdots$ & ${8.47}^{+0.32}_{-0.34}$ \\
3 & MACS0451\_1.1 & $\leq 7.95$ & PP04-N2 & ${8.06}^{+0.52}_{-0.23}$ & ${7.56}^{+0.13}_{-0.11}$ & $\cdots$ \\
4 & 2 & $7.81\pm0.07$ & direct & $\cdots$ & ${7.91}^{+0.03}_{-0.03}$ & ${7.98}^{+0.04}_{-0.03}$ \\
4 & 36 & $7.92\pm0.04$ & direct & $\cdots$ & $\cdots$ & $\cdots$ \\
4 & 80 & $8.24\pm0.06$ & direct & $\cdots$ & $\cdots$ & $\cdots$ \\
4 & 82 & $7.91\pm0.04$ & direct & $\cdots$ & ${8.12}^{+0.02}_{-0.02}$ & ${8.23}^{+0.01}_{-0.01}$ \\
4 & 110 & $8.17\pm0.08$ & direct & $\cdots$ & $\cdots$ & $\cdots$ \\
4 & 111 & $7.81\pm0.08$ & direct & $\cdots$ & $\cdots$ & $\cdots$ \\
4 & 179 & $8.35\pm0.07$ & direct & $\cdots$ & $\cdots$ & $\cdots$ \\
4 & 182 & $8.01\pm0.04$ & direct & $\cdots$ & ${7.94}^{+0.00}_{-0.01}$ & ${8.36}^{+0.02}_{-0.02}$ \\
4 & 191 & $8.30\pm0.07$ & direct & $\cdots$ & $\cdots$ & $\cdots$ \\
4 & 198 & $8.48\pm0.11$ & direct & $\cdots$ & $\cdots$ & $\cdots$ \\
5 & Q2343-BX418 & $7.90\pm0.20$ & direct & $\cdots$ & ${7.54}^{+0.07}_{-0.07}$ & $\cdots$ \\
6 & A1689\_31.1 & $7.69\pm0.13$ & direct & $\cdots$ & $\cdots$ & ${8.69}^{+0.52}_{-0.39}$ \\
7 & SL2SJ0217 & $\geq {7.50}$ & direct & ${7.99}^{+0.02}_{-0.03}$ & ${7.60}^{+0.01}_{-0.01}$ & ${8.64}^{+0.06}_{-0.04}$ \\
8 & stack & $8.14\pm0.04$ & direct & ${8.05}^{+0.25}_{-0.13}$ & ${8.16}^{+0.22}_{-0.03}$ & $\cdots$ \\
9 & SGASJ105039 & $8.26\pm0.20$ & KK04-R23-l & ${8.24}^{+0.59}_{-0.25}$ & ${7.91}^{+0.31}_{-0.14}$ & $\cdots$ \\
10 & HS1442+4250 & $7.65\pm0.04$ & direct & $\cdots$ & ${7.77}^{+0.03}_{-0.03}$ & ${8.19}^{+0.00}_{-0.00}$ \\
10 & J0405-3648 & $7.56\pm0.07$ & direct & $\cdots$ & $\cdots$ & $\cdots$ \\
10 & J0940+2935 & $7.63\pm0.14$ & direct & $\cdots$ & $\cdots$ & $\cdots$ \\
10 & J1119+5130 & $7.51\pm0.07$ & direct & $\cdots$ & $\cdots$ & $\cdots$ \\
10 & SBSG1129+576 & $7.47\pm0.06$ & direct & $\cdots$ & $\cdots$ & $\cdots$ \\
10 & UM133 & $7.70\pm0.09$ & direct & $\cdots$ & ${7.89}^{+0.05}_{-0.04}$ & $\cdots$
\enddata
\tablecomments{(1) \citet{Berg+2016}; (2) \citet{Rigby+2018a}; (3) \citet{Stark+2014}; (4) \citet{Senchyna+2017}; (5) \citet{Erb+2010}; (6) \citet{Christensen+2012}; (7) \citet{Berg+2018}; (8) \citet{Steidel+2016}, a stack of 30 galaxy spectra; (9) \citet{Bayliss+2014}; (10) \citet{Senchyna+2019}.}
\end{deluxetable*}

\section{UV-Optical abundance comparisons}\label{sec:ZZ}

In this section, we compare the metallicities derived with UV emission lines (\S\ref{sec:Z:UV}) to those derived with optical emission lines, to evaluate the utility of UV diagnostic diagrams as metallicity indicators.

Our UV-optical sample requires three significant emission line detections in the UV. Across the sample, \ciii$\,\lambda$1906,1909 and \oiii$\,\lambda$1666 were the most commonly detected emission lines. This is not surprising, given that the \ciii$\,\lambda$1906,1909 doublet is the brightest emission line in the UV spectra of star-forming galaxies after Ly-$\alpha$. As such, a number of authors have suggested emission line diagnostics that make use of the \ciii$\,\lambda$1906,1909 doublet \citep[e.g.,][]{Feltre+2016, Jaskot+2016, Byler+2018, Hirschmann+2019}.

In \S\ref{sec:ZZ:Si}-\ref{sec:ZZ:CIV}, we highlight three UV diagnostic diagrams that use the \ciii$\,\lambda$1906,1909 and \oiii$\,\lambda$1666 lines paired with a third emission line: \SiuIII$\,\lambda$1893 (\S\ref{sec:ZZ:Si}), \heii$\,\lambda$1640 (\S\ref{sec:ZZ:He}), and the \civ$\,\lambda$1548,1550 doublet (\S\ref{sec:ZZ:CIV}).

\subsection{\SiuIII$\,\lambda$1883,1893}\label{sec:ZZ:Si}

In \citetalias{Byler+2018}, we highlighted the potential of the \SiuIII$\,\lambda$1883 / \ciii$\,\lambda$1906 (Si3C3) \vs \oiii$\,\lambda$1666 / \ciii$\,\lambda$1906 (O3C3) diagnostic diagram. These emission lines are relatively bright and easy to detect, and are closely spaced in wavelength to minimize dust extinction errors.

\begin{figure*}
  \begin{center}
    \includegraphics[width=\linewidth]{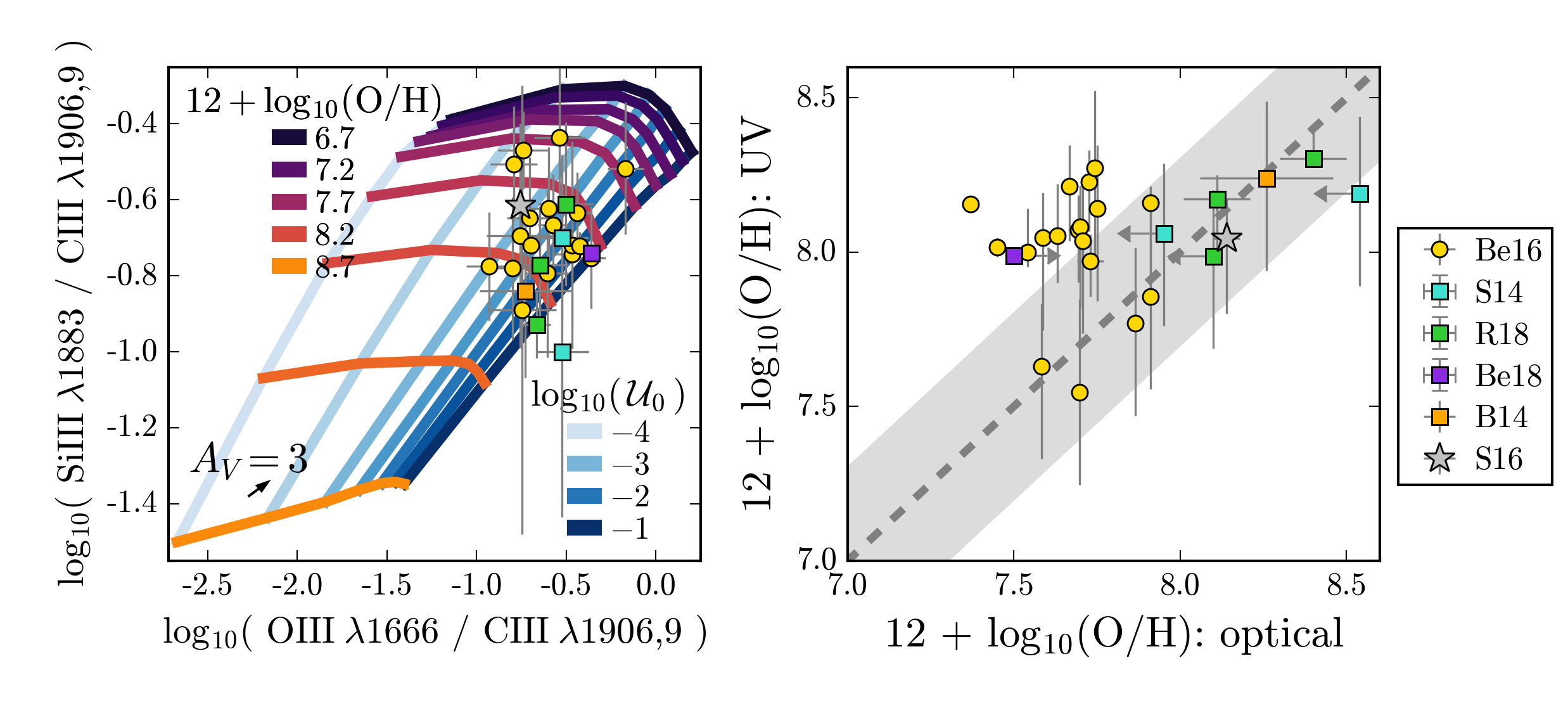}
    \caption{\emph{Left:} Si3-O3C3 UV diagnostic diagram, as presented in \citetalias{Byler+2018}. The blue lines connect models of constant ionization parameter, from \logUeq{-1} (dark blue) to \logUeq{-4} (light blue). Models of constant metallicity are shown from $12+\logOH=6.69$ (\logZeq{-2}; purple) to $12+\logOH=8.69$ (\logZeq{0}; orange). \emph{Right:} Metallicity derived using the Si3-O3C3 UV diagnostic ($y$-axis) compared to the optically-derived metallicity ($x$-axis). The dashed line shows a one-to-one relationship, and a 0.3 dex spread is shown by the grey shaded region. Identical symbols are used for the galaxy sample in both panels. Low-redshift galaxies from \citet{Berg+2016} are shown with gold circles. Moderate-redshift galaxies are shown with squares, from \citet{Stark+2014} (cyan), \citet{Bayliss+2014} (orange), \citet{Rigby+2018b} (green), and \citet{Berg+2018} (purple). The gray star is the stacked spectrum of moderate-redshift galaxies from \citet{Steidel+2016}.}
    \label{fig:UVSi}
  \end{center}
\end{figure*}

In the left panel of Fig.~\ref{fig:UVSi} we show the Si3-O3C3 diagnostic diagram as presented in \citetalias{Byler+2018}. We compare the model grid with the observed galaxy sample, where the nearby galaxies are shown with circular markers and moderate-redshift galaxies are shown with square markers, including \citealt{Berg+2016} (Be16; gold circles), \citealt{Rigby+2018b} (R18; green squares), \citealt{Stark+2014} (S14; cyan squares), and \citealt{Berg+2018} (Be18; purple square). The stacked spectrum from \citealt{Steidel+2016} (S16) is shown with the gray star. As noted in \citetalias{Byler+2018}, the model grid is able to reproduce the observed range of line ratios in samples of both low and moderate-redshift galaxies.

The right panel of Fig.~\ref{fig:UVSi} shows the comparison between optical metallicity ($x$-axis) and UV metallicity ($y$-axis) derived with the Si3-O3C3 diagnostic. Galaxy observations are shown with the same marker shapes and colors as in the left panel, and the black dashed line shows a one-to-one correlation between UV and optical metallicities. The grey shaded region shows a $0.3$ dex spread from the one-to-one relationship. In some cases, the optical metallicity errors may be underestimated, and $\pm 0.3$ dex represents the typical systematic errors inherent in optical strong line methods \citep{Kewley+2019ARAA}.

The UV and optical metallicities agree within error for 13 of the 26 galaxies ($50\%$). In general, the UV metallicity is biased toward higher values, with a median offset of 0.35\,dex from optical metallicities. The UV metallicities show significant scatter and are only marginally positively correlated with optical measurements, with a Spearman correlation coefficient of 0.26. Additionally, there are several objects (S14, Be18) where the UV metallicity catastrophically fails to match the optical metallicity. However, these failures do not seem to have an overall bias toward higher or lower metallicities.

In general, we do not necessarily expect the UV and optical abundances to exactly match, because many bright UV emission lines (e.g., \ciii) are relatively high ionization species. These high ionization lines may reflect conditions in the inner part of the nebula where the gas is more highly ionized than in the regions probed by optical lines, leading to an offset between UV and optical metallicity. In the case of the direct-\Te{} method, the metallicity offset is driven by temperature differences; for strong line methods, the offset would be driven by variations in ionization parameter that are not accounted for in the optical metallicity diagnostic. However, most of the galaxies in this sample are relatively high excitation objects, with temperatures dominated by the high-excitation zone. Moreover, the \emph{scatter} in UV-derived abundances for a comparatively narrow range in optical abundance is surprising, and suggests the source of metallicity discrepancy has a different origin. It is possible that the scatter in Si3-O3C3 metallicities is the result of variation in elemental abundances (i.e., carbon or silicon relative to oxygen) or in the dust depletion factors, as both carbon and silicon are expected to be heavily depleted from the gas phase onto dust grains.

We discuss the Si3-O3C3 metallicities in more detail in \S\ref{sec:discussion}, where we use rest-optical spectroscopy from the \citet{Berg+2016} sample to better understand the source of the scatter.

\subsection{\heii$\,\lambda$1640}\label{sec:ZZ:He}

\heii$\,\lambda$1640 emission has been detected locally and at high-redshift, and is one of the brighter UV emission lines. The \heii$\,\lambda$1640 line has been used in several proposed emission line diagnostics \citep[e.g.,][]{Jaskot+2016, Feltre+2016}, including the He2-O3C3 diagnostic presented in \citetalias{Byler+2018}, which uses the \heii$\,\lambda$1640, \ciii$\,\lambda$1906,9 and \oiii$\,\lambda$1666 emission lines. In \citetalias{Byler+2018}, however, we noted that observations of \heii$\,\lambda$1640 emission can include contributions from both nebular emission and stellar wind emission, making it a potentially problematic metallicity tracer.

\begin{figure*}
  \begin{center}
    \includegraphics[width=\linewidth]{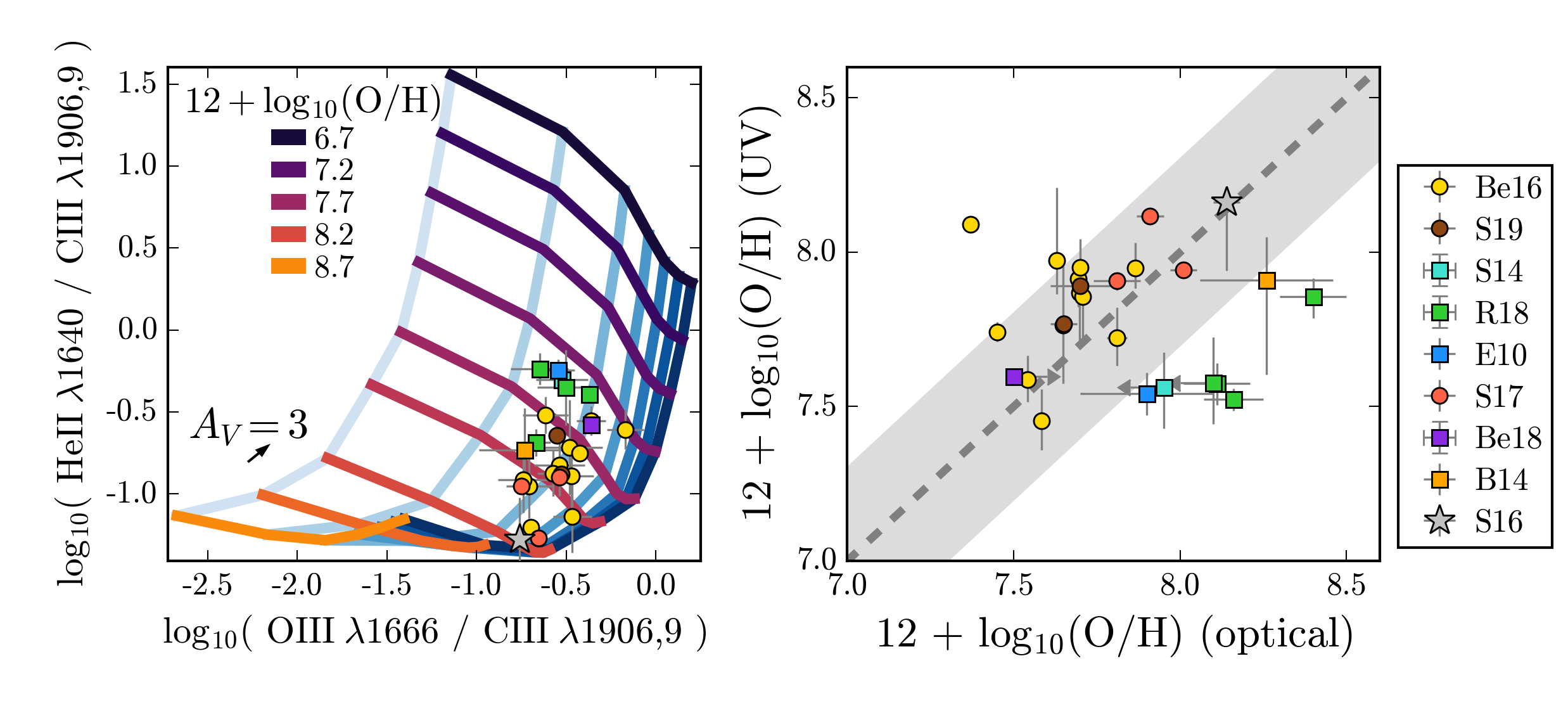}
    \caption{\emph{Left:} The He2-O3C3 diagnostic diagram, with the model grid as described in the caption of Fig~\ref{fig:UVSi}. \emph{Right:} Metallicity derived with the He2-O3C3 diagnostic ($y$-axis) compared to the optical metallicity ($x$-axis), where the dashed line shows a one-to-one relationship. The UV and optical metallicities are well-matched for most of the galaxies, particularly the metal-poor objects. For the comparatively metal-rich \mage galaxies (green squares), the UV diagnostic significantly under-predicts optical metallicities.}
    \label{fig:UVHe}
  \end{center}
\end{figure*}

We assess the He2-O3C3 diagnostic in Fig.~\ref{fig:UVHe}, following the format of Fig.~\ref{fig:UVSi}, where the left panel shows the diagnostic diagram and the right panel shows the comparison between UV and optical metallicity.

In general, the agreement between the UV and optical metallicities with the He2-O3C3 diagnostic is improved over the the Si3-O3C3 diagnostic. The UV metallicities in the right panel of Fig.~\ref{fig:UVHe} show a stronger correlation with the optical metallicity (with a Spearman correlation coefficient of 0.3) and have a median offset of 0.1\,dex. However, there are still three objects where the UV metallicity is entirely at odds with the optical abundance. Specifically, for the E10, B14, and R18 galaxies (blue, orange, and green squares, respectively), the UV abundance is systematically lower than the optical abundance by 0.3-0.6\,dex.

The source of \ion{He}{2} emission is a subject of ongoing debate, and \ion{He}{2} emission is difficult to produce with current models, both stellar and nebular. Narrow \ion{He}{2} emission is generally interpreted as having a nebular origin, and requires significant numbers of high energy photons. With currently available stellar models, very hard ionizing spectra requires the presence of stellar multiplicity, stellar rotation, or very massive stars \citep[e.g.,][]{Stark+2014, Steidel+2016, Byler+2017, Choi+2017}. 

Broad \ion{He}{2} emission in galaxies is commonly interpreted as an indication of the presence of W-R stars \citep[e.g., ][]{Kunth+1985, Conti+1991, Schaerer+1999, Brinchmann+2008}. W-R stars are more common in metal-rich stellar populations ($12+\logOH \gtrsim 8$), with the strongest \ion{He}{2} emission associated with populations at solar metallicity or higher. 

With the exception of the \citet{Senchyna+2017} and \citet{Berg+2018} spectra (red circles and purple square, respectively), none of the spectra in the sample has fit separate components for broad and narrow \ion{He}{2} emission. Most of the \citet{Berg+2016} galaxies (gold circles) have low enough metallicities that the ``contamination'' from stellar emission should be small (${\sim}$25\% or less of the nebular emission flux), and there is no evidence that the \ion{He}{2} emission is any broader than the other nebular emission lines. If broad \ion{He}{2} were responsible for artificially inflating the observed \heii$\,\lambda1640$ fluxes, we might expect that the contamination would be worse for the relatively metal-rich \mage galaxies (R18, green squares).

We note that the stacked spectrum from \citet{Steidel+2016} has relatively high metallicity but does not seem to suffer from under-predicted UV metallicities like the \mage galaxies. However, as a composite spectrum, it is difficult to extrapolate how light-weighted changes from the stellar continuum and nebular components ultimately impacts the relative line strengths.

We discuss the nature of the \ion{He}{2} emission at length in \S\ref{sec:discussion:HeII}, where we compare the predictions from rotating stellar populations (used in this section) to predictions from binary stellar models, which have harder ionizing spectra at older ages. We also assess the level of contamination from broad stellar emission using a ``wind-contaminated'' emission line grid.

The He2-O3C3 diagram shows promise as an oxygen abundance diagnostic, especially at low metallicities where the stellar contribution is minimal ($12+\logOH \lesssim 8$). However, a more detailed understanding of the various mechanisms responsible for \ion{He}{2} photon production is required before the He3-O3C3 diagnostic can be applied to large samples with confidence.

\subsection{\civ$\,\lambda$1548,1550}\label{sec:ZZ:CIV}

\begin{figure*}
  \begin{center}
    \includegraphics[width=\linewidth]{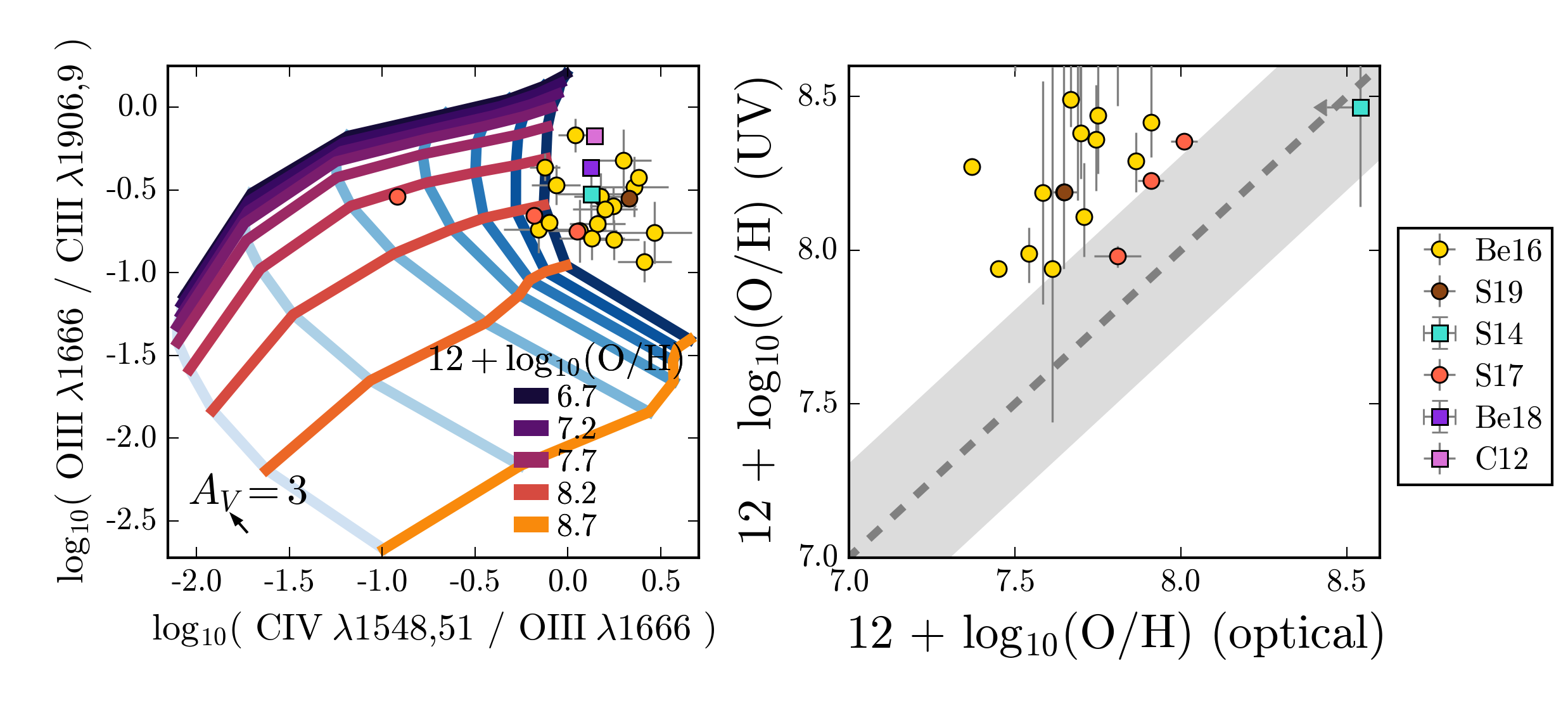}
    \caption{\emph{Left:} The C4-O3C3 diagnostic diagram, with the model grid as described in the caption of Fig~\ref{fig:UVSi}. Many of the galaxies have larger-than-predicted C4O3 ratios, in some cases due to contamination from stellar wind emission to the \civ{} emission line, though harder ionizing spectra could also help explain the offset. \emph{Right:} Metallicities predicted by the C4-O3C3 diagnostic compared to optically-derived metallicities. The dashed line shows a one-to-one relationship. The UV diagnostic predicts metallicities that are poorly correlated with optical metallicities, and offset by 0.3-0.5 dex.}
    \label{fig:UVCIV}
  \end{center}
\end{figure*}

In this section we assess the utility of the C4-O3C3 diagnostic, which uses the \civ$\,\lambda$ 1548,1550 / \oiii $\,\lambda$\,1666 and \oiii $\,\lambda$\,1666 / \ciii $\,\lambda$ 1906,1909 emission line ratios. The left panel of Fig.~\ref{fig:UVCIV} shows the model grid and observed emission line ratios. Unlike Figs.~\ref{fig:UVSi}~\&~\ref{fig:UVHe}, the model grid is unable to reproduce the range of observed emission line ratios, as noted in \citetalias{Byler+2018}. Here, 70\% of the galaxies have larger C4O3 ratios than predicted by the models.

The right panel of Fig.~\ref{fig:UVCIV} shows the comparison between UV and optical metallicities for the C4-O3C3 diagnostic. Unsurprisingly, the mismatch between observed and model line ratios translates to poor agreement between the UV and optical metallicities in the right panel. For nearly all objects, the UV metallicity is larger than the optical metallicity, by 0.1-0.7\,dex, with an average offset of 0.5\,dex. Only two objects have UV metallicities consistent with their optical estimates within errors. Additionally, there is some evidence for a metallicity-dependent offset, where the metal-rich objects show somewhat more scatter.

\ion{C}{4} emission is one of the more difficult spectral features to interpret, due to the competing effects of nebular emission at 1548 and 1550\ang, stellar wind emission at 1550\ang (often broad, with a strong P-Cygni profile) and interstellar absorption between $1545-1550$\ang. Generally, the strength of the nebular \ion{C}{4} emission peaks at low metallicity ($12+\logOH \sim 7$) and high ionization parameter (\logU$\gtrsim-2$). In contrast, stellar emission is wind-driven and strongest at higher metallicities, solar-like and above. However, even at $12+\logOH = 7.5$ (\logZ${\sim}-1$), stellar \ion{C}{4} emission can account for as much as 30\% of the total \ion{C}{4} emission \citepalias{Byler+2018}. Interstellar absorption plays an important role at higher metallicities (solar-like and above), where it can dominate the composite \ion{C}{4} spectral feature. ISM absorption must be accounted for when making integrated line index measurements \citep[e.g.,][]{Vidal-Garcia+2017}, however, $R>1000$ is generally sufficient to distinguish the narrow interstellar absorption from the broad P-Cygni absorption \citep[e.g.,][]{Crowther+2006, Chisholm+2019}.

At the metallicities associated with the nearby galaxy sample ($12+\logOH \sim 7.5-8$; yellow and red circles), the {\tt MIST+wind} \ion{C}{4} emission models from \citetalias{Byler+2018} predict that stellar contamination can change the C4O3 ratio by 0.1-0.3\,dex. We note that contamination from stellar emission should less of an issue for the \citet{Senchyna+2017} galaxies (red circles), because these observations fit for both broad and narrow \ion{C}{4} components. The \ion{C}{4} flux used in the line ratios from Fig.~\ref{fig:UVCIV} is that of the narrow, nebular \ion{C}{4} only. Encouragingly, the \citet{Senchyna+2017} galaxies do show the smallest offset, potentially due to the fact that the \ion{C}{4} emission line fluxes are a better representation of the uncontaminated nebular flux.

We note the absence of the \mage galaxies in Fig.~\ref{fig:UVCIV}. Inspection of the \ion{C}{4} emission feature in these comparatively metal-rich objects reveals strong stellar emission with the signature P-Cygni profile, and little or no evidence for nebular emission\footnote{with the exception of RCS0327-G, which does not have a matched optical metallicity}\citep{Chisholm+2019}.

In \S\ref{sec:discussion} we test whether harder ionizing spectra or stellar wind contamination can ameliorate the mismatch between model and data \ion{C}{4} line strengths.

\subsection{Abundance determination equations}\label{sec:ZZ:poly}
As described in \S\ref{sec:Z}, the UV oxygen abundances presented in Table~\ref{tab:logOH} are calculated by interpolating the model grid. For users who wish to replicate this process on their own data, the model emission line ratios were published in \citetalias{Byler+2018} and are publicly available.

To facilitate abundance determinations for the UV diagnostics discussed in this work, we also provide a simple functional form for the diagnostic diagrams shown in Figs.~\ref{fig:UVSi} \&~\ref{fig:UVHe}. Using {\tt Polynomial2D} from {\tt astropy.models}, we fit a 3\textsuperscript{rd} degree 2D polynomial to the model grid surface with a Levenberg-Marquardt algorithm and least squares statistic from {\tt astropy.fitting}. These fits are valid for objects with $6.2 \leq 12+\logOH \leq 9.2$, and $-4.0 \leq \logU \leq -1.0$. These fits should only be applied to objects with observed line ratios that are well-described by the model grid (i.e., extrapolation to off-grid data points is not valid).

For the Si3-O3C3 diagnostic (\S\ref{sec:ZZ:Si}), the fit yields:
\begin{equation}\begin{aligned}\label{eq:polySi}
    12+&\logOH = 3.09\;+\\
    &0.09x -1.71x^2 -0.73x^{3}\\
    &-16.51y -19.84y^{2} -6.26y^{3}\\
    &4.79xy -0.28xy^{2} +1.67x^{2}y,
\end{aligned}\end{equation}
where $x$ is $\log_{10}($\oiii$\,\lambda1666$/\ciii$\,\lambda1906,9)$ and $y$ is $\log_{10}($\SiuIII$\,\lambda1883$/\ciii$\,\lambda1906,9)$. Typical statistical errors are $\pm 0.14$\,dex.

For the He2-O3C3 diagnostic (\S\ref{sec:ZZ:He}), the fit yields:
\begin{equation}\begin{aligned}\label{eq:polyHe}
    12+&\logOH = 6.88 +\\
    &-1.13x -0.46x^2 -0.03x^{3}\\
    &-0.61y +0.02y^{2} -0.04y^{3}\\
    &-0.32xy +0.03xy^{2} -0.21x^{2}y,
\end{aligned}\end{equation}
where $x$ is $\log_{10}($\oiii$\,\lambda1666$/\ciii$\,\lambda1906,9)$ and $y$ is $\log_{10}($\heii$\,\lambda1640$/\ciii$\,\lambda1906,9)$. Typical statistical errors are $\pm 0.08$\,dex.

We do not provide an equation for the C4-O3C3 diagnostic (\S\ref{sec:ZZ:CIV}), because the model grid does not satisfactorily cover observed parameter space, which we discuss in \S\ref{sec:discussion:CIV}.

We note that the oxygen abundances obtained with Eqs.~\ref{eq:polySi}~\&~\ref{eq:polyHe} will not be identical to those obtained from the direct interpolation of the model grid. However, the oxygen abundances from the two methods are tightly correlated, with $0.11$ and $0.03$ dex scatter for Si3-O3C3 (Eq.~\ref{eq:polySi}) and He2-O3C3 (Eq.~\ref{eq:polyHe}), respectively. For Si3-O3C3 diagnostic, this scatter can be reduced to $0.06$ dex by using a 4\textsuperscript{th} degree 2D polynomial, which increases the number of terms in the equation from 10 to a more unwieldy 15. We provide the coefficients for all fits in Tables~\ref{tab:poly3deg}~\&~\ref{tab:poly4deg}, following the form for a general polynomial of degree n\footnote{\emph{Quick-start for Python users:} Create a dictionary of coefficient names and values from either Table~\ref{tab:poly3deg} or \ref{tab:poly4deg}, ``{\tt coeffs}''. Input this dictionary to {\tt Polynomial2D(degree, **coeffs)} from {\tt astropy.modeling.models}.}: 
\begin{equation}\begin{aligned}\label{eq:genPoly}
P(x,y) &= c_{00} + c_{10}x + \dotsb \\
& + c_{n0}x^n + c_{01}y + \dotsb \\
& + c_{0n}y^n + c_{11}xy + c_{12}xy^2 + \dotsb \\
& + c_{1(n-1)}xy^{n-1}+ \dotsb \\
& + c_{(n-1)1}x^{n-1}y. 
\end{aligned}\end{equation}

\begin{figure*}
  \begin{center}
    \includegraphics[width=\linewidth]{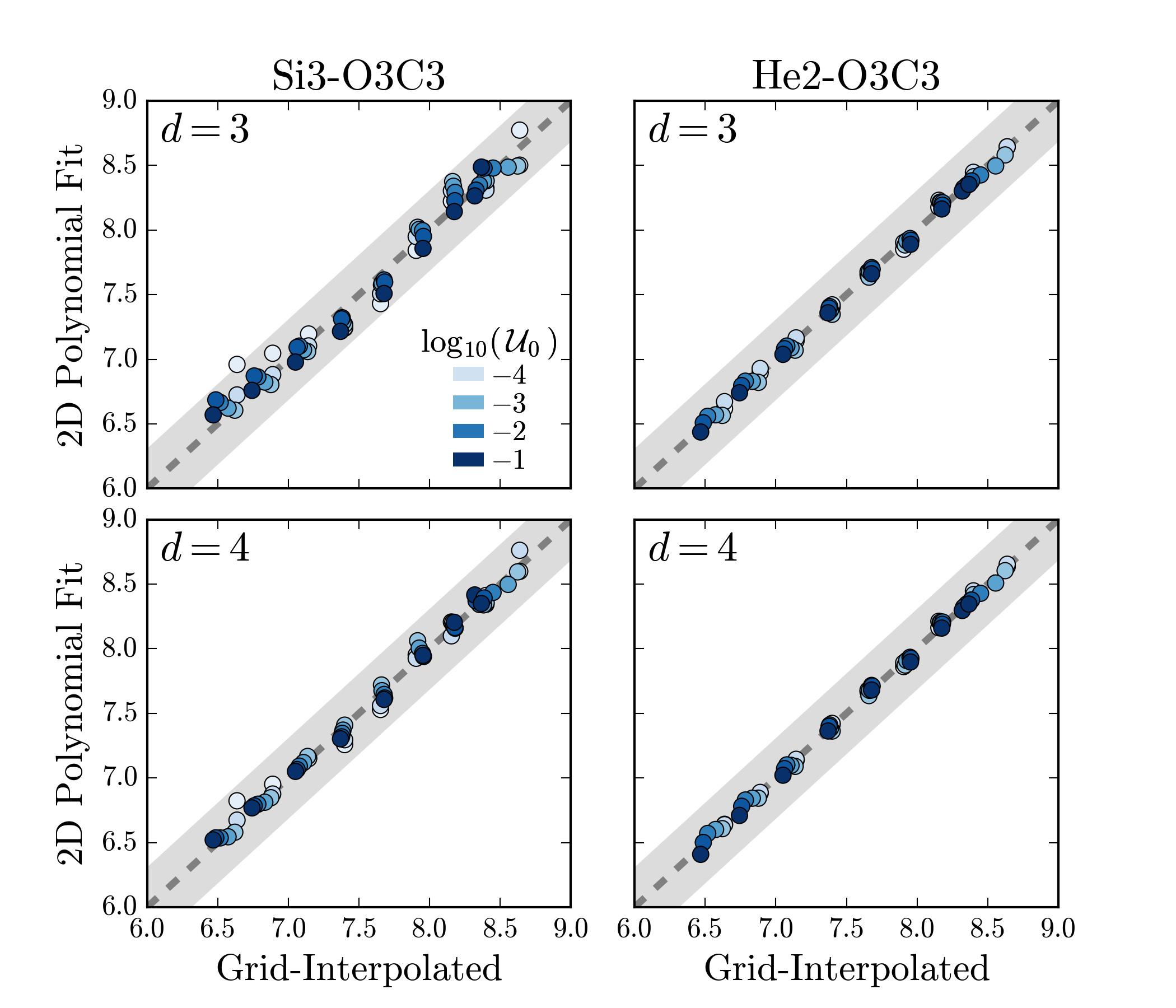}
    \caption{Comparison of oxygen abundances derived from the 2D polynomial fits ($y$-axis) and direct model grid interpolation ($x$-axis). The grey shaded region shows 0.3 dex scatter around a one-to-one relationship (dashed line). \emph{Left column:} The Si3-O3C3 diagnostic, for the 3\textsuperscript{rd} degree 2D polynomial (top; Eq.~\ref{eq:polySi} or Table~\ref{tab:poly3deg}) and the 4\textsuperscript{th} degree 2D polynomial (bottom; Eq.~\ref{eq:genPoly} or Table~\ref{tab:poly4deg}). \emph{Right column:} The He2-O3C3 diagnostic, for the degree=3 2D polynomial (top; Eq.~\ref{eq:polyHe} or Table~\ref{tab:poly3deg}) and the degree=4 2D polynomial (bottom; Eq.~\ref{eq:genPoly} or Table~\ref{tab:poly4deg}). }
    \label{fig:polyfit}
  \end{center}
\end{figure*}

A comparison of the oxygen abundances derived with Eqs.~\ref{eq:polySi}~\&~\ref{eq:polyHe} is shown in Fig.~\ref{fig:polyfit}. The abundance derived from the direct interpolation of the model grid is shown on the $x$-axis and the oxygen abundance derived from the polynomial fit is shown on the $y$-axis. The Si3-O3C3 diagnostic is shown in the left column and the He2-O3C3 diagnostic is shown in the right column. For each diagnostic, the 3\textsuperscript{rd} degree 2D polynomial is shown on the top (Table~\ref{tab:poly3deg}) and the 4\textsuperscript{th} degree 2D polynomial is shown on the bottom (Table~\ref{tab:poly4deg}).

\begin{deluxetable}{lcc}
\tabletypesize{\footnotesize}
\tablecolumns{3}
\tablecaption{Coefficients for the 3\textsuperscript{rd} degree 2D polynomial fit to the model grid surfaces.\label{tab:poly3deg}}
\tablehead{
\colhead{ } &
\colhead{Si3-O3C3} &
\colhead{He2-O3C3}
}
\startdata
c0\_0 & $3.09$ & $6.88$ \\
c1\_0 & $0.09$ & $-1.13$ \\
c2\_0 & $-1.71$ & $-0.46$ \\
c3\_0 & $-0.73$ & $-0.03$ \\
c0\_1 & $-16.51$ & $-0.61$ \\
c0\_2 & $-19.84$ & $0.02$ \\
c0\_3 & $-6.26$ & $-0.04$ \\
c1\_1 & $4.79$ & $-0.32$ \\
c1\_2 & $-0.28$ & $0.03$ \\
c2\_1 & $1.67$ & $-0.21$ \\
\enddata
\end{deluxetable}

\begin{deluxetable}{lcc}
\tabletypesize{\footnotesize}
\tablecolumns{3}
\tablecaption{Coefficients for the 4\textsuperscript{th} degree 2D polynomial fit to the model grid surfaces.\label{tab:poly4deg}}
\tablehead{
\colhead{ } &
\colhead{Si3-O3C3} &
\colhead{He2-O3C3}
}
\startdata
c0\_0 & $-0.29$ & $6.88$ \\
c1\_0 & $1.23$ & $-1.28$ \\
c2\_0 & $-5.11$ & $-0.74$ \\
c3\_0 & $-2.58$ & $-0.14$ \\
c4\_0 & $1.01$ & $0.01$ \\
c0\_1 & $-40.48$ & $-0.59$ \\
c0\_2 & $-80.57$ & $0.06$ \\
c0\_3 & $-65.69$ & $-0.04$ \\
c0\_4 & $-13.78$ & $0.03$ \\
c1\_1 & $17.89$ & $-0.30$ \\
c1\_2 & $15.28$ & $0.33$ \\
c1\_3 & $-11.48$ & $0.00$ \\
c2\_1 & $1.67$ & $-0.44$ \\
c2\_2 & $18.33$ & $0.20$ \\
c3\_1 & $-8.27$ & $-0.18$ \\
\enddata
\end{deluxetable}

\section{Discussion}\label{sec:discussion}

\subsection{The silicon discrepancy}\label{sec:discussion:Si}

The galaxies in the Berg sample have a fairly narrow range in optically-derived nebular properties, with high ionization parameters ($-2.8 \lesssim \logU \lesssim -1.8$) and low metallicities ($7.5 \lesssim 12 + \logOH \lesssim 8 $). We would thus expect UV-derived metallicities for these objects to reflect the similarity in gas properties. However, the  metallicities derived using the \SiuIII$\lambda$1883 line show 0.2 dex larger scatter and appear to have a bimodal distribution (Fig.~\ref{fig:UVSi}). 

The spread in UV-derived metallicities could be explained by variations in gas-phase silicon abundance relative to oxygen within the Berg sample. Silicon is an $\alpha$ element, and such variations could be the result of chemical evolution driven by star formation. However, silicon is also one of the main components of cosmic dust, which complicates abundance determinations. 

Moreover, the chemical composition of dust can evolve as grains lose atoms to the gas phase through high energy processes that occur in the supernova-generated shock waves in the ISM. High energy collisions between grains can cause erosion on the surface of dust, transferring elements (in particular Mg, Si, and Fe) from the grain surface to the gas phase. Notably, the fraction of silicon that is transferred back to the gas phase \emph{increases} with shock velocity \citep[see review on depletion patterns and dust evolution in][]{Jones+2000}.

Our model assumes that the gas phase abundance of silicon scales with the oxygen abundance, and that a fixed fraction of silicon is depleted onto dust grains (Table~\ref{tab:solarAbunds}), such that the ratio of between silicon and oxygen is constant in all models, $\log_{10}$(Si/O)$=-1.9$. This is lower than the average $\log_{10}$(Si/O)$=-1.6$ measured in extragalactic \hii{} regions by \citet{Garnett+1995} and the lensed galaxy from \citet{Berg+2018}, but similar to the $\log_{10}$(Si/O)$=-1.8$ measured from the \citet{Steidel+2016} stack of $z\sim2$ galaxies.

A total of 21 of the 26 galaxies in the Berg sample have significant detections of the collisionally excited intercombination \SiuIII$\,\lambda$1883,1892 doublet, which can be used to calculate the abundance of Si relative to C, and then combined with the C/O ratio to estimate the Si/O abundance ratio, as described in \citet{Berg+2018}. We note, however, that if the fraction of Si depleted onto dust grains varies significantly across the sample (i.e., if the depletion fraction varies with metallicity), the calculated Si/O ratios will be incorrect. A full analysis of Si/O abundances will be presented in Berg et al. (\emph{in preparation}).

\begin{figure}
  \begin{center}
    \includegraphics[width=\linewidth]{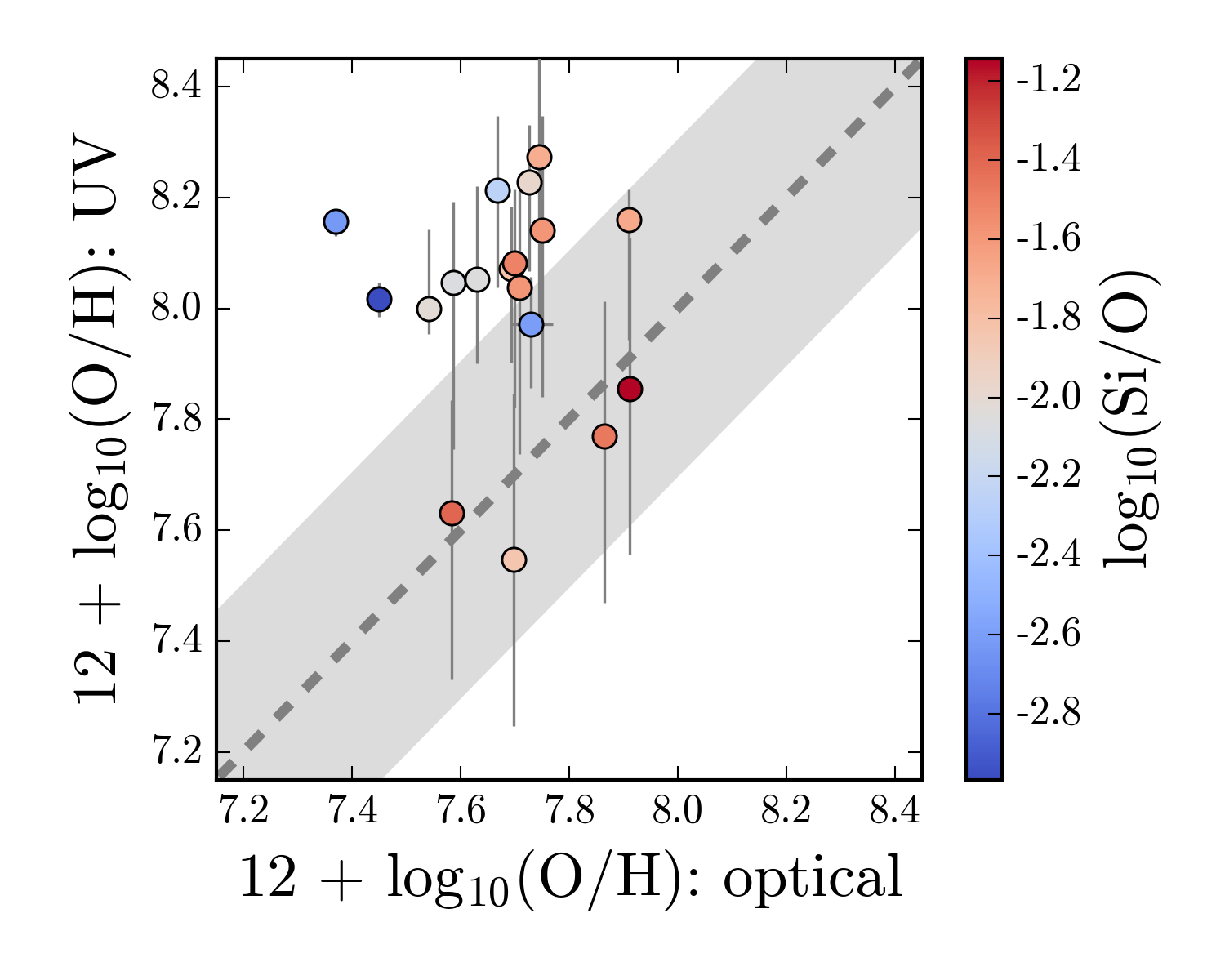}
    \caption{UV metallicities ($y$-axis), derived from the Si3-O3C3 diagnostic, compared to the optical metallicities ($x$-axis) for the \citet{Berg+2016} galaxies. Objects are color-coded by $\log_{10}$(Si/O). We suggest that the large scatter in the metallicities derived with the Si3-O3C3 diagnostic is driven by variations in the gas-phase silicon abundance.}
    \label{fig:SiO}
  \end{center}
\end{figure}

In Fig.~\ref{fig:SiO}, we again show the comparison between UV and optical metallicities for the Berg sample, derived using the Si3-O3C3 diagnostic. Now, each point is color-coded by $\log_{10}$(Si/O). The four objects offset from the rest of the \citet{Berg+2016} galaxies also have the largest $\log_{10}$(Si/O) abundance ratios, between $-1.8 \leq \log_{10}$(Si/O)$ \leq-1.1$. The four objects (J171236, J132347, J025346, J092055) have an average silicon abundance of $\log_{10}$(Si/O)$=-1.46\pm0.25$, which is more than 0.4\,dex larger than the average silicon abundance of the full sample, $\log_{10}$(Si/O)$=-1.89\pm0.48$, and the silicon abundance assumed in the model, $\log_{10}$(Si/O)$=-1.9$. It is interesting to note that these four offset objects also show the best agreement between UV and optical metallicities in the Si3-O3C3 diagnostic.

We conclude that the elevated Si/O abundance ratios in these four objects may be responsible for driving the large scatter in UV metallicities, though it is not clear what underlying physical process is responsible. If shocks driven by intense SF are responsible for returning additional silicon to the gas phase, we might expect the four offset objects to have larger ionization parameters or higher specific SFRs than the rest of the Berg sample, which they do not. The four objects have an average $\logUeq{-2.4} \pm 0.3$ and \logten(sSFR) $=-8.1 \pm 0.2$, compared to the averages for the full sample of $\logUeq{-2.3} \pm 0.3$ and \logten(sSFR) $=-8.1 \pm 0.3$.

Dust studies suggest that the amount of Si dust increases with reddening (e.g., \citealt{Haris+2016}, but see also \citealt{Mishra+2017}). We do not find a significant correlation between $E(B-V)$ and $\log_{10}$(Si/O) for these objects; however, the dust content in the BCD sample is generally quite low ($E(B-V)<0.18$). A more in-depth investigation of gas phase silicon abundances will be presented in future work.

\subsection{Broad or narrow emission?} \label{sec:discussion:broad}

Thus far, we have assumed that the \heii$\,\lambda$1640 and \civ$\,\lambda$1548,1551 emission is solely nebular in nature. However, \heii$\,\lambda$1640 and \civ$\,\lambda$1548,1551 emission can also be produced in stellar photospheres, artificially inflating the measured nebular emission line flux. In practice, it can be difficult to disentangle the narrow nebular and broad stellar components, especially at low signal-to-noise and moderate spectral resolution. Before discussing sources for the narrow, nebular emission, we briefly assess the level of ``contamination'' from stellar wind emission to the total line flux.

Stellar photospheric emission is produced in the winds of hot, young, stars. However, the \civ$\,\lambda$1548,1551 and \heii$\,\lambda$1640 lines are produced in very different types of stars. We should thus expect that the stellar contribution for each of these lines will scale differently with stellar metallicity and operate on different timescales.

\civ{} is produced in the atmospheres of massive main sequence stars, via line-driven winds. \civ{} emission is strongest at young ages (3-5\,\Myr and younger) and at high stellar metallicity (at or above solar metallicity) \citep{Walborn+1987, Pauldrach+1990, Leitherer+1995, Walborn+2002}.

Current research suggests that only W-R stars or Very Massive Stars (VMS) should produce significant \heii{} wind emission \citep[e.g.,][]{Crowther+2016, Leitherer+2018}. These stars are short-lived and we do not expect them to dominate the \heii{} emission in all but extremely young star bursts. \heii{} emission from W-R stars should be strongest at later times ($4-6\,$\Myr) and at high stellar metallicity \citep[e.g.,][]{Schaerer+1998, Vink+2005}. 

We note that our understanding of the physical mechanisms that drive the various W-R evolutionary pathways is still incomplete. Binary interactions enhance mass-loss and provide additional pathways to strip the outer hydrogen envelope from a star \citep[e.g.,][]{Eldridge+2017}. At low metallicities, rotational mixing can dredge up significant amounts of helium to the stellar surface, which can also produce broad \heii{} emission \citep[e.g., ][]{Yoon+2005, Cantiello+2007, Eldridge+2011, Choi+2017, Eldridge+2017}. Recent theoretical work suggests that chemical dredge-up in hydrogen-burning main sequence stars can produce surface enhancements in He and N consistent with W-R spectral classification \citep{Roy+2019}.

In the MIST models, broad \ion{He}{2} emission is produced by traditional W-R stars at high metallicity (solar-like and above) and rotational mixing at low metallicity (10\% solar and below; sometimes called quasi-homogeneous evolution or QHE). To quantify the relative importance of stellar and nebular emission for the \civ{} and \heii{} spectral features, we calculate the flux from both the stellar and nebular components. We refer to this model as the {\tt MIST+wind} model, which was first presented in \citetalias{Byler+2018}. A full description of the process is found in \citetalias{Byler+2018}; briefly, the ``total'' \civ{} or \heii{} emission flux is calculated by summing the flux from both the broad and narrow emission components.

\begin{figure*}
  \begin{center}
    \includegraphics[width=\linewidth]{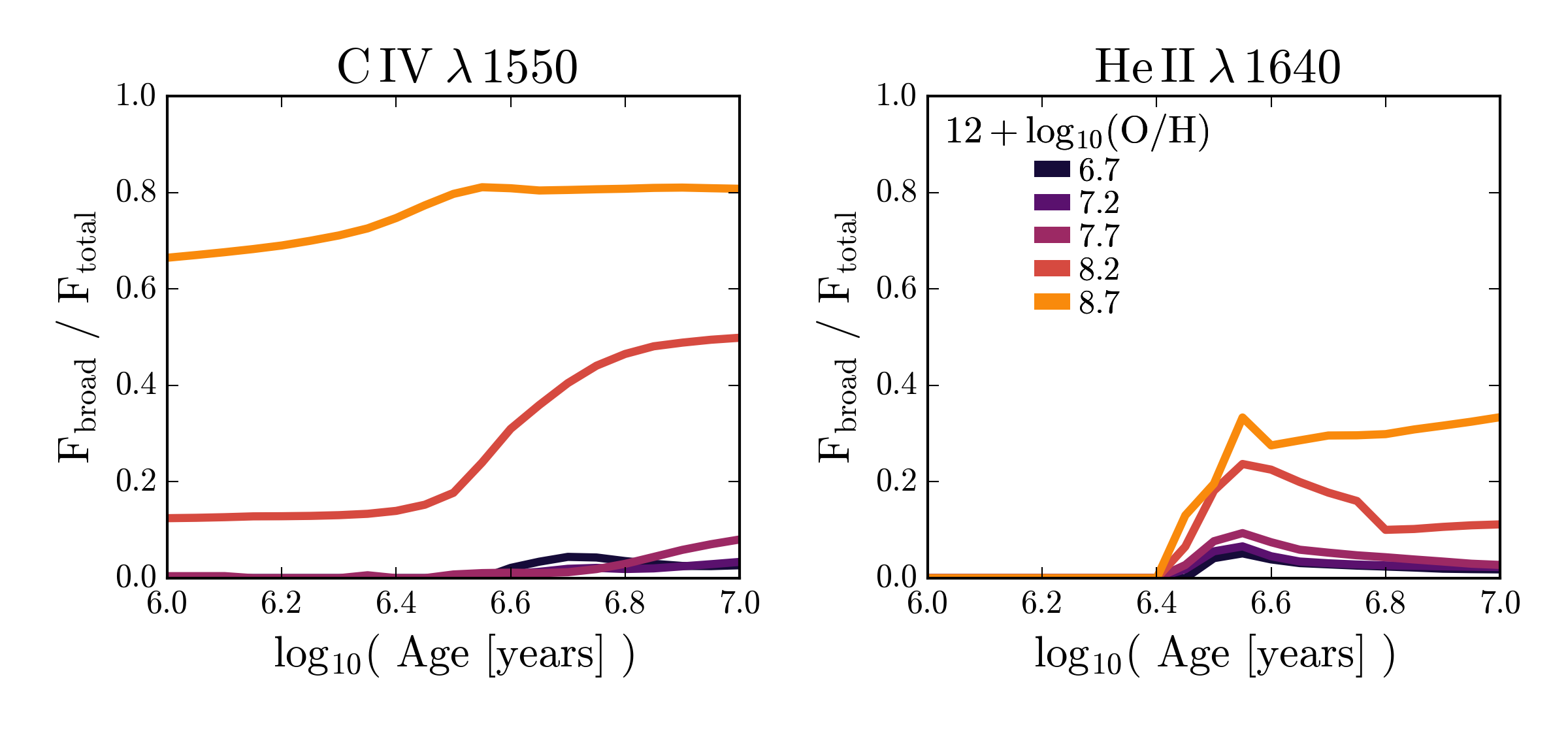}
    \caption{\emph{Left:} The fraction of the total \civ{} flux (the sum of broad wind and narrow nebular) from the broad component as a function of time.  \emph{Right:} The fraction of the total \heii{} flux (the sum of broad wind and narrow nebular) from the broad component as a function of time. In both panels, we use a constant SFR model with \logUeq{-2.5}. Lines are color-coded by metallicity, from $12+\logOH=6.7$ (\logZeq{-2}; purple) to $12+\logOH=8.7$ (\logZeq{0}; orange).}
    \label{fig:FracBroad}
  \end{center}
\end{figure*}

In Fig.~\ref{fig:FracBroad}, we show the fraction of the total \civ{} flux (left) and \heii{} flux (right) that is contained within the broad, stellar component ($F_{\mathrm{broad}}$/$F_{\mathrm{total}}$) as a function of model age, assuming \logUeq{-2.5}. The lines are color-coded by metallicity, from $12+\logOH=6.7$ (\logZeq{-2}; purple) to $12+\logOH=8.7$ (\logZeq{0}; orange). $F_{\mathrm{broad}}$/$F_{\mathrm{total}}$ initially increases as the population of young main sequence stars builds. $F_{\mathrm{broad}}$/$F_{\mathrm{total}}$ eventually plateaus as the rate of stars being formed reaches an equilibrium with the rate of stars leaving the main sequence.

For both \civ{} and \heii{}, $F_{\mathrm{broad}}$/$F_{\mathrm{total}}$ is highest in the solar metallicity models, and decreases with decreasing metallicity. For \civ{} (left panel), stellar emission contributes ${\sim}80\%$ of the total \civ{} flux at solar metallicity, decreasing to 10\% at $12+\logOH=7.7$ (0.1\,Z$_{\odot}$). For \heii{} (right panel), stellar emission contributes ${\sim}40\%$ of the total \heii{} flux at solar metallicity, and contributes less than a few percent of the total \heii{} flux at $12+\logOH=7.7$ (0.1\,Z$_{\odot}$).

For both lines, the relative strength of the stellar emission at high metallicity is further enhanced by the paucity of narrow emission at these metallicities, a by-product of cooler nebular temperatures and softer radiation fields. Similarly, the broad contribution is more modest at low metallicities, partially driven by fewer traditional W-R stars (\heii{}) and weaker line-driven winds (\civ{}). The narrow, nebular emission is also stronger in these models, driven by higher nebular temperatures and harder radiation fields.

We note that the broad contribution to the total \heii{} flux is difficult to interpret due to the brevity of the W-R phase, and will depend strongly on the SFH of the system. In star-bursting systems, the broad \heii{} flux contribution from W-R stars can be as large as 80\% at solar metallicity (\citetalias{Byler+2018}). Thus, the CSFR models presented here represent one of the limiting SFH scenarios.

\begin{figure*}
  \begin{center}
    \includegraphics[width=\linewidth]{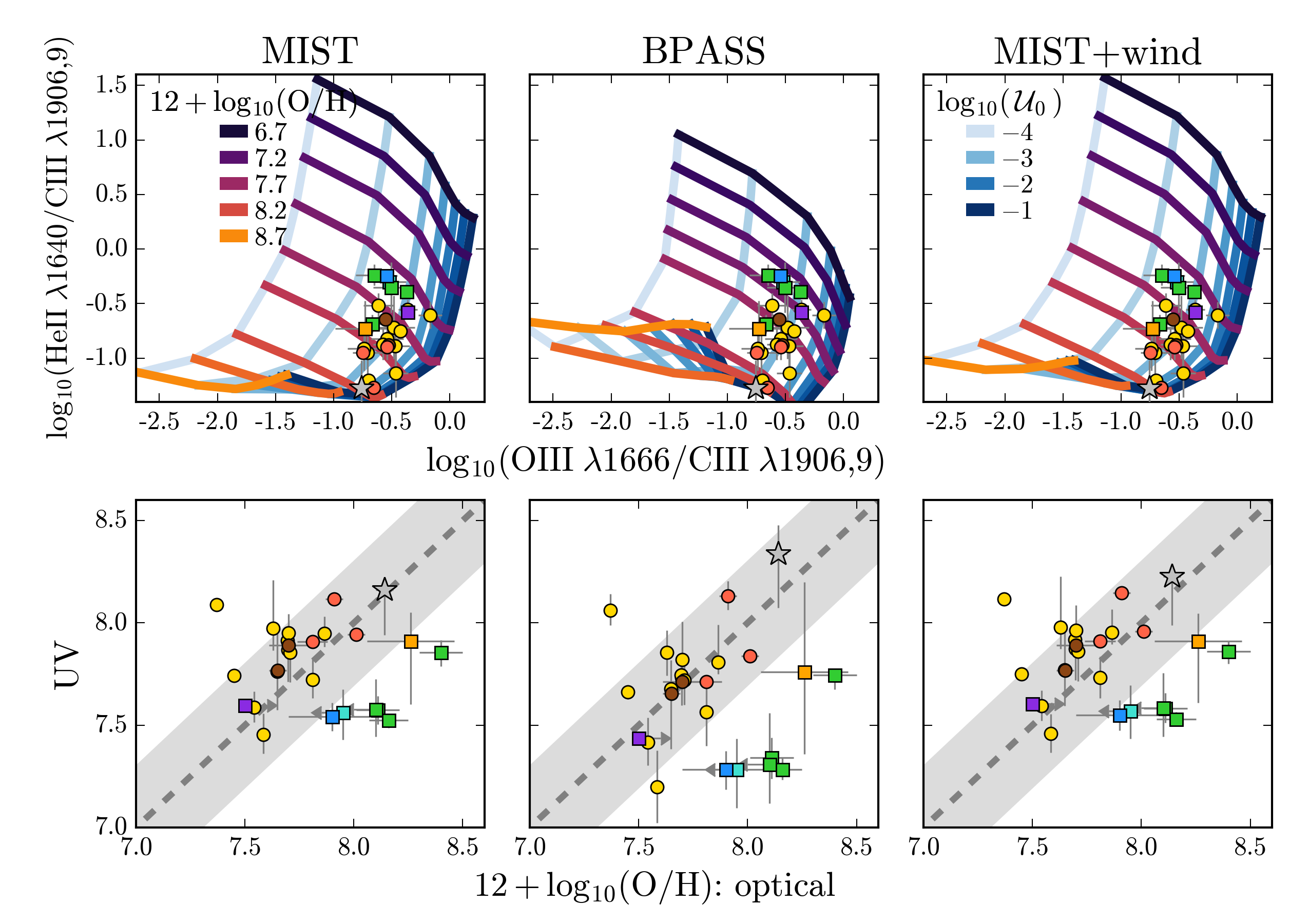}
    \caption{\emph{Top row:} He2-O3C3 diagnostic diagrams using the MIST (left), BPASS (center), and {\tt MIST+wind} (right) stellar models. \emph{Bottom row:} UV metallicities ($y$-axis) derived from the He2-O3C3 diagnostic compared to the optical oxygen abundance ($x$-axis) for models using MIST (left), BPASS (center), and {\tt MIST+wind} (right) stellar models. The harder ionizing spectra from the BPASS models do not significantly improve the agreement between UV and optical metallicities. Some objects show clear improvement with either the BPASS or {\tt MIST+wind} models.}
    \label{fig:HeIIdds}
  \end{center}
\end{figure*}

\subsection{The source of narrow \ion{He}{2}$\,\lambda$1640 emission} \label{sec:discussion:HeII}

Significant nebular \heii{} emission requires high energy photons, and current stellar models have difficulty producing the hard ionizing spectra required without invoking binary populations or rotating stars \citep[e.g.,][]{Stark+2014, Steidel+2016, Choi+2017, Byler+2017}. Harder ionizing spectra would produce stronger \heii{} emission and create an extended partial ionization zone in the nebula, changing emission line ratios.

We can test the sensitivity of the derived metallicities to the hardness of the ionizing spectrum using the \CloudyFSPS model integrated within \FSPS \citep{Byler+2017}, which includes self-consistent nebular emission predictions for all isochrone sets available within FSPS: Padova, MIST, PARSEC, and BPASS. The nebular inputs (i.e., gas phase abundances, geometry) are identical across all stellar models, which provides a clean test of the sensitivity of \heii{} to the ionizing spectrum\footnote{We compare the MIST and BPASS models using models with a constant SFR. To ensure that we are comparing truly ``equilibrated'' populations, we assume a constant SFR over 10\,\Myr for the MIST models and 100\,\Myr for the BPASS models, as suggested by \citet{Xiao+2018}.}. A comprehensive comparison of the hydrogen- and helium-ionizing properties of the MIST and BPASS models can be found in \citet{Choi+2017}.

We compare the UV diagnostic diagrams for the MIST model (which includes the effects of stellar rotation) and the BPASS model (which includes the effects of stellar multiplicity) in the top row of Fig.~\ref{fig:HeIIdds} (left and center panels). The two model grids are similar in shape, but there are visible differences, especially at high metallicity ($8.0 \leq 12+\logOH \leq 8.7$). At these metallicities, the harder ionizing spectra from the BPASS models produce more high energy photons and relatively more \heii{} emission, elevating the predicted He2C3 ratios.

The bottom row of Fig.~\ref{fig:HeIIdds} shows the UV-derived metallicities ($y$-axis) compared to the optical metallicity ($x$-axis) for the MIST (left) and BPASS (center) models. Despite the use of harder ionizing spectra, the agreement between UV and optical metallicities has not noticeably improved. With the MIST models, 11 of the 24 galaxies (46\%) have UV metallicities that agree with optical metallicities, within error. For the BPASS models, that number is increased to 12 (50\%). 

To understand the scatter between UV and optical metallicities, we calculate the average offset of the data points from a one-to-one relationship, and the percentage of galaxies that fall within $0.3$ dex of the one-to-one relationship, shown by the grey shaded region in Fig.~\ref{fig:HeIIdds}. As discussed earlier, $\pm 0.3$ dex represents the typical systematic errors inherent in optical strong line methods.

For the MIST models, the UV metallicities show an average offset of $0.11 \pm 0.24$ dex from the optical, while 65\% of the galaxies fall within 0.3 dex of the grey dashed line with slope unity. For the BPASS models, the UV metallicities have an average offset of $-0.12 \pm 0.30$ dex, and 60\% of the galaxies fall within 0.3 dex of the grey dashed line with slope unity.

Thus, we do not find any statistically significant improvement in the offset or scatter of the UV metallicities with the BPASS models. We note that the use of harder ionizing spectra does not improve metallicity estimates for the \mage galaxies (green squares).

The right column of Fig.~\ref{fig:HeIIdds} shows the {\tt MIST+wind} grid (top) and the resulting UV-optical metallicity comparison (bottom). The inclusion of stellar \heii{} emission in the {\tt MIST+wind} grid increases the model He2C3 ratios by 0.1-0.3\,dex ($y$-axis) at metallicities above $12+\logOH=8$.

When the metallicity is calculated from the {\tt MIST+wind} grid and compared to the metallicity derived from optical emission lines (bottom right panel), the agreement between UV- and optically-derived metallicities is not noticeably improved. The {\tt wind} grid shows a mean offset of $0.1\pm 0.23$, statistically indistinguishable from the MIST grid. We also note that the use of the {\tt wind} grid does not improve UV metallicity estimates for the \mage galaxies (green squares), which are still more than 0.8 dex smaller than those derived using optical emission lines. 

Despite the complications associated with modelling \heii{}, the He2-O3C3 diagnostic yields metallicity measurements that agree well with optical measurements, especially at low metallicities ($12+\logOH \lesssim 8$). However, on a galaxy by galaxy basis, improvement between UV and optical metallicities changes whether the BPASS models or the {\tt MIST+wind} models are used. Put differently, some objects are better fit with the BPASS models, while other objects are better fit with the {\tt MIST+wind} models. This could indicate that multiple competing processes are at work, impacting the observed \heii{} fluxes. Separating and characterizing these different processes will likely require a joint analysis of ISM properties and the local massive star populations in these objects.

\begin{figure*}
  \begin{center}
    \includegraphics[width=\linewidth]{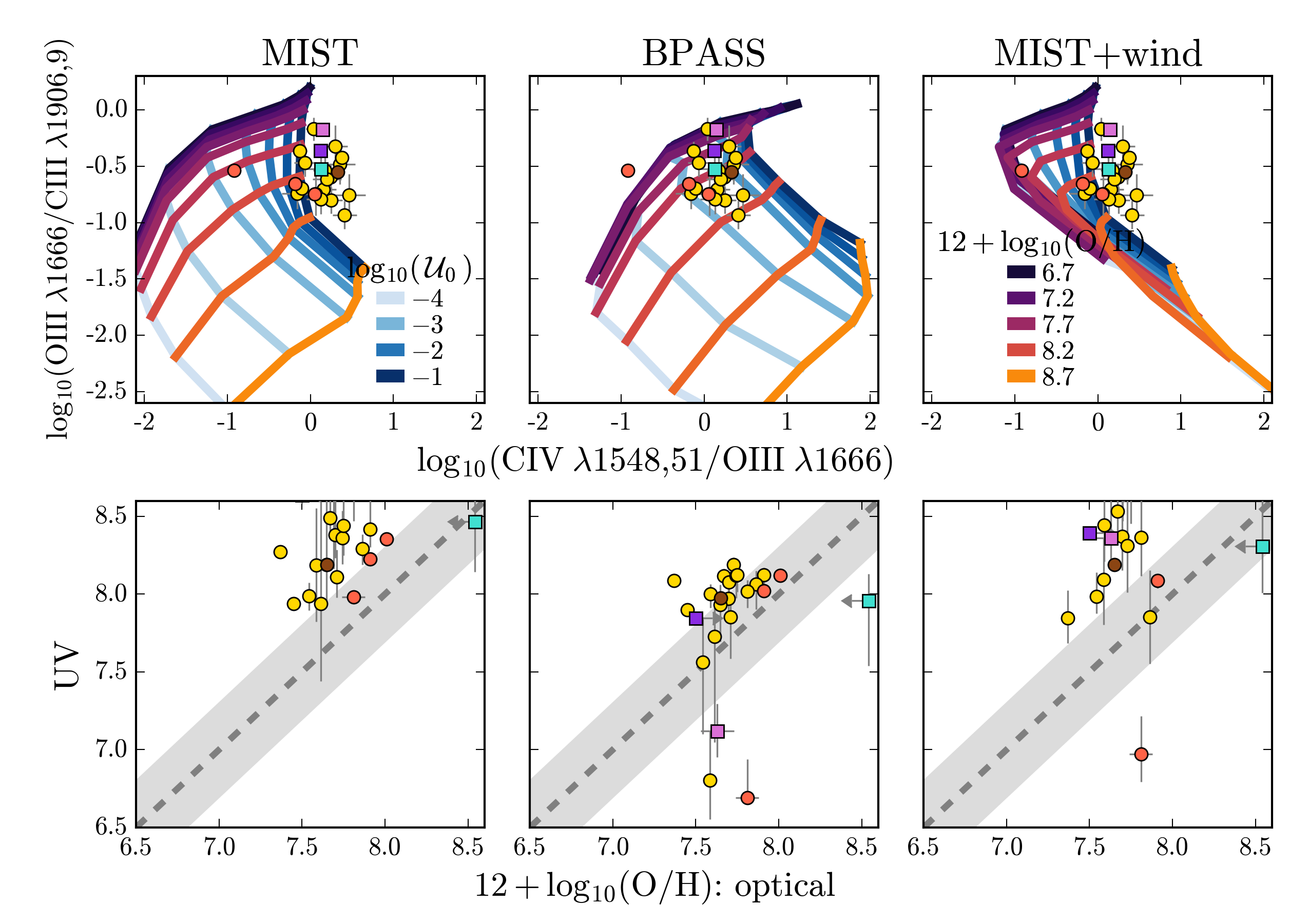}
    \caption{\emph{Top row:} C4-O3C3 diagnostic diagrams using the MIST (left), BPASS (center), and {\tt MIST+wind} (right) stellar models. \emph{Bottom row:} UV metallicities ($y$-axis) derived from the He2-O3C3 diagnostic compared to the optical oxygen abundance ($x$-axis) for models using MIST (left), BPASS (center), and {\tt MIST+wind} (right) stellar models. The harder ionizing spectra from the BPASS models significantly improve the agreement between UV and optical oxygen abundance.}
    \label{fig:CIVdds}
  \end{center}
\end{figure*}

\subsection{The source of narrow \civ$\lambda$1550 emission} \label{sec:discussion:CIV}

Similar to \heii{}, \civ{} is a relatively high excitation line. As in the previous section, we repeat the experiment with the MIST and BPASS models for the C4-O3C3 diagnostic to determine if a harder radiation field can improve the disagreement between model and observed \civ{} strengths.

Fig.~\ref{fig:CIVdds} shows the resulting MIST (left), BPASS (center), and {\tt MIST+wind} (right) comparisons. The harder ionizing spectra found in the BPASS models produce larger C4O3 ratios, showing clear improvement when compared to observed C4O3 ratios, and the BPASS grid is able to reproduce most of the observed line ratios, with the exception of one of the \citet{Senchyna+2017} galaxies (red circle). It is thus unsurprising that the metallicities derived using the BPASS models show a clear improvement over the MIST models. The galaxies with UV metallicities that agree with optical estimates increases from 8\% (MIST) to 17\% (BPASS). While both grids still overestimate the UV metallicity, the offset is decreased with the BPASS models. For the MIST models, the average offset is $0.57 \pm 0.3$ dex. This offset is decreased to $0.24 \pm 0.4$ with the BPASS models.

The right column of Fig.~\ref{fig:CIVdds} shows the {\tt MIST+wind} grid (top) and the resulting UV-optical metallicity comparison (bottom). Model C4O3 ratios increase when stellar emission is included in the model, especially for models at high metallicity and low ionization parameters. The agreement between model grid and data is still poor, and unsurprisingly the agreement between UV and optical metallicities shows little improvement over the standard MIST model. The UV metallicities are still larger than in the optical, with an average offset of $0.62 \pm 0.5$.

The lack of improvement with the {\tt MIST+wind} grid does not imply that wind contamination cannot be responsible for inflating measured \civ{} fluxes, rather that the stellar emission as implemented in this model does not improve metallicity constraints. We note that wind predictions vary substantially from model to model and are poorly constrained at low metallicity.

Clearly there is still work to be done to fully understand \civ{} emission from galaxies, locally and at high-redshift. The BPASS grid provides an improved interpretation of \civ{} line strengths, but still does not reproduce the full range of observed C4O3 line ratios. Moreover, \emph{none} of the \civ{} grids show a positive correlation between UV and optical metallicity. Caution should be used when interpreting \civ{} line strengths, especially at high redshift, where different ISM conditions may prevail.

\subsection{C/O variations} \label{sec:discussion:CIII}

The \citetalias{Byler+2018} model uses a polynomial equation to describe the increase of [N/H] with [O/H] and [C/H] with [O/H] (\S\ref{sec:model:neb}). The relationship accounts for the additional production of N and C at high metallicity, and is matched to observations of local star-forming galaxies. These empirical relationships are used to describe the broad behavior of the galaxy population, but individual objects can have abundance patterns for C, N, and O that deviate from these relationships.

The C/O relationship used in this work was derived to match the \citet{Berg+2016} galaxies, and as such, most of the galaxies in the Berg sample have C/O ratios that are well-matched to our model. However, there are a handful of objects with C/O ratios that deviate significantly from our C/O relationship. It is possible that the \ciii{} line strengths are too sensitive to the specific C and O abundances to be a useful metallicity indicator. \citet{PerezMontero+2017} presented an analysis of metallicities derived using \ciii{} lines, and found that it was essential to estimate the C/O ratio before calculating the metallicity. 

The emissivity of \oiii$\,\lambda$1666 is much more sensitive to \Te (and thus the gas-phase oxygen abundance) than the emissivity of \ciii$\,\lambda$1909. However, photoionization models have shown that \ciii{} line strengths are more sensitive to \Te (and thus the gas-phase oxygen abundance) than to the absolute gas-phase carbon abundance \citep{Jaskot+2016, Byler+2018}. Put differently, \ciii{} line strengths vary more strongly with changes to the gas phase oxygen abundance than with changes to the gas-phase C/O ratio. As such, to first order, both the \oiii{} and \ciii{} emission lines trace the gas-phase oxygen abundance.

\begin{figure*}
  \begin{center}
    \includegraphics[width=\linewidth]{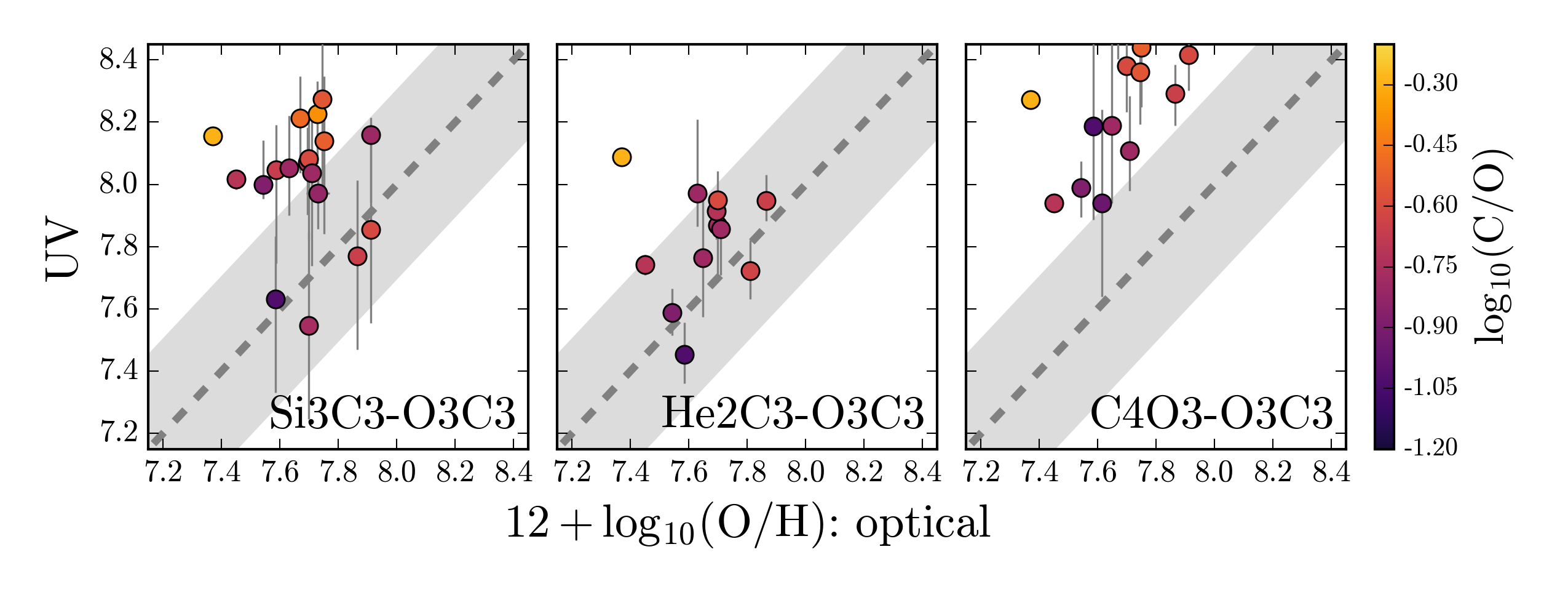}
    \caption{Comparison between UV metallicity ($y$-axis) and optical metallicity ($x$-axis) for the \citet{Berg+2016} galaxies, using the Si3-O3C3 diagnostic (left), He2-O3C3 diagnostic (middle), and C4-O3C3 diagnostic (right). Objects are color-coded by the measured C/O ratio. While the UV metallicity is weakly correlated with the C/O ratio, variations in C and O abundances alone cannot be responsible for driving the large scatter in UV metallicities.}
    \label{fig:CO}
  \end{center}
\end{figure*}

In Fig.~\ref{fig:CO}, we show the comparison between UV and optical metallicities for the Berg sample, for the Si3-O3C3 diagnostic (left), He2-O3C3 diagnostic (middle), and C4-O3C3 diagnostic (right). In all panels, the points are color-coded by the C/O ratio. There is a weak trend between C/O and UV-metallicity, where larger C/O ratios are found in higher-metallicity objects. For two objects with identical optical metallicities but a factor of three difference in C/O ratio, the difference in derived UV metallicity is less than 0.1 dex, which is smaller than the statistical errors calculated here. This suggests that metallicities derived from the \ciii$\,\lambda$1906,1909 and \oiii$\,\lambda$1661,1666 lines will not be dominated by uncertainties driven by C/O variations.

Despite recent progress in building samples of objects with rest-UV and rest-optical spectra, there is still much work to be done to interpret rest-frame UV spectra. In particular, it is critical that we understand the behavior of UV emission lines in the context of optically-derived ISM properties so that we can fully harness their diagnostic power in preparation for JWST.

\section{Conclusions}\label{sec:conclusions}
We have derived gas phase oxygen abundances for galaxies with rest-UV spectra using different combinations of UV emission lines (Table~\ref{tab:logOH}). For a sample of galaxies with both rest-UV and rest-optical spectra, we have compared UV and optical abundances to identify useful UV metallicity diagnostics. Our conclusions are as follows.

\begin{enumerate}
    \item Metallicities derived using the \SiuIII$\,\lambda$1893 emission line do not reliably correlate with optical metallicities and show a comparatively large scatter, with an average offset of $0.35 \pm 0.28$ dex from the optical (Fig.~\ref{fig:UVSi}). We suggest that this is likely driven by variations in the silicon abundance relative to oxygen, either from variable dust depletion factors or from enhanced silicon abundances in the gas phase caused by the erosion of Si from the surface of dust grains by shocks (Fig.~\ref{fig:SiO}).
    \item UV diagnostics that include the \heii$\,\lambda$1640 emission line are reliable metallicity indicators at metallicities below $12+\logOH \sim 8$. At higher metallicities ($12+\logOH \sim 8.5$), discrepant abundances may arise due to contamination by stellar \heii{} emission (Fig.~\ref{fig:UVHe}). Consistent gas phase abundances are found regardless of stellar model choice (rotating or binary, Fig.~\ref{fig:HeIIdds}), with an average offset of $0.11 \pm 0.24$ dex for the rotating MIST models and $-0.12 \pm 0.30$ dex for the binary BPASS models.
    \item UV diagnostics that include the \civ$\,\lambda$ 1548,1550 emission lines are unreliable metallicity indicators. None of the models tested in this work showed a positive correlation between UV and optical metallicity when the \civ{} diagnostic was used. The harder radiation fields found in the BPASS models significantly improve agreement between model and observed C4O3 line ratios, reducing the offset between UV and optical metallicities by 0.3 dex. The C4-O3C3 abundances are still offset from optical abundances by 0.2 dex, even with the BPASS models, which suggests that stellar models do not produce hard enough ionizing spectra. Some of the remaining offset may be driven by contamination from stellar emission, though our interpretation is further complicated by strong interstellar absorption just blueward of the emission feature, which could complicate continuum estimation in low-resolution, low-S/N observations (Fig.~\ref{fig:UVCIV}). We note that the disagreement could also be driven by contamination from shocks or AGN in some cases, as \civ{} is also sensitive to these power sources. 
    \item We provide new oxygen abundance diagnostics for the Si3-O3C3 and He2-O3C3 diagrams based on polynomial fits to the model grid surface (\S~\ref{sec:ZZ:poly}). These fits are valid for $6.2 \leq 12+\logOH \leq 9.2$, and $-4.0 \leq \logU \leq -1.0$.
    \item We calculate the relative flux from stellar and nebular emission for the \heii{} and \civ{} spectral features for the constant SFR models used in this work. For \civ{}, at high metallicities ($12+\logOH>8.2$) the broad stellar emission contributes more than 50\% of the total emission flux. However, at low metallicities ($12+\logOH\leq8$), stellar emission contributes at most 10-20\% of the total \civ{} flux (Fig.~\ref{fig:FracBroad}). Compared to the stellar \civ{} emission, the flux contribution from stellar \heii{} emission is smaller (at most 40\% of the total flux), because in the models used in this work, the broad \heii{} emission is produced by short-lived W-R stars.
    \item In Appendix~\ref{appdx:UVdirectTe}, we calculate direct-\Te{} oxygen abundances for our photoionization models using the UV \oii{}$\;\lambda2471$, \oii{}$\;\lambda3729$, \oiii{}$\;\lambda1666$, and \oiii{}$\;\lambda2321$ emission lines. We compare the UV direct-\Te{} abundances with optical direct-\Te abundances and find that the two abundances are tightly correlated. The UV direct-\Te abundances are systematically 0.1\,dex lower than the optical direct-\Te abundances, driven by temperature differences, as the UV emission lines are higher exitation transitions. 
\end{enumerate}

Currently planned studies of $z>6$ galaxies with NIRSpec on JWST will rely on UV emission line diagnostics to calculate metallicities. Obtaining rest-optical \nii{}/\ha{} measurements with MIRI for even a small subset of these objects will be expensive, without NIRSpec's multiplexing ability. For those objects observed with NIRSpec and MIRI, our results suggest that metallicities calculated from NIRSPec will be quite low ($6.5\lesssim 12+\logOH \lesssim 7.5$), while metallicities calculated from MIRI will be moderately high ($7.5\lesssim 12+\logOH \lesssim 8.5$).

While the proposed UV metallicity diagnostics show theoretical promise, in practice, UV-derived abundances show considerable scatter and do not always recover optical abundances. Unfortunately, the combination of fragmented wavelength coverage and the inhomogenous sample selection make it difficult to determine the source of the discrepancy. A first step in understanding these diagnostics requires a detailed comparison of gas properties in a small set of galaxies with high S/N, high resolution, aperture-matched rest-frame UV and optical spectroscopy. With these observations we can detect the faint auroral emission lines and ensure the identification of stellar and shock emission, if present, and quantify the relationship between UV and optical measures of temperature, density, ionization parameter, and C, N, O, and Si abundances. As it stands, the current discrepancies between UV and optical diagnostics critically limit our ability to study chemical enrichment across cosmic time.

\acknowledgments

We would like to thank the anonymous referee for thorough and constructive feedback that greatly improved this work. Parts of this research were supported by the Australian Research Council Centre of Excellence for All Sky Astrophysics in 3 Dimensions (ASTRO 3D), through project number CE170100013. L.J.K. gratefully acknowledges the support of an ARC Laureate Fellowship (FL150100113). This research has made use of NASA's Astrophysics Data System Bibliographic Services. This research made use of Astropy,\footnote{http://www.astropy.org} a community-developed core Python package for Astronomy \citep{astropy:2013, astropy:2018}.

\bibliographystyle{aasjournal}
\bibliography{main}

\begin{thebibliography}{}
\expandafter\ifx\csname natexlab\endcsname\relax\def\natexlab#1{#1}\fi
\providecommand{\url}[1]{\href{#1}{#1}}
\providecommand{\dodoi}[1]{doi:~\href{http://doi.org/#1}{\nolinkurl{#1}}}
\providecommand{\doeprint}[1]{\href{http://ascl.net/#1}{\nolinkurl{http://ascl.net/#1}}}
\providecommand{\doarXiv}[1]{\href{https://arxiv.org/abs/#1}{\nolinkurl{https://arxiv.org/abs/#1}}}

\bibitem[{{Abazajian} {et~al.}(2009){Abazajian}, {Adelman-McCarthy},
  {Ag{\"u}eros}, {Allam}, {Allende Prieto}, {An}, {Anderson}, {Anderson},
  {Annis}, {Bahcall}, \& et~al.}]{Abazajian+2009}
{Abazajian}, K.~N., {Adelman-McCarthy}, J.~K., {Ag{\"u}eros}, M.~A., {et~al.}
  2009, \apjs, 182, 543, \dodoi{10.1088/0067-0049/182/2/543}

\bibitem[{{Acharyya} {et~al.}(2019){Acharyya}, {Kewley}, {Rigby}, {Bayliss},
  {Bian}, {Nicholls}, {Federrath}, {Kaasinen}, {Florian}, \&
  {Blanc}}]{Acharyya+2019}
{Acharyya}, A., {Kewley}, L.~J., {Rigby}, J.~R., {et~al.} 2019, \mnras, 1930,
  \dodoi{10.1093/mnras/stz1987}

\bibitem[{{Aggarwal} \& {Keenan}(1999)}]{Aggarwal+1999}
{Aggarwal}, K.~M., \& {Keenan}, F.~P. 1999, \apjs, 123, 311,
  \dodoi{10.1086/313232}

\bibitem[{{Asplund} {et~al.}(2009){Asplund}, {Grevesse}, {Sauval}, \&
  {Scott}}]{Asplund+2009}
{Asplund}, M., {Grevesse}, N., {Sauval}, A.~J., \& {Scott}, P. 2009, \araa, 47,
  481, \dodoi{10.1146/annurev.astro.46.060407.145222}

\bibitem[{{Astropy Collaboration} {et~al.}(2013){Astropy Collaboration},
  {Robitaille}, {Tollerud}, {Greenfield}, {Droettboom}, {Bray}, {Aldcroft},
  {Davis}, {Ginsburg}, {Price-Whelan}, {Kerzendorf}, {Conley}, {Crighton},
  {Barbary}, {Muna}, {Ferguson}, {Grollier}, {Parikh}, {Nair}, {Unther},
  {Deil}, {Woillez}, {Conseil}, {Kramer}, {Turner}, {Singer}, {Fox}, {Weaver},
  {Zabalza}, {Edwards}, {Azalee Bostroem}, {Burke}, {Casey}, {Crawford},
  {Dencheva}, {Ely}, {Jenness}, {Labrie}, {Lim}, {Pierfederici}, {Pontzen},
  {Ptak}, {Refsdal}, {Servillat}, \& {Streicher}}]{astropy:2013}
{Astropy Collaboration}, {Robitaille}, T.~P., {Tollerud}, E.~J., {et~al.} 2013,
  \aap, 558, A33, \dodoi{10.1051/0004-6361/201322068}

\bibitem[{{Baldwin} {et~al.}(1981){Baldwin}, {Phillips}, \& {Terlevich}}]{BPT}
{Baldwin}, J.~A., {Phillips}, M.~M., \& {Terlevich}, R. 1981, \pasp, 93, 5,
  \dodoi{10.1086/130766}

\bibitem[{{Bayliss} {et~al.}(2014){Bayliss}, {Rigby}, {Sharon}, {Wuyts},
  {Florian}, {Gladders}, {Johnson}, \& {Oguri}}]{Bayliss+2014}
{Bayliss}, M.~B., {Rigby}, J.~R., {Sharon}, K., {et~al.} 2014, \apj, 790, 144,
  \dodoi{10.1088/0004-637X/790/2/144}

\bibitem[{{Belfiore} {et~al.}(2017){Belfiore}, {Maiolino}, {Tremonti},
  {S{\'a}nchez}, {Bundy}, {Bershady}, {Westfall}, {Lin}, {Drory}, {Boquien},
  {Thomas}, \& {Brinkmann}}]{Belfiore+2017b}
{Belfiore}, F., {Maiolino}, R., {Tremonti}, C., {et~al.} 2017, \mnras, 469,
  151, \dodoi{10.1093/mnras/stx789}

\bibitem[{{Berg} {et~al.}(2018){Berg}, {Erb}, {Auger}, {Pettini}, \&
  {Brammer}}]{Berg+2018}
{Berg}, D.~A., {Erb}, D.~K., {Auger}, M.~W., {Pettini}, M., \& {Brammer}, G.~B.
  2018, \apj, 859, 164, \dodoi{10.3847/1538-4357/aab7fa}

\bibitem[{{Berg} {et~al.}(2019){Berg}, {Erb}, {Henry}, {Skillman}, \&
  {McQuinn}}]{Berg+2019}
{Berg}, D.~A., {Erb}, D.~K., {Henry}, R.~B.~C., {Skillman}, E.~D., \&
  {McQuinn}, K.~B.~W. 2019, arXiv e-prints.
\newblock \doarXiv{1901.08160}

\bibitem[{{Berg} {et~al.}(2015){Berg}, {Skillman}, {Croxall}, {Pogge},
  {Moustakas}, \& {Johnson-Groh}}]{Berg+2015}
{Berg}, D.~A., {Skillman}, E.~D., {Croxall}, K.~V., {et~al.} 2015, \apj, 806,
  16, \dodoi{10.1088/0004-637X/806/1/16}

\bibitem[{{Berg} {et~al.}(2016){Berg}, {Skillman}, {Henry}, {Erb}, \&
  {Carigi}}]{Berg+2016}
{Berg}, D.~A., {Skillman}, E.~D., {Henry}, R.~B.~C., {Erb}, D.~K., \& {Carigi},
  L. 2016, \apj, 827, 126, \dodoi{10.3847/0004-637X/827/2/126}

\bibitem[{{Berg} {et~al.}(2012){Berg}, {Skillman}, {Marble}, {van Zee},
  {Engelbracht}, {Lee}, {Kennicutt}, {Calzetti}, {Dale}, \&
  {Johnson}}]{Berg+2012}
{Berg}, D.~A., {Skillman}, E.~D., {Marble}, A.~R., {et~al.} 2012, \apj, 754,
  98, \dodoi{10.1088/0004-637X/754/2/98}

\bibitem[{{Bian} {et~al.}(2017){Bian}, {Kewley}, {Dopita}, \&
  {Blanc}}]{Bian+2017}
{Bian}, F., {Kewley}, L.~J., {Dopita}, M.~A., \& {Blanc}, G.~A. 2017, \apj,
  834, 51, \dodoi{10.3847/1538-4357/834/1/51}

\bibitem[{{Bian} {et~al.}(2010){Bian}, {Fan}, {Bechtold}, {McGreer}, {Just},
  {Sand}, {Green}, {Thompson}, {Peng}, {Seifert}, {Ageorges}, {Juette},
  {Knierim}, \& {Buschkamp}}]{Bian+2010}
{Bian}, F., {Fan}, X., {Bechtold}, J., {et~al.} 2010, \apj, 725, 1877,
  \dodoi{10.1088/0004-637X/725/2/1877}

\bibitem[{{Bresolin}(2007)}]{Bresolin+2007}
{Bresolin}, F. 2007, \apj, 656, 186, \dodoi{10.1086/510380}

\bibitem[{{Brinchmann} {et~al.}(2008){Brinchmann}, {Pettini}, \&
  {Charlot}}]{Brinchmann+2008}
{Brinchmann}, J., {Pettini}, M., \& {Charlot}, S. 2008, \mnras, 385, 769,
  \dodoi{10.1111/j.1365-2966.2008.12914.x}

\bibitem[{{Byler}(2018)}]{cloudyFSPSv1}
{Byler}, N. 2018, cloudyFSPS, 1.0.0,  Zenodo, \dodoi{10.5281/zenodo.1156412}.
\newblock \url{https://doi.org/10.5281/zenodo.1156412}

\bibitem[{{Byler} {et~al.}(2017){Byler}, {Dalcanton}, {Conroy}, \&
  {Johnson}}]{Byler+2017}
{Byler}, N., {Dalcanton}, J.~J., {Conroy}, C., \& {Johnson}, B.~D. 2017, \apj,
  840, 44, \dodoi{10.3847/1538-4357/aa6c66}

\bibitem[{{Byler} {et~al.}(2018){Byler}, {Dalcanton}, {Conroy}, {Johnson},
  {Levesque}, \& {Berg}}]{Byler+2018}
{Byler}, N., {Dalcanton}, J.~J., {Conroy}, C., {et~al.} 2018, \apj, 863, 14,
  \dodoi{10.3847/1538-4357/aacd50}

\bibitem[{{Cantiello} {et~al.}(2007){Cantiello}, {Yoon}, {Langer}, \&
  {Livio}}]{Cantiello+2007}
{Cantiello}, M., {Yoon}, S.~C., {Langer}, N., \& {Livio}, M. 2007, in American
  Institute of Physics Conference Series, Vol. 948, Unsolved Problems in
  Stellar Physics: A Conference in Honor of Douglas Gough, ed. R.~J.
  {Stancliffe}, G.~{Houdek}, R.~G. {Martin}, \& C.~A. {Tout}, 413--418

\bibitem[{{Cardelli} {et~al.}(1989){Cardelli}, {Clayton}, \&
  {Mathis}}]{Cardelli+1989}
{Cardelli}, J.~A., {Clayton}, G.~C., \& {Mathis}, J.~S. 1989, \apj, 345, 245,
  \dodoi{10.1086/167900}

\bibitem[{{Chisholm} {et~al.}(2019){Chisholm}, {Rigby}, {Bayliss}, {Berg},
  {Dahle}, {Gladders}, \& {Sharon}}]{Chisholm+2019}
{Chisholm}, J., {Rigby}, J.~R., {Bayliss}, M., {et~al.} 2019, arXiv e-prints.
\newblock \doarXiv{1905.04314}

\bibitem[{{Choi} {et~al.}(2017){Choi}, {Conroy}, \& {Byler}}]{Choi+2017}
{Choi}, J., {Conroy}, C., \& {Byler}, N. 2017, \apj, 838, 159,
  \dodoi{10.3847/1538-4357/aa679f}

\bibitem[{{Choi} {et~al.}(2016){Choi}, {Dotter}, {Conroy}, {Cantiello},
  {Paxton}, \& {Johnson}}]{Choi+2016}
{Choi}, J., {Dotter}, A., {Conroy}, C., {et~al.} 2016, \apj, 823, 102,
  \dodoi{10.3847/0004-637X/823/2/102}

\bibitem[{{Christensen} {et~al.}(2012){Christensen}, {Laursen}, {Richard},
  {Hjorth}, {Milvang-Jensen}, {Dessauges-Zavadsky}, {Limousin}, {Grillo}, \&
  {Ebeling}}]{Christensen+2012}
{Christensen}, L., {Laursen}, P., {Richard}, J., {et~al.} 2012, \mnras, 427,
  1973, \dodoi{10.1111/j.1365-2966.2012.22007.x}

\bibitem[{{Conroy} \& {Gunn}(2010)}]{Conroy+2010}
{Conroy}, C., \& {Gunn}, J.~E. 2010, \apj, 712, 833,
  \dodoi{10.1088/0004-637X/712/2/833}

\bibitem[{{Conroy} {et~al.}(2009){Conroy}, {Gunn}, \& {White}}]{Conroy+2009}
{Conroy}, C., {Gunn}, J.~E., \& {White}, M. 2009, \apj, 699, 486,
  \dodoi{10.1088/0004-637X/699/1/486}

\bibitem[{{Conti}(1991)}]{Conti+1991}
{Conti}, P.~S. 1991, \apj, 377, 115, \dodoi{10.1086/170340}

\bibitem[{{Crowther} {et~al.}(2006){Crowther}, {Prinja}, {Pettini}, \&
  {Steidel}}]{Crowther+2006}
{Crowther}, P.~A., {Prinja}, R.~K., {Pettini}, M., \& {Steidel}, C.~C. 2006,
  \mnras, 368, 895, \dodoi{10.1111/j.1365-2966.2006.10164.x}

\bibitem[{{Crowther} {et~al.}(2016){Crowther}, {Caballero-Nieves}, {Bostroem},
  {Ma{\'{\i}}z Apell{\'a}niz}, {Schneider}, {Walborn}, {Angus}, {Brott},
  {Bonanos}, {de Koter}, {de Mink}, {Evans}, {Gr{\"a}fener}, {Herrero},
  {Howarth}, {Langer}, {Lennon}, {Puls}, {Sana}, \& {Vink}}]{Crowther+2016}
{Crowther}, P.~A., {Caballero-Nieves}, S.~M., {Bostroem}, K.~A., {et~al.} 2016,
  \mnras, 458, 624, \dodoi{10.1093/mnras/stw273}

\bibitem[{{Dopita} {et~al.}(2013){Dopita}, {Sutherland}, {Nicholls}, {Kewley},
  \& {Vogt}}]{Dopita+2013}
{Dopita}, M.~A., {Sutherland}, R.~S., {Nicholls}, D.~C., {Kewley}, L.~J., \&
  {Vogt}, F.~P.~A. 2013, \apjs, 208, 10, \dodoi{10.1088/0067-0049/208/1/10}

\bibitem[{{Dotter}(2016)}]{Dotter+2016}
{Dotter}, A. 2016, \apjs, 222, 8, \dodoi{10.3847/0067-0049/222/1/8}

\bibitem[{{Du} {et~al.}(2017){Du}, {Shapley}, {Martin}, \& {Coil}}]{Du+2017}
{Du}, X., {Shapley}, A.~E., {Martin}, C.~L., \& {Coil}, A.~L. 2017, \apj, 838,
  63, \dodoi{10.3847/1538-4357/aa64cf}

\bibitem[{{Eldridge} {et~al.}(2011){Eldridge}, {Langer}, \&
  {Tout}}]{Eldridge+2011}
{Eldridge}, J.~J., {Langer}, N., \& {Tout}, C.~A. 2011, \mnras, 414, 3501,
  \dodoi{10.1111/j.1365-2966.2011.18650.x}

\bibitem[{{Eldridge} {et~al.}(2017){Eldridge}, {Stanway}, {Xiao}, {McClelland},
  {Taylor}, {Ng}, {Greis}, \& {Bray}}]{Eldridge+2017}
{Eldridge}, J.~J., {Stanway}, E.~R., {Xiao}, L., {et~al.} 2017, \pasa, 34,
  e058, \dodoi{10.1017/pasa.2017.51}

\bibitem[{{Erb} {et~al.}(2010){Erb}, {Pettini}, {Shapley}, {Steidel}, {Law}, \&
  {Reddy}}]{Erb+2010}
{Erb}, D.~K., {Pettini}, M., {Shapley}, A.~E., {et~al.} 2010, \apj, 719, 1168,
  \dodoi{10.1088/0004-637X/719/2/1168}

\bibitem[{{Erb} {et~al.}(2006){Erb}, {Shapley}, {Pettini}, {Steidel}, {Reddy},
  \& {Adelberger}}]{Erb+2006}
{Erb}, D.~K., {Shapley}, A.~E., {Pettini}, M., {et~al.} 2006, \apj, 644, 813,
  \dodoi{10.1086/503623}

\bibitem[{{Evans}(1999)}]{Evans+1999}
{Evans}, II, N.~J. 1999, \araa, 37, 311, \dodoi{10.1146/annurev.astro.37.1.311}

\bibitem[{{Feltre} {et~al.}(2016){Feltre}, {Charlot}, \&
  {Gutkin}}]{Feltre+2016}
{Feltre}, A., {Charlot}, S., \& {Gutkin}, J. 2016, \mnras, 456, 3354,
  \dodoi{10.1093/mnras/stv2794}

\bibitem[{{Ferland} {et~al.}(2013){Ferland}, {Porter}, {van Hoof}, {Williams},
  {Abel}, {Lykins}, {Shaw}, {Henney}, \& {Stancil}}]{Ferland+2013}
{Ferland}, G.~J., {Porter}, R.~L., {van Hoof}, P.~A.~M., {et~al.} 2013, \rmxaa,
  49, 137.
\newblock \doarXiv{1302.4485}

\bibitem[{{Fitzpatrick}(1999)}]{Fitzpatrick+1999}
{Fitzpatrick}, E.~L. 1999, \pasp, 111, 63, \dodoi{10.1086/316293}

\bibitem[{{Foreman-Mackey} {et~al.}(2014){Foreman-Mackey}, {Sick}, \&
  {Johnson}}]{pythonFSPSdfm}
{Foreman-Mackey}, D., {Sick}, J., \& {Johnson}, B.~D. 2014, python-fsps: Python
  bindings to FSPS, 0.1.1,  Zenodo, \dodoi{10.5281/zenodo.12157}.
\newblock \url{https://doi.org/10.5281/zenodo.12157}

\bibitem[{{Garnett}(1992)}]{Garnett+1992}
{Garnett}, D.~R. 1992, \aj, 103, 1330, \dodoi{10.1086/116146}

\bibitem[{{Garnett} {et~al.}(1995){Garnett}, {Dufour}, {Peimbert},
  {Torres-Peimbert}, {Shields}, {Skillman}, {Terlevich}, \&
  {Terlevich}}]{Garnett+1995}
{Garnett}, D.~R., {Dufour}, R.~J., {Peimbert}, M., {et~al.} 1995, \apjl, 449,
  L77, \dodoi{10.1086/309620}

\bibitem[{{Grevesse} {et~al.}(2010){Grevesse}, {Asplund}, {Sauval}, \&
  {Scott}}]{Grevesse+2010}
{Grevesse}, N., {Asplund}, M., {Sauval}, A.~J., \& {Scott}, P. 2010, \apss,
  328, 179, \dodoi{10.1007/s10509-010-0288-z}

\bibitem[{{Hamann} \& {Gr{\"a}fener}(2003)}]{Hamann+2003}
{Hamann}, W.~R., \& {Gr{\"a}fener}, G. 2003, \aap, 410, 993,
  \dodoi{10.1051/0004-6361:20031308}

\bibitem[{{Haris} {et~al.}(2016){Haris}, {Parvathi}, {Gudennavar}, {Bubbly},
  {Murthy}, \& {Sofia}}]{Haris+2016}
{Haris}, U., {Parvathi}, V.~S., {Gudennavar}, S.~B., {et~al.} 2016, \aj, 151,
  143, \dodoi{10.3847/0004-6256/151/6/143}

\bibitem[{{Heckman} {et~al.}(1998){Heckman}, {Robert}, {Leitherer}, {Garnett},
  \& {van der Rydt}}]{Heckman+1998}
{Heckman}, T.~M., {Robert}, C., {Leitherer}, C., {Garnett}, D.~R., \& {van der
  Rydt}, F. 1998, \apj, 503, 646, \dodoi{10.1086/306035}

\bibitem[{{Hillier} \& {Lanz}(2001)}]{Hillier+2001}
{Hillier}, D.~J., \& {Lanz}, T. 2001, in Astronomical Society of the Pacific
  Conference Series, Vol. 247, Spectroscopic Challenges of Photoionized
  Plasmas, ed. G.~{Ferland} \& D.~W. {Savin}, 343

\bibitem[{{Hirschmann} {et~al.}(2019){Hirschmann}, {Charlot}, {Feltre}, {Naab},
  {Somerville}, \& {Choi}}]{Hirschmann+2019}
{Hirschmann}, M., {Charlot}, S., {Feltre}, A., {et~al.} 2019, \mnras, 487, 333,
  \dodoi{10.1093/mnras/stz1256}

\bibitem[{{James} {et~al.}(2014){James}, {Aloisi}, {Heckman}, {Sohn}, \&
  {Wolfe}}]{James+2014}
{James}, B.~L., {Aloisi}, A., {Heckman}, T., {Sohn}, S.~T., \& {Wolfe}, M.~A.
  2014, \apj, 795, 109, \dodoi{10.1088/0004-637X/795/2/109}

\bibitem[{{Jaskot} \& {Ravindranath}(2016)}]{Jaskot+2016}
{Jaskot}, A.~E., \& {Ravindranath}, S. 2016, \apj, 833, 136,
  \dodoi{10.3847/1538-4357/833/2/136}

\bibitem[{{Jones}(2000)}]{Jones+2000}
{Jones}, A.~P. 2000, \jgr, 105, 10257, \dodoi{10.1029/1999JA900264}

\bibitem[{{Kauffmann} {et~al.}(2003){Kauffmann}, {Heckman}, {Tremonti},
  {Brinchmann}, {Charlot}, {White}, {Ridgway}, {Brinkmann}, {Fukugita}, {Hall},
  {Ivezi{\'c}}, {Richards}, \& {Schneider}}]{Kauffmann+2003b}
{Kauffmann}, G., {Heckman}, T.~M., {Tremonti}, C., {et~al.} 2003, \mnras, 346,
  1055, \dodoi{10.1111/j.1365-2966.2003.07154.x}

\bibitem[{{Kennicutt} {et~al.}(2003){Kennicutt}, {Bresolin}, \&
  {Garnett}}]{Kennicutt+2003}
{Kennicutt}, Robert~C., J., {Bresolin}, F., \& {Garnett}, D.~R. 2003, \apj,
  591, 801, \dodoi{10.1086/375398}

\bibitem[{{Kewley} \& {Dopita}(2002)}]{Kewley+2002}
{Kewley}, L.~J., \& {Dopita}, M.~A. 2002, \apjs, 142, 35,
  \dodoi{10.1086/341326}

\bibitem[{{Kewley} {et~al.}(2013{\natexlab{a}}){Kewley}, {Dopita}, {Leitherer},
  {Dav{\'e}}, {Yuan}, {Allen}, {Groves}, \& {Sutherland}}]{Kewley+2013b}
{Kewley}, L.~J., {Dopita}, M.~A., {Leitherer}, C., {et~al.} 2013{\natexlab{a}},
  \apj, 774, 100, \dodoi{10.1088/0004-637X/774/2/100}

\bibitem[{{Kewley} {et~al.}(2001){Kewley}, {Dopita}, {Sutherland}, {Heisler},
  \& {Trevena}}]{Kewley+2001}
{Kewley}, L.~J., {Dopita}, M.~A., {Sutherland}, R.~S., {Heisler}, C.~A., \&
  {Trevena}, J. 2001, \apj, 556, 121, \dodoi{10.1086/321545}

\bibitem[{{Kewley} \& {Ellison}(2008)}]{Kewley+2008}
{Kewley}, L.~J., \& {Ellison}, S.~L. 2008, \apj, 681, 1183,
  \dodoi{10.1086/587500}

\bibitem[{{Kewley} {et~al.}(2013{\natexlab{b}}){Kewley}, {Maier}, {Yabe},
  {Ohta}, {Akiyama}, {Dopita}, \& {Yuan}}]{Kewley+2013a}
{Kewley}, L.~J., {Maier}, C., {Yabe}, K., {et~al.} 2013{\natexlab{b}}, \apjl,
  774, L10, \dodoi{10.1088/2041-8205/774/1/L10}

\bibitem[{{Kewley} {et~al.}(2019{\natexlab{a}}){Kewley}, {Nicholls},
  {Sutherland}, {Rigby}, {Acharya}, {Dopita}, \& {Bayliss}}]{Kewley+2019}
{Kewley}, L.~J., {Nicholls}, D.~C., {Sutherland}, R., {et~al.}
  2019{\natexlab{a}}, \apj, 880, 16, \dodoi{10.3847/1538-4357/ab16ed}

\bibitem[{{Kewley} {et~al.}(2019{\natexlab{b}}){Kewley}, {Nicholls}, \&
  {Sutherland}}]{Kewley+2019ARAA}
{Kewley}, L.~J., {Nicholls}, D.~C., \& {Sutherland}, R.~S. 2019{\natexlab{b}},
  \araa, 57, 511, \dodoi{10.1146/annurev-astro-081817-051832}

\bibitem[{{Kinney} {et~al.}(1993){Kinney}, {Bohlin}, {Calzetti}, {Panagia}, \&
  {Wyse}}]{Kinney+1993}
{Kinney}, A.~L., {Bohlin}, R.~C., {Calzetti}, D., {Panagia}, N., \& {Wyse},
  R.~F.~G. 1993, \apjs, 86, 5, \dodoi{10.1086/191771}

\bibitem[{{Kobulnicky} \& {Kewley}(2004)}]{Kobulnicky+2004}
{Kobulnicky}, H.~A., \& {Kewley}, L.~J. 2004, \apj, 617, 240,
  \dodoi{10.1086/425299}

\bibitem[{{Kroupa}(2001)}]{Kroupa+2001}
{Kroupa}, P. 2001, \mnras, 322, 231, \dodoi{10.1046/j.1365-8711.2001.04022.x}

\bibitem[{{Kunth} \& {Joubert}(1985)}]{Kunth+1985}
{Kunth}, D., \& {Joubert}, M. 1985, \aap, 142, 411

\bibitem[{{Kurucz}(2005)}]{Kurucz+2005}
{Kurucz}, R.~L. 2005, Memorie della Societa Astronomica Italiana Supplementi,
  8, 14

\bibitem[{{Lee} {et~al.}(2006){Lee}, {Skillman}, {Cannon}, {Jackson}, {Gehrz},
  {Polomski}, \& {Woodward}}]{Lee+2006}
{Lee}, H., {Skillman}, E.~D., {Cannon}, J.~M., {et~al.} 2006, \apj, 647, 970,
  \dodoi{10.1086/505573}

\bibitem[{{Leitherer} {et~al.}(2018){Leitherer}, {Byler}, {Lee}, \&
  {Levesque}}]{Leitherer+2018}
{Leitherer}, C., {Byler}, N., {Lee}, J.~C., \& {Levesque}, E.~M. 2018, \apj,
  865, 55, \dodoi{10.3847/1538-4357/aada84}

\bibitem[{{Leitherer} {et~al.}(1995){Leitherer}, {Robert}, \&
  {Heckman}}]{Leitherer+1995}
{Leitherer}, C., {Robert}, C., \& {Heckman}, T.~M. 1995, \apjs, 99, 173,
  \dodoi{10.1086/192183}

\bibitem[{{Leitherer} {et~al.}(2011){Leitherer}, {Tremonti}, {Heckman}, \&
  {Calzetti}}]{Leitherer+2011}
{Leitherer}, C., {Tremonti}, C.~A., {Heckman}, T.~M., \& {Calzetti}, D. 2011,
  \aj, 141, 37, \dodoi{10.1088/0004-6256/141/2/37}

\bibitem[{{Luridiana} {et~al.}(2013){Luridiana}, {Morisset}, \& {Shaw}}]{PyNeb}
{Luridiana}, V., {Morisset}, C., \& {Shaw}, R.~A. 2013, {PyNeb: Analysis of
  emission lines}, Astrophysics Source Code Library.
\newblock \doeprint{1304.021}

\bibitem[{{Maiolino} \& {Mannucci}(2018)}]{Maiolino+2018}
{Maiolino}, R., \& {Mannucci}, F. 2018, arXiv e-prints, arXiv:1811.09642.
\newblock \doarXiv{1811.09642}

\bibitem[{{Mamon} {et~al.}(2019){Mamon}, {Trevisan}, {Thuan}, {Gallazzi}, \&
  {Dav{\'e}}}]{Mamon+2019}
{Mamon}, G.~A., {Trevisan}, M., {Thuan}, T.~X., {Gallazzi}, A., \& {Dav{\'e}},
  R. 2019, \mnras, 3173, \dodoi{10.1093/mnras/stz3556}

\bibitem[{{Masters} {et~al.}(2014){Masters}, {McCarthy}, {Siana}, {Malkan},
  {Mobasher}, {Atek}, {Henry}, {Martin}, {Rafelski}, {Hathi}, {Scarlata},
  {Ross}, {Bunker}, {Blanc}, {Bedregal}, {Dom{\'{\i}}nguez}, {Colbert},
  {Teplitz}, \& {Dressler}}]{Masters+2014}
{Masters}, D., {McCarthy}, P., {Siana}, B., {et~al.} 2014, \apj, 785, 153,
  \dodoi{10.1088/0004-637X/785/2/153}

\bibitem[{{McGaugh}(1991)}]{McGaugh+1991}
{McGaugh}, S.~S. 1991, \apj, 380, 140, \dodoi{10.1086/170569}

\bibitem[{{McKee} \& {Ostriker}(1977)}]{McKee+1977}
{McKee}, C.~F., \& {Ostriker}, J.~P. 1977, \apj, 218, 148,
  \dodoi{10.1086/155667}

\bibitem[{{Mishra} \& {Li}(2017)}]{Mishra+2017}
{Mishra}, A., \& {Li}, A. 2017, \apj, 850, 138,
  \dodoi{10.3847/1538-4357/aa937a}

\bibitem[{{Nomoto} {et~al.}(2013){Nomoto}, {Kobayashi}, \&
  {Tominaga}}]{Nomoto+2013}
{Nomoto}, K., {Kobayashi}, C., \& {Tominaga}, N. 2013, Annual Review of
  Astronomy and Astrophysics, 51, 457,
  \dodoi{10.1146/annurev-astro-082812-140956}

\bibitem[{{Pagel} {et~al.}(1979){Pagel}, {Edmunds}, {Blackwell}, {Chun}, \&
  {Smith}}]{Pagel+1979}
{Pagel}, B.~E.~J., {Edmunds}, M.~G., {Blackwell}, D.~E., {Chun}, M.~S., \&
  {Smith}, G. 1979, \mnras, 189, 95, \dodoi{10.1093/mnras/189.1.95}

\bibitem[{{Pagel} \& {Tautvaisiene}(1995)}]{Pagel+1995}
{Pagel}, B.~E.~J., \& {Tautvaisiene}, G. 1995, \mnras, 276, 505,
  \dodoi{10.1093/mnras/276.2.505}

\bibitem[{{Patr{\'{\i}}cio} {et~al.}(2019){Patr{\'{\i}}cio}, {Richard},
  {Carton}, {P{\'e}roux}, {Contini}, {Brinchmann}, {Schaye}, {Weilbacher},
  {Nanayakkara}, {Maseda}, {Mahler}, \& {Wisotzki}}]{Patricio+2019}
{Patr{\'{\i}}cio}, V., {Richard}, J., {Carton}, D., {et~al.} 2019, \mnras, 489,
  224, \dodoi{10.1093/mnras/stz2114}

\bibitem[{{Pauldrach} {et~al.}(2001){Pauldrach}, {Hoffmann}, \&
  {Lennon}}]{Pauldrach+2001}
{Pauldrach}, A.~W.~A., {Hoffmann}, T.~L., \& {Lennon}, M. 2001, \aap, 375, 161,
  \dodoi{10.1051/0004-6361:20010805}

\bibitem[{{Pauldrach} {et~al.}(1990){Pauldrach}, {Kudritzki}, {Puls}, \&
  {Butler}}]{Pauldrach+1990}
{Pauldrach}, A.~W.~A., {Kudritzki}, R.~P., {Puls}, J., \& {Butler}, K. 1990,
  \aap, 228, 125

\bibitem[{{Paxton} {et~al.}(2011){Paxton}, {Bildsten}, {Dotter}, {Herwig},
  {Lesaffre}, \& {Timmes}}]{Paxton+2011}
{Paxton}, B., {Bildsten}, L., {Dotter}, A., {et~al.} 2011, \apjs, 192, 3,
  \dodoi{10.1088/0067-0049/192/1/3}

\bibitem[{{Paxton} {et~al.}(2013){Paxton}, {Cantiello}, {Arras}, {Bildsten},
  {Brown}, {Dotter}, {Mankovich}, {Montgomery}, {Stello}, {Timmes}, \&
  {Townsend}}]{Paxton+2013}
{Paxton}, B., {Cantiello}, M., {Arras}, P., {et~al.} 2013, \apjs, 208, 4,
  \dodoi{10.1088/0067-0049/208/1/4}

\bibitem[{{Paxton} {et~al.}(2015){Paxton}, {Marchant}, {Schwab}, {Bauer},
  {Bildsten}, {Cantiello}, {Dessart}, {Farmer}, {Hu}, {Langer}, {Townsend},
  {Townsley}, \& {Timmes}}]{Paxton+2015}
{Paxton}, B., {Marchant}, P., {Schwab}, J., {et~al.} 2015, \apjs, 220, 15,
  \dodoi{10.1088/0067-0049/220/1/15}

\bibitem[{{Peimbert} {et~al.}(2017){Peimbert}, {Peimbert}, \&
  {Delgado-Inglada}}]{Peimbert+2017}
{Peimbert}, M., {Peimbert}, A., \& {Delgado-Inglada}, G. 2017, Publications of
  the Astronomical Society of the Pacific, 129, 082001,
  \dodoi{10.1088/1538-3873/aa72c3}

\bibitem[{{P{\'e}rez-Montero} \& {Amor{\'{\i}}n}(2017)}]{PerezMontero+2017}
{P{\'e}rez-Montero}, E., \& {Amor{\'{\i}}n}, R. 2017, \mnras, 467, 1287,
  \dodoi{10.1093/mnras/stx186}

\bibitem[{{Pettini} \& {Pagel}(2004)}]{Pettini+2004}
{Pettini}, M., \& {Pagel}, B.~E.~J. 2004, \mnras, 348, L59,
  \dodoi{10.1111/j.1365-2966.2004.07591.x}

\bibitem[{{Pettini} {et~al.}(2000){Pettini}, {Steidel}, {Adelberger},
  {Dickinson}, \& {Giavalisco}}]{Pettini+2000}
{Pettini}, M., {Steidel}, C.~C., {Adelberger}, K.~L., {Dickinson}, M., \&
  {Giavalisco}, M. 2000, \apj, 528, 96, \dodoi{10.1086/308176}

\bibitem[{{Pilyugin} \& {Thuan}(2005)}]{Pilyugin+2005}
{Pilyugin}, L.~S., \& {Thuan}, T.~X. 2005, \apj, 631, 231,
  \dodoi{10.1086/432408}

\bibitem[{{Plat} {et~al.}(2019){Plat}, {Charlot}, {Bruzual}, {Feltre},
  {Vidal-Garc{\'\i}a}, {Morisset}, {Chevallard}, \& {Todt}}]{Plat+2019}
{Plat}, A., {Charlot}, S., {Bruzual}, G., {et~al.} 2019, \mnras, 490, 978,
  \dodoi{10.1093/mnras/stz2616}

\bibitem[{{Price-Whelan} {et~al.}(2018){Price-Whelan}, {Sip{\H{o}}cz},
  {G{\"u}nther}, {Lim}, {Crawford}, {Conseil}, {Shupe}, {Craig}, {Dencheva},
  {Ginsburg}, {VanderPlas}, {Bradley}, {P{\'e}rez-Su{\'a}rez}, {de Val-Borro},
  {Paper Contributors}, {Aldcroft}, {Cruz}, {Robitaille}, {Tollerud},
  {Coordination Committee}, {Ardelean}, {Babej}, {Bach}, {Bachetti}, {Bakanov},
  {Bamford}, {Barentsen}, {Barmby}, {Baumbach}, {Berry}, {Biscani}, {Boquien},
  {Bostroem}, {Bouma}, {Brammer}, {Bray}, {Breytenbach}, {Buddelmeijer},
  {Burke}, {Calderone}, {Cano Rodr{\'\i}guez}, {Cara}, {Cardoso}, {Cheedella},
  {Copin}, {Corrales}, {Crichton}, {D{\textquoteright}Avella}, {Deil},
  {Depagne}, {Dietrich}, {Donath}, {Droettboom}, {Earl}, {Erben}, {Fabbro},
  {Ferreira}, {Finethy}, {Fox}, {Garrison}, {Gibbons}, {Goldstein}, {Gommers},
  {Greco}, {Greenfield}, {Groener}, {Grollier}, {Hagen}, {Hirst}, {Homeier},
  {Horton}, {Hosseinzadeh}, {Hu}, {Hunkeler}, {Ivezi{\'c}}, {Jain}, {Jenness},
  {Kanarek}, {Kendrew}, {Kern}, {Kerzendorf}, {Khvalko}, {King}, {Kirkby},
  {Kulkarni}, {Kumar}, {Lee}, {Lenz}, {Littlefair}, {Ma}, {Macleod},
  {Mastropietro}, {McCully}, {Montagnac}, {Morris}, {Mueller}, {Mumford},
  {Muna}, {Murphy}, {Nelson}, {Nguyen}, {Ninan}, {N{\"o}the}, {Ogaz}, {Oh},
  {Parejko}, {Parley}, {Pascual}, {Patil}, {Patil}, {Plunkett}, {Prochaska},
  {Rastogi}, {Reddy Janga}, {Sabater}, {Sakurikar}, {Seifert}, {Sherbert},
  {Sherwood-Taylor}, {Shih}, {Sick}, {Silbiger}, {Singanamalla}, {Singer},
  {Sladen}, {Sooley}, {Sornarajah}, {Streicher}, {Teuben}, {Thomas},
  {Tremblay}, {Turner}, {Terr{\'o}n}, {van Kerkwijk}, {de la Vega}, {Watkins},
  {Weaver}, {Whitmore}, {Woillez}, {Zabalza}, \& {Contributors}}]{astropy:2018}
{Price-Whelan}, A.~M., {Sip{\H{o}}cz}, B.~M., {G{\"u}nther}, H.~M., {et~al.}
  2018, \aj, 156, 123, \dodoi{10.3847/1538-3881/aabc4f}

\bibitem[{{Rigby} {et~al.}(2014){Rigby}, {Bayliss}, {Gladders}, {Sharon},
  {Wuyts}, \& {Dahle}}]{Rigby+2014}
{Rigby}, J.~R., {Bayliss}, M.~B., {Gladders}, M.~D., {et~al.} 2014, \apj, 790,
  44, \dodoi{10.1088/0004-637X/790/1/44}

\bibitem[{{Rigby} {et~al.}(2011){Rigby}, {Wuyts}, {Gladders}, {Sharon}, \&
  {Becker}}]{Rigby+2011}
{Rigby}, J.~R., {Wuyts}, E., {Gladders}, M.~D., {Sharon}, K., \& {Becker},
  G.~D. 2011, \apj, 732, 59, \dodoi{10.1088/0004-637X/732/1/59}

\bibitem[{{Rigby} {et~al.}(2018{\natexlab{a}}){Rigby}, {Bayliss}, {Sharon},
  {Gladders}, {Chisholm}, {Dahle}, {Johnson}, {Paterno-Mahler}, {Wuyts}, \&
  {Kelson}}]{Rigby+2018a}
{Rigby}, J.~R., {Bayliss}, M.~B., {Sharon}, K., {et~al.} 2018{\natexlab{a}},
  \aj, 155, 104, \dodoi{10.3847/1538-3881/aaa2ff}

\bibitem[{{Rigby} {et~al.}(2018{\natexlab{b}}){Rigby}, {Bayliss}, {Chisholm},
  {Bordoloi}, {Sharon}, {Gladders}, {Johnson}, {Paterno-Mahler}, {Wuyts},
  {Dahle}, \& {Acharyya}}]{Rigby+2018b}
{Rigby}, J.~R., {Bayliss}, M.~B., {Chisholm}, J., {et~al.} 2018{\natexlab{b}},
  \apj, 853, 87, \dodoi{10.3847/1538-4357/aaa2fc}

\bibitem[{{Rivera-Thorsen} {et~al.}(2017){Rivera-Thorsen}, {Dahle}, {Gronke},
  {Bayliss}, {Rigby}, {Simcoe}, {Bordoloi}, {Turner}, \&
  {Furesz}}]{Rivera+2017}
{Rivera-Thorsen}, T.~E., {Dahle}, H., {Gronke}, M., {et~al.} 2017, \aap, 608,
  L4, \dodoi{10.1051/0004-6361/201732173}

\bibitem[{{Roy} {et~al.}(2019){Roy}, {Sutherland}, {Krumholz}, {Heger}, \&
  {Dopita}}]{Roy+2019}
{Roy}, A., {Sutherland}, R.~S., {Krumholz}, M.~R., {Heger}, A., \& {Dopita},
  M.~A. 2019, arXiv e-prints, arXiv:1907.07666.
\newblock \doarXiv{1907.07666}

\bibitem[{{Sanders} {et~al.}(2016){Sanders}, {Shapley}, {Kriek}, {Reddy},
  {Freeman}, {Coil}, {Siana}, {Mobasher}, {Shivaei}, {Price}, \& {de
  Groot}}]{Sanders+2016}
{Sanders}, R.~L., {Shapley}, A.~E., {Kriek}, M., {et~al.} 2016, \apj, 816, 23,
  \dodoi{10.3847/0004-637X/816/1/23}

\bibitem[{{Sargent} \& {Searle}(1970)}]{Sargent+1970}
{Sargent}, W.~L.~W., \& {Searle}, L. 1970, \apjl, 162, L155,
  \dodoi{10.1086/180644}

\bibitem[{{Schaerer} {et~al.}(1999){Schaerer}, {Contini}, \&
  {Pindao}}]{Schaerer+1999}
{Schaerer}, D., {Contini}, T., \& {Pindao}, M. 1999, \aaps, 136, 35,
  \dodoi{10.1051/aas:1999197}

\bibitem[{{Schaerer} \& {Vacca}(1998)}]{Schaerer+1998}
{Schaerer}, D., \& {Vacca}, W.~D. 1998, \apj, 497, 618, \dodoi{10.1086/305487}

\bibitem[{{Schlafly} \& {Finkbeiner}(2011)}]{Schlafly+2011}
{Schlafly}, E.~F., \& {Finkbeiner}, D.~P. 2011, \apj, 737, 103,
  \dodoi{10.1088/0004-637X/737/2/103}

\bibitem[{{Schlegel} {et~al.}(1998){Schlegel}, {Finkbeiner}, \&
  {Davis}}]{Schlegel+1998}
{Schlegel}, D.~J., {Finkbeiner}, D.~P., \& {Davis}, M. 1998, \apj, 500, 525,
  \dodoi{10.1086/305772}

\bibitem[{{Searle} \& {Sargent}(1972)}]{Searle+1972}
{Searle}, L., \& {Sargent}, W. L.~W. 1972, \apj, 173, 25,
  \dodoi{10.1086/151398}

\bibitem[{{Senchyna} {et~al.}(2019){Senchyna}, {Stark}, {Chevallard},
  {Charlot}, {Jones}, \& {Vidal Garc{\'{\i}}a}}]{Senchyna+2019}
{Senchyna}, P., {Stark}, D.~P., {Chevallard}, J., {et~al.} 2019, arXiv
  e-prints.
\newblock \doarXiv{1904.01615}

\bibitem[{{Senchyna} {et~al.}(2017){Senchyna}, {Stark}, {Vidal-Garc{\'{\i}}a},
  {Chevallard}, {Charlot}, {Mainali}, {Jones}, {Wofford}, {Feltre}, \&
  {Gutkin}}]{Senchyna+2017}
{Senchyna}, P., {Stark}, D.~P., {Vidal-Garc{\'{\i}}a}, A., {et~al.} 2017,
  \mnras, 472, 2608, \dodoi{10.1093/mnras/stx2059}

\bibitem[{{Shapley}(2011)}]{Shapley+2011}
{Shapley}, A.~E. 2011, Annual Review of Astronomy and Astrophysics, 49, 525,
  \dodoi{10.1146/annurev-astro-081710-102542}

\bibitem[{{Shapley} {et~al.}(2003){Shapley}, {Steidel}, {Pettini}, \&
  {Adelberger}}]{Shapley+2003}
{Shapley}, A.~E., {Steidel}, C.~C., {Pettini}, M., \& {Adelberger}, K.~L. 2003,
  \apj, 588, 65, \dodoi{10.1086/373922}

\bibitem[{{Smith} {et~al.}(2002){Smith}, {Norris}, \& {Crowther}}]{Smith+2002}
{Smith}, L.~J., {Norris}, R.~P.~F., \& {Crowther}, P.~A. 2002, \mnras, 337,
  1309, \dodoi{10.1046/j.1365-8711.2002.06042.x}

\bibitem[{{Stark} {et~al.}(2014){Stark}, {Richard}, {Siana}, {Charlot},
  {Freeman}, {Gutkin}, {Wofford}, {Robertson}, {Amanullah}, {Watson}, \&
  {Milvang-Jensen}}]{Stark+2014}
{Stark}, D.~P., {Richard}, J., {Siana}, B., {et~al.} 2014, \mnras, 445, 3200,
  \dodoi{10.1093/mnras/stu1618}

\bibitem[{{Stasi{\'n}ska}(2005)}]{Stasinska+2005}
{Stasi{\'n}ska}, G. 2005, \aap, 434, 507, \dodoi{10.1051/0004-6361:20042216}

\bibitem[{{Steidel} {et~al.}(2016){Steidel}, {Strom}, {Pettini}, {Rudie},
  {Reddy}, \& {Trainor}}]{Steidel+2016}
{Steidel}, C.~C., {Strom}, A.~L., {Pettini}, M., {et~al.} 2016, \apj, 826, 159,
  \dodoi{10.3847/0004-637X/826/2/159}

\bibitem[{{Steidel} {et~al.}(2014){Steidel}, {Rudie}, {Strom}, {Pettini},
  {Reddy}, {Shapley}, {Trainor}, {Erb}, {Turner}, {Konidaris}, {Kulas}, {Mace},
  {Matthews}, \& {McLean}}]{Steidel+2014}
{Steidel}, C.~C., {Rudie}, G.~C., {Strom}, A.~L., {et~al.} 2014, \apj, 795,
  165, \dodoi{10.1088/0004-637X/795/2/165}

\bibitem[{{Storey} {et~al.}(2014){Storey}, {Sochi}, \& {Badnell}}]{Storey+2014}
{Storey}, P.~J., {Sochi}, T., \& {Badnell}, N.~R. 2014, \mnras, 441, 3028,
  \dodoi{10.1093/mnras/stu777}

\bibitem[{{Strom} {et~al.}(2017){Strom}, {Steidel}, {Rudie}, {Trainor},
  {Pettini}, \& {Reddy}}]{Strom+2017}
{Strom}, A.~L., {Steidel}, C.~C., {Rudie}, G.~C., {et~al.} 2017, \apj, 836,
  164, \dodoi{10.3847/1538-4357/836/2/164}

\bibitem[{{Tremonti} {et~al.}(2004){Tremonti}, {Heckman}, {Kauffmann},
  {Brinchmann}, {Charlot}, {White}, {Seibert}, {Peng}, {Schlegel}, {Uomoto},
  {Fukugita}, \& {Brinkmann}}]{Tremonti+2004}
{Tremonti}, C.~A., {Heckman}, T.~M., {Kauffmann}, G., {et~al.} 2004, \apj, 613,
  898, \dodoi{10.1086/423264}

\bibitem[{{Vidal-Garc{\'{\i}}a} {et~al.}(2017){Vidal-Garc{\'{\i}}a}, {Charlot},
  {Bruzual}, \& {Hubeny}}]{Vidal-Garcia+2017}
{Vidal-Garc{\'{\i}}a}, A., {Charlot}, S., {Bruzual}, G., \& {Hubeny}, I. 2017,
  \mnras, 470, 3532, \dodoi{10.1093/mnras/stx1324}

\bibitem[{{Vink} \& {de Koter}(2005)}]{Vink+2005}
{Vink}, J.~S., \& {de Koter}, A. 2005, \aap, 442, 587,
  \dodoi{10.1051/0004-6361:20052862}

\bibitem[{{Walborn} {et~al.}(2002){Walborn}, {Fullerton}, {Crowther},
  {Bianchi}, {Hutchings}, {Pellerin}, {Sonneborn}, \& {Willis}}]{Walborn+2002}
{Walborn}, N.~R., {Fullerton}, A.~W., {Crowther}, P.~A., {et~al.} 2002, \apjs,
  141, 443, \dodoi{10.1086/340571}

\bibitem[{{Walborn} \& {Nichols-Bohlin}(1987)}]{Walborn+1987}
{Walborn}, N.~R., \& {Nichols-Bohlin}, J. 1987, \pasp, 99, 40,
  \dodoi{10.1086/131954}

\bibitem[{{Whitaker} {et~al.}(2014){Whitaker}, {Rigby}, {Brammer}, {Gladders},
  {Sharon}, {Teng}, \& {Wuyts}}]{Whitaker+2014}
{Whitaker}, K.~E., {Rigby}, J.~R., {Brammer}, G.~B., {et~al.} 2014, \apj, 790,
  143, \dodoi{10.1088/0004-637X/790/2/143}

\bibitem[{{Wuyts} {et~al.}(2012){Wuyts}, {Rigby}, {Gladders}, {Gilbank},
  {Sharon}, {Gralla}, \& {Bayliss}}]{Wuyts+2012a}
{Wuyts}, E., {Rigby}, J.~R., {Gladders}, M.~D., {et~al.} 2012, \apj, 745, 86,
  \dodoi{10.1088/0004-637X/745/1/86}

\bibitem[{{Wuyts} {et~al.}(2014){Wuyts}, {Rigby}, {Gladders}, \&
  {Sharon}}]{Wuyts+2014}
{Wuyts}, E., {Rigby}, J.~R., {Gladders}, M.~D., \& {Sharon}, K. 2014, \apj,
  781, 61, \dodoi{10.1088/0004-637X/781/2/61}

\bibitem[{{Xiao} {et~al.}(2018){Xiao}, {Stanway}, \& {Eldridge}}]{Xiao+2018}
{Xiao}, L., {Stanway}, E., \& {Eldridge}, J.~J. 2018, ArXiv e-prints.
\newblock \doarXiv{1801.07068}

\bibitem[{{Yoon} \& {Langer}(2005)}]{Yoon+2005}
{Yoon}, S.~C., \& {Langer}, N. 2005, \aap, 443, 643,
  \dodoi{10.1051/0004-6361:20054030}

\bibitem[{{Zahid} {et~al.}(2014){Zahid}, {Dima}, {Kudritzki}, {Kewley},
  {Geller}, {Hwang}, {Silverman}, \& {Kashino}}]{Zahid+2014}
{Zahid}, H.~J., {Dima}, G.~I., {Kudritzki}, R.-P., {et~al.} 2014, \apj, 791,
  130, \dodoi{10.1088/0004-637X/791/2/130}

\bibitem[{{Zaritsky} {et~al.}(1994){Zaritsky}, {Kennicutt}, \&
  {Huchra}}]{Zaritsky+1994}
{Zaritsky}, D., {Kennicutt}, Robert~C., J., \& {Huchra}, J.~P. 1994, \apj, 420,
  87, \dodoi{10.1086/173544}

\bibitem[{{Zetterlund} {et~al.}(2015){Zetterlund}, {Levesque}, {Leitherer}, \&
  {Danforth}}]{Zetterlund+2015}
{Zetterlund}, E., {Levesque}, E.~M., {Leitherer}, C., \& {Danforth}, C.~W.
  2015, \apj, 805, 151, \dodoi{10.1088/0004-637X/805/2/151}

\end{thebibliography}
\appendix

\section{Polynomial fits to offset between theoretical and measured oxygen abundances.}\label{sec:appdx:poly}

Table~\ref{tab:polyA} includes the coefficients for the third order polynomial used to fit the \Cloudy and direct-\Te method oxygen abundances. The fits are computed for both CSFR and instantaneous burst models at each ionization parameter. Table~\ref{tab:polyB} includes the coefficients for the linear function used to fit the \Cloudy and PP04-N2 method oxygen abundances. As before, fits are computed for both CSFR and instantaneous burst models at each ionization parameter.

For three objects in the sample we were unable to calculate direct-\Te or PP04-N2 oxygen abundances, and instead used the KK04-R23 method. Table~\ref{tab:polyC} includes the third order polynomial used to fit the \Cloudy and KK04-R23 oxygen abundances. The fits are computed for a 10 Myr CSFR model at each ionization parameter.

\startlongtable
\begin{deluxetable}{ccccccc}
\tabletypesize{\footnotesize}
\tablecaption{Polynomial fits to the oxygen abundance offset between \Cloudy and the direct-\Te method.\label{tab:polyA}}
\tablehead{\colhead{SFH} & \colhead{Age (Myr)} & \colhead{\logU} & \colhead{$a$} & \colhead{$b$} & \colhead{$c$} & \colhead{$d$}}
\startdata
CSFR & 10 & -4.0 & -0.01 & 0.30 & -1.15 & 5.09 \\
 &  & -3.5 & -0.03 & 0.75 & -4.48 & 13.10 \\
 &  & -3.0 & -0.09 & 2.06 & -13.97 & 35.98 \\
 &  & -2.5 & -0.18 & 4.02 & -28.25 & 70.38 \\
 &  & -2.0 & -0.22 & 4.90 & -34.60 & 85.52 \\
 &  & -1.5 & -0.23 & 5.11 & -36.14 & 89.13 \\
 &  & -1.0 & -0.23 & 5.14 & -36.38 & 89.78 \\
Burst & 1 & -4.0 & -0.01 & 0.31 & -1.23 & 5.32 \\
 &  & -3.5 & -0.03 & 0.71 & -4.16 & 12.36 \\
 &  & -3.0 & -0.08 & 1.83 & -12.33 & 32.02 \\
 &  & -2.5 & -0.17 & 3.67 & -25.70 & 64.24 \\
 &  & -2.0 & -0.22 & 4.75 & -33.58 & 83.17 \\
 &  & -1.5 & -0.23 & 5.06 & -35.81 & 88.47 \\
 &  & -1.0 & -0.23 & 5.09 & -36.06 & 89.14 \\
Burst & 2 & -4.0 & -0.01 & 0.31 & -1.23 & 5.31 \\
 &  & -3.5 & -0.03 & 0.75 & -4.48 & 13.12 \\
 &  & -3.0 & -0.09 & 1.96 & -13.25 & 34.28 \\
 &  & -2.5 & -0.17 & 3.80 & -26.68 & 66.63 \\
 &  & -2.0 & -0.21 & 4.63 & -32.69 & 80.97 \\
 &  & -1.5 & -0.22 & 4.79 & -33.81 & 83.56 \\
 &  & -1.0 & -0.22 & 4.79 & -33.81 & 83.62 \\
Burst & 3 & -4.0 & -0.01 & 0.25 & -0.79 & 4.22 \\
 &  & -3.5 & -0.03 & 0.75 & -4.44 & 13.04 \\
 &  & -3.0 & -0.08 & 1.77 & -11.91 & 31.07 \\
 &  & -2.5 & -0.13 & 2.82 & -19.55 & 49.41 \\
 &  & -2.0 & -0.18 & 3.92 & -27.54 & 68.61 \\
 &  & -1.5 & -0.20 & 4.43 & -31.18 & 77.27 \\
 &  & -1.0 & -0.21 & 4.59 & -32.39 & 80.21 \\
Burst & 4 & -4.0 & -0.02 & 0.39 & -1.81 & 6.66 \\
 &  & -3.5 & -0.05 & 1.11 & -7.04 & 19.26 \\
 &  & -3.0 & -0.13 & 2.78 & -19.23 & 48.63 \\
 &  & -2.5 & -0.19 & 4.27 & -30.04 & 74.46 \\
 &  & -2.0 & -0.24 & 5.34 & -37.71 & 92.78 \\
 &  & -1.5 & -0.27 & 5.98 & -42.41 & 104.08 \\
 &  & -1.0 & -0.29 & 6.27 & -44.51 & 109.22 \\
Burst & 5 & -4.0 & -0.03 & 0.61 & -3.44 & 10.55 \\
 &  & -3.5 & -0.08 & 1.68 & -11.21 & 29.32 \\
 &  & -3.0 & -0.14 & 3.05 & -21.21 & 53.44 \\
 &  & -2.5 & -0.19 & 4.24 & -29.85 & 74.18 \\
 &  & -2.0 & 0.00 & 0.03 & 0.25 & 2.05 \\
 &  & -1.5 & 0.00 & 0.03 & 0.25 & 2.04 \\
 &  & -1.0 & 0.00 & 0.03 & 0.25 & 2.04
\enddata
\tablecomments{Fits are of the form $y = ax^3 + bx^2 + cx + d$, where $y \equiv 12+\logOH$ (\Cloudy) and $x \equiv 12+\logOH$ (direct-\Te).}
\end{deluxetable}

\startlongtable
\begin{deluxetable}{ccccc}
\tabletypesize{\footnotesize}
\tablecaption{Polynomial fits to the oxygen abundance offset between \Cloudy and the PP04-N2 method.\label{tab:polyB}}
\tablehead{\colhead{SFH} & \colhead{Age (Myr)} & \colhead{\logU} & \colhead{$a$} & \colhead{$b$}}
\startdata
CSFR & 10 & -4.0 & 0.53 & 4.00 \\
 &  & -3.5 & 0.52 & 4.00 \\
 &  & -3.0 & 0.53 & 3.81 \\
 &  & -2.5 & 0.54 & 3.61 \\
 &  & -2.0 & 0.55 & 3.46 \\
 &  & -1.5 & 0.55 & 3.31 \\
 &  & -1.0 & 0.56 & 3.15 \\
Burst & 1 & -4.0 & 0.54 & 3.90 \\
 &  & -3.5 & 0.54 & 3.91 \\
 &  & -3.0 & 0.54 & 3.75 \\
 &  & -2.5 & 0.55 & 3.58 \\
 &  & -2.0 & 0.55 & 3.46 \\
 &  & -1.5 & 0.55 & 3.34 \\
 &  & -1.0 & 0.56 & 3.21 \\
Burst & 2 & -4.0 & 0.53 & 3.98 \\
 &  & -3.5 & 0.53 & 3.98 \\
 &  & -3.0 & 0.54 & 3.79 \\
 &  & -2.5 & 0.54 & 3.59 \\
 &  & -2.0 & 0.55 & 3.43 \\
 &  & -1.5 & 0.56 & 3.29 \\
 &  & -1.0 & 0.57 & 3.13 \\
Burst & 3 & -4.0 & 0.51 & 4.15 \\
 &  & -3.5 & 0.51 & 4.09 \\
 &  & -3.0 & 0.53 & 3.82 \\
 &  & -2.5 & 0.55 & 3.56 \\
 &  & -2.0 & 0.56 & 3.39 \\
 &  & -1.5 & 0.57 & 3.22 \\
 &  & -1.0 & 0.58 & 3.03 \\
Burst & 4 & -4.0 & 0.51 & 4.08 \\
 &  & -3.5 & 0.51 & 4.07 \\
 &  & -3.0 & 0.52 & 3.88 \\
 &  & -2.5 & 0.53 & 3.68 \\
 &  & -2.0 & 0.53 & 3.55 \\
 &  & -1.5 & 0.54 & 3.44 \\
 &  & -1.0 & 0.54 & 3.30 \\
Burst & 5 & -4.0 & 0.40 & 4.87 \\
 &  & -3.5 & 0.42 & 4.71 \\
 &  & -3.0 & 0.47 & 4.27 \\
 &  & -2.5 & 0.51 & 3.83 \\
 &  & -2.0 & 0.54 & 3.52 \\
 &  & -1.5 & 0.56 & 3.28 \\
 &  & -1.0 & 0.59 & 3.04
\enddata
\tablecomments{Fits are of the form $y = ax + b$, where $y \equiv 12+\logOH$ (\Cloudy) and $x \equiv 12+\logOH$ (PP04-N2).}
\end{deluxetable}

\begin{deluxetable}{ccccccc}
\tabletypesize{\footnotesize}
\tablecaption{Polynomial fits to the oxygen abundance offset between \Cloudy and the KK04-R23 method (upper and lower branches, respectively).\label{tab:polyC}}
\tablehead{\colhead{SFH} & \colhead{Age (Myr)} & \colhead{logU} & \colhead{a} & \colhead{b} & \colhead{c} & \colhead{d}}
\startdata
CSFR & 10 & -4.0 & -0.13 & 2.93 & -20.80 & 54.22 \\
 &  & -3.5 & -0.23 & 5.33 & -39.97 & 105.36 \\
 &  & -3.0 & -0.41 & 9.44 & -71.22 & 184.30 \\
 &  & -2.5 & -0.61 & 13.92 & -104.75 & 267.70 \\
 &  & -2.0 & -0.78 & 17.70 & -132.93 & 337.41 \\
 &  & -1.5 & -0.91 & 20.74 & -155.59 & 393.48 \\
 &  & -1.0 & -1.04 & 23.46 & -175.90 & 443.74 \\
 \hline
CSFR & 10 & -4.0 & -1.76 & 47.32 & -421.84 & 1259.10 \\
 &  & -3.5 & -1.39 & 37.30 & -331.78 & 989.16 \\
 &  & -3.0 & 0.03 & -0.42 & 0.75 & 12.31 \\
 &  & -2.5 & 2.05 & -53.56 & 467.45 & -1353.25 \\
 &  & -2.0 & 3.87 & -101.48 & 886.79 & -2575.71 \\
 &  & -1.5 & 5.45 & -142.88 & 1248.10 & -3626.04 \\
 &  & -1.0 & 6.85 & -179.54 & 1567.75 & -4554.40
\enddata
\tablecomments{Fits are of the form $y = ax^3 + bx^2 + cx + d$, where $y \equiv 12+\logOH$ (\Cloudy) and $x \equiv 12+\logOH$ (KK04-R23).}
\end{deluxetable}

\section{Comparison of UV and optical direct-\Te abundances}\label{appdx:UVdirectTe}

We have already discussed the established practice for determining direct-\Te abundances using emission lines in the optical part of the spectrum. Here, we briefly explore comparable measurements using emission lines in the UV.

The direct-temperature method relies on the use of collisionally excited transitions with a difference in energy of order ${\sim}k T_{e}$. The two transitions are populated by electron-ion collisions, and the relative population in each energy level will thus depend directly on the electron temperature of the gas. In the optical, the temperature leverage comes from the combination of the ${}^1 \mathrm{S}_0$ to ${}^1 \mathrm{D}_2$ transition ($\lambda$4363) and the  ${}^1 \mathrm{D}_2$ to ${}^3 \mathrm{P}_2$ ($\lambda$5007) transition (often with the addition of the ${}^1\mathrm{D}_2$ to $^3\mathrm{P}_1$ transition at $\lambda$4959).

Similar transitions exist in the UV wavelength regime, making it possible to derive direct method metallicities using only UV emission lines. For the O$^{++}$ zone, the temperature leverage comes from the combination of the ${}^5 \mathrm{S}_2 \rightarrow {}^3 \mathrm{P}_2$ ($\lambda$1666) and the ${}^1 \mathrm{S}_0 \rightarrow {}^3 \mathrm{P}_1$ ($\lambda$2321) transitions. For the O$^{+}$ zone, the temperature-sensitive transitions are the ${}^2 \mathrm{P}_{1/2} \rightarrow {}^4 \mathrm{S}_{3/2}$ ($\lambda$2471) and the ${}^2 \mathrm{D}_{3/2} \rightarrow {}^4 \mathrm{S}_{3/2}$ ($\lambda$3727) transitions. It is worth noting that the the \oii$\,\lambda2471$ and \oiii$\,\lambda2321$ lines are very weak and have only been detected in a handful of galaxies; \oii$\,\lambda2471$ is ${\sim}6$\% of the \oii$\lambda$3727 line strength while \oiii$\,\lambda2321$ is ${\sim}2$\% of the \oiii$\lambda$4959 line strength.

To follow the methodology used in the optical, we must find temperatures and densities for the O$^{+}$ and O$^{++}$ zones. While there are a number of density-sensitive line combinations in the UV, the brightest line combinations (\ciii$\lambda$1906,1909 and \SiuIII$\lambda$1883,1892) are not sensitive to densities below $\sim10^3\,$cm$^{-3}$ \citep{Berg+2019}. We instead assume a constant $n_{\mathrm{e}} = 100$ cm$^{-3}$, which is the density used in the \Cloudy models\footnote{The use of densities calculated from \ciii{} and \SiuIII{} ratios changes the resultant oxygen abundances by less than 0.1 dex, and is thus not a significant source of error.}.

The temperature of the O$^{+}$ zone is calculated using the (\oii$\,\lambda$2471) / (\oii$\,\lambda 3727$ + \oii$\,\lambda 3729$) ratio. The ionic abundance for O$^{+}$ is calculated with {\tt PyNeb} using:
\begin{equation}
    \left[ \frac{\mathrm{O}^{++}}{\mathrm{H}^{+}} \right] = \frac{\mathrm{I}\,\lambda 2471 +\mathrm{I}\,\lambda 3727 + \mathrm{I}\,\lambda 3729}{\mathrm{I}\,\mathrm{H}\beta} \cdot \frac{j_{\mathrm{H}\beta}}{j_{\lambda}}.
\end{equation}

The temperature of the O$^{++}$ zone is calculated using \oiii$\,\lambda1666$ / \oiii$\,\lambda2321$ ratio. The total ionic abundance for O$^{++}$ is calculated with {\tt PyNeb} using:
\begin{equation}
    \left[ \frac{\mathrm{O}^{+}}{\mathrm{H}^{+}} \right] = \frac{\mathrm{I}\,\lambda 1666 +\mathrm{I}\,\lambda 2321}{\mathrm{I}\,\mathrm{H}\beta} \cdot \frac{j_{\mathrm{H}\beta}}{j_{\lambda}}.
\end{equation}

The total oxygen abundance is then calculated as the sum of the ionic abundances:
\begin{equation}
    \log_{10}(\mathrm{O}/\mathrm{H}) = \left[ \frac{\mathrm{O}^{++}}{\mathrm{H}^{+}} \right] + \left[ \frac{\mathrm{O}^{+}}{\mathrm{H}^{+}} \right],
\end{equation}
again, assuming a negligible contribution from O$^{+++}$ ions.

\begin{figure}
  \begin{center}
    \includegraphics[width=\linewidth]{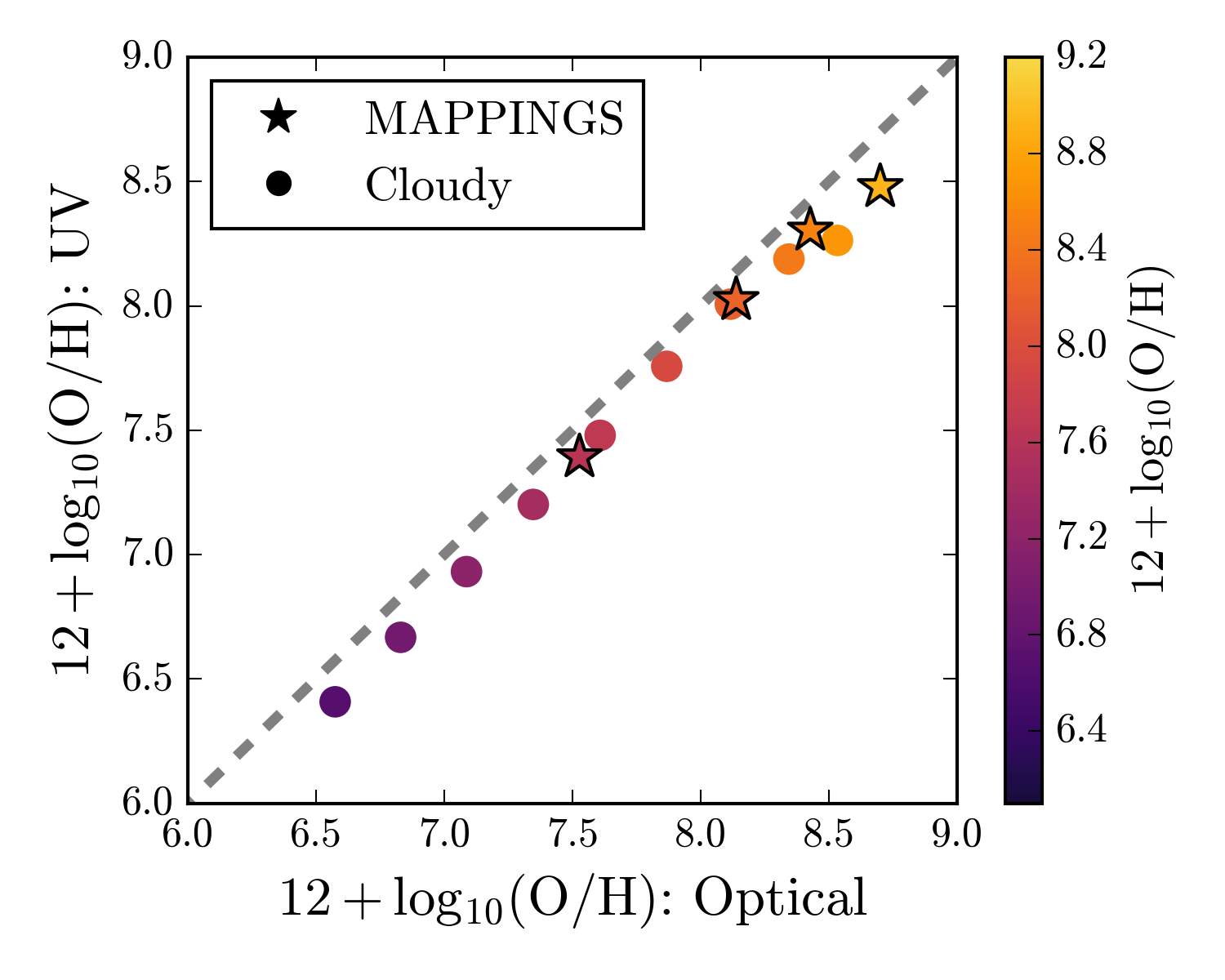}
    \caption{Comparison of direct-\Te metallicities using optical emission lines ($x$-axis) and UV emission lines ($y$-axis). Circular markers show predictions from the \Cloudy photoionization model used in this work, while star-shaped markers show predictions from the {\sc Mappings} photoionization code \citep{Kewley+2019}. Color indicates the input gas-phase oxygen abundance and the black dashed line shows a one-to-one relationship. Direct-\Te abundances using UV and optical emission lines are correlated, but the UV-derived abundances are systematically lower than the optically-derived abundances by $\sim0.2$\,dex.}
    \label{fig:UVoptZ}
  \end{center}
\end{figure}

In Fig.~\ref{fig:UVoptZ}, we show direct-temperature metallicities from optical emission lines ($x$-axis) and UV emission lines ($y$-axis) from two different photoionization models, \Cloudy \citep[circles;][]{Byler+2018} and {\sc Mappings} \citep[squares;][]{Kewley+2019}. The UV and optical abundances are correlated, and the two different photoionization models produce very similar predictions for both UV and optical metallicities. However, the UV direct-\Te abundances are systematically lower than the optical direct-\Te abundances by $\sim0.2$\,dex. This offset is likely driven by temperature differences in the UV and optical direct-\Te calculation; the oxygen lines in the UV are higher energy transitions and are produced in the inner regions of the nebula where temperatures are higher.

We note that the \oii$\,\lambda$2471 and \oiii$\,\lambda2321$ lines are quite weak and difficult to observe, and thus large samples of direct-method metallicities in the UV are unlikely. Unfortunately, the weak \oii$\,\lambda$2471 and \oiii$\,\lambda2321$ lines are the only oxygen-based temperature anchors in the UV. For objects with both rest-UV and rest-optical spectroscopy, it is common to combine the \oiii$\,\lambda$1661,1666 with the optical \oiii$\,\lambda$5007,4959 lines for temperature determinations. Combining UV and Optical emission lines introduces additional uncertainties, with flux calibration and aperture matching between UV and optical observations.

\end{document}